\documentclass[a4paper,fleqn,usenatbib]{mnras}

\usepackage{newtxtext,newtxmath}

\usepackage[T1]{fontenc}
\usepackage{ae,aecompl}

\usepackage{graphicx}	
\usepackage{amsmath}	
\usepackage{amssymb}	

\DeclareMathOperator{\atan}{atan}

\title[A study of spiral structure in the optical]{A multi-wavelength study of spiral structure in galaxies. I. General characteristics in the optical}

\author[S. Savchenko et al.]{
Sergey Savchenko,$^{1,3}$\thanks{E-mail: s.s.savchenko@spbu.ru}
Alexander Marchuk,$^{1}$
Aleksandr Mosenkov,$^{2}$
\newauthor
Konstantin Grishunin$^{1}$
\\
$^{1}$Saint Petersburg State University, Department of Astrophysics, St. Petersburg 198504, Russia\\
$^{2}$Pulkovo Observatory of the Russian Academy of Sciences, St. Petersburg 196140, Russia\\
$^{3}$Special Astrophysical Observatory, Russian Academy of Sciences, Nizhnii Arkhyz, 369167 Russia\\
}

\date{Accepted XXX. Received YYY; in original form ZZZ}

\pubyear{2019}

\begin{document}
\label{firstpage}
\pagerange{\pageref{firstpage}--\pageref{lastpage}}
\maketitle

\begin{abstract}
Different spiral generation mechanisms are expected to produce different morphological and kinematic features.
In this first paper in a series we carefully study the parameters of spiral structure in 155 face-on spiral galaxies, selected from the Sloan Digital Sky Survey, in the three $gri$ bands. We use a method for deriving a set of parameters of spiral structure, such as the width of the spiral arms, their fraction to the total galaxy luminosity and their colour, which have not been properly studied before. Our method is based on an analysis of a set of photometric cuts perpendicular to the direction of a spiral arm.
Based on the results of our study, we compare the main three classes of spirals: grand design, multi-armed and flocculent. We conclude that:
i) for the vast majority of galaxies (86\%) we observe an increase of their arm width with galactocentric distance;
ii) more luminous spirals in grand design galaxies exhibit smaller variations of the pitch angle with radius than less luminous grand design spirals;
iii) grand design galaxies show less difference between the pitch angles of individual arms than multi-armed galaxies.
Apart from these distinctive features, all three spiral classes do not differ significantly by their pitch angle, arm width, width asymmetry, and environment. Wavelength dependence is found only for the arm fraction. Therefore, observationally we find no strong difference (except for the view and number of arms) between grand design, multi-armed and flocculent spirals in the sample galaxies.
\end{abstract}

\begin{keywords}
galaxies: spiral -- galaxies: structure -- methods: data analysis
\end{keywords}



\section{Introduction}
\label{sec:intro}
Spiral arms of discoidal galaxies are remarkable structures with regions of ongoing star formation \citep[see e.g.][ and references therein]{2005ApJ...633..871C,2011EAS....51...19E,2012A&A...542A..39G} which are embedded into a smooth stellar disc. There are different tracers of spiral structure in galaxies (including our own Milky Way), such as H{\sc ii} regions \citep{1976A&A....49...57G,2009A&A...499..473H,2015ApJ...800...53H}, OB associations \citep{1993AJ....105..499R}, population I Cepheids \citep{2015AstL...41..489D,2019Sci...365..478S}, giant molecular clouds \citep{1986ApJS...60..695C,1988Natur.334..402V,1990A&A...232L..11W}, enhanced gas density \citep{1987ApJ...315..122G,2003ApJS..149..343E,2017A&A...603A.113S}, and dust clouds \citep{2005A&A...444..109H}.

The study of spiral galaxies is of great importance as these galaxies present a significant part (about 75\% of galaxies brighter than $M_B=-20$) of the local Universe (\citealp{2006MNRAS.373.1389C}). Spiral structure is thus an almost ubiquitous feature in the present disc galaxies and can also be detected in many distant galaxies up to redshift $z\sim 2$ \citep[see e.g.][]{2005ApJ...631...85E,2012Natur.487..338L,2015llg..book..455E}. Studying the properties of spiral structure, along with other galaxy characteristics, enables astronomers to establish a theory to explain this structure and use simulations to reproduce these quantified properties and, thus, check (confirm or reject) and refine the theory proposed.

The simple classification of spirals includes three types (we will call them classes to distinguish from galaxy morphological types, see \citealt{1990NYASA.596...40E}): flocculent spiral galaxies (with many short arms, such
as NGC\,2841), multi-armed spirals (e.g. M\,33), and grand design galaxies (with two main spiral arms, e.g. M\,81). Each class, as suggested, should have its own dominant mechanism which produces the observed structure (see below).

After more than half a century of research, the physical explanation of observed spiral arms in galaxies is still debated \citep[for a concise description of all theories, see the general review by][]{2014PASA...31...35D}.
There are several theories that attempt to explain spiral structure in disc galaxies. The dominant quasi-stationary density wave theory \citep{1964ApJ...140..646L,1989ApJ...338...78B,1973PASAu...2..174K,1989ApJ...338..104B} is able to successfully explain many observed properties of spiral structures. The swing amplification theory  \citep{1966ApJ...146..810J,1978ApJ...223..129G,1978ApJ...222..850G,1979ApJ...233...56S,1981seng.proc..111T} implies that spiral arms are transient but recurrent structures due to local instabilities, perturbations or noise which are swing amplified into flocculent spiral arms. In the Manifold theory \citep{1996A&A...309..381K,2009MNRAS.394.1605H,2009MNRAS.394...67A,2009MNRAS.400.1706A,2010MNRAS.407.1433A,2012MNRAS.426L..46A}, spiral structure is the result of stars formed near the ends of a galactic bar moving into chaotic, highly eccentric orbits which nevertheless cause the stars to move along relatively narrow tubes called manifolds (bar driven arms). Tidal interactions were shown to produce two-armed spiral galaxies \citep{1941ApJ....94..385H,1969ApJ...158..899T,1972ApJ...178..623T,1991A&A...244...52E}. Each of these mechanisms has significant support, at least for certain types of spiral galaxies. It has even been suggested that different classes (for instance, grand design versus flocculent) have different mechanisms explaining their spiral structure. It is quite likely that the mechanism differs from galaxy to galaxy and, in some cases, is represented as a combination of different models.

The overall appearance of the spiral structure may differ in different spiral galaxies proving the measurement of its parameters to be a hard task.
In recent years, approaches for investigating galactic spirals have changed dramatically since observational data is now being mostly obtained via all-sky survey missions.
Being a source of uniform observational data, the surveys allow for studying galactic morphology by assembling statistically significant samples.

Usually, in the literature only a few geometric parameters, which describe the spiral pattern, are considered: the number of spiral arms \citep{2005AJ....130..569V,2016MNRAS.461.3663H}, the pitch angle \citep[see e.g.][]{1981AJ.....86.1847K,2002A&A...388..389M,1998MNRAS.299..685S,2011MNRAS.414..538K,2013MNRAS.436.1074S,2019MNRAS.487.1808M,2019ApJ...871..194Y}, and the relative amplitude of the spiral pattern \citep{2008MNRAS.387.1007K}.

The number of arms $N_\mathrm{arms}$ is an important characteristic of spiral pattern which points to a possible driving mechanism of spiral structure \citep{2016MNRAS.461.3663H,2017MNRAS.468.1850H} and is linked to star formation
 \citep{2010MNRAS.405..783M}.

The pitch angle $\psi$ is defined as the angle between the tangent of the spiral and azimuthal direction \citep{2008gady.book.....B} and is a measure of the tightness of spiral structure. A number of
properties of spiral galaxies have now been found to correlate to the pitch angle. It was found that this parameter correlates with the galaxy morphology \citep{1981AJ.....86.1847K}, though this trend is not as good as expected: one of the criteria for galaxy classification is the tightness of its spiral pattern \citep{1936rene.book.....H}, therefore this correlation should arise automatically (see discussion in \citealp{2019MNRAS.487.1808M}). Other correlations, which are often discussed in the literature, are the dependence of the pitch angle on the bulge size \citep{1970ApJ...160..811F}, the rotation properties of galaxies \citep{1981AJ.....86.1847K,2005MNRAS.359.1065S,2006ApJ...645.1012S}, the velocity dispersion in the central region of the galaxy \citep{2019ApJ...871..194Y}, and the mass of the central supermassive black hole \citep{2008ApJ...678L..93S,2017MNRAS.471.2187D}. The results of these studies are often contradictory, and our understanding of spiral structure in galaxies is still incomplete.

The amplitude of spiral pattern is often expressed in terms of an amplitude of Fourier modes fitted into an azimuthal profile of a galaxy \citep{2008MNRAS.387.1007K,
2011MNRAS.414..538K, 2018ApJ...862...13Y}, or using direct measurements of maximal and minimal flux values along an azimuthal profile \citep{2011ApJ...737...32E}.
This parameter is found to be a good estimator of the galaxy type: galaxies of later Elmegreen classes \citep[][see Sect.~\ref{sec:classification}]{1987ApJ...314....3E} tend to have
a spiral structure with a larger amplitude. Also, more massive galaxies harbour more prominent spirals \citep{2015MNRAS.446.4155K}.

Another parameter, which defines a spiral arm and is hard to determine, is its width $w$.
The width, in contrast to the multiple studies on pitch angle, had not been paid much attention to until recently.
It has been investigated by \citet{2014ApJ...783..130R} for the Milky Way and by \citet{2015ApJ...800...53H} for M\,51, M\,74, NGC\,1232, and NGC\,3184, using trigonometric parallaxes of masers
and positions of H{\sc ii} regions, respectively. In both works it has been shown that the width increases with distance from the galactic center. A similar behaviour was noted
 by \citet{1970IAUS...38...26L} for the width of dust lanes in spirals of ten nearby galaxies. \citet{2018MNRAS.476.2384F} describe a method, which identifies spirals in
hydrodynamic simulations of discs, and allows one to measure the widths of individual arms.

So far, the width for a statistically significant sample of appropriate (observed face-on and large enough to explore) spiral galaxies has not been
investigated. As such, a detailed study of the overall spiral structure, described by the above listed parameters, appears lacking in the literature.

This is the first paper in a series where we perform a detailed study of spiral pattern for a relatively large sample of spiral galaxies by utilising observational data from modern sky surveys in a broad wavelength range from the ultraviolet (UV) to far-infrared (FIR). We aim to systematically study the properties of the spiral structure (pitch angle, arm width, number of arms, class of spiral structure, etc.), their relation to the general galaxy quantities, as well as structural composition, in dependence on wavelength. This will allow us to study, for the first time, the multi-band photometry of spiral galaxies by taking into account their main structural components (bulge, disc, bar) and asymmetric spiral pattern, and, most importantly, their interplay. The results of this study should point to possible different mechanisms which are responsible for the observed spiral pattern in galaxies.

In this first study, we investigate in the optical the general properties of spiral structure in galaxies:  the pitch angle and its variation with radius, the width of the spirals and its dependence on radius, their contrast in comparison with the overall disc component (what is the fraction of the spiral structure to the total galaxy luminosity in the optical?), and colours of the spiral arms. Also, the dependence of these parameters on the general galaxy properties is investigated.
To accomplish this study, we use a rather straightforward method for measuring the mentioned quantities of spiral structure and apply it to a relatively large sample of spiral galaxies.

The paper is organised as follows. We outline the sample selection, data preparation, and general properties of the sample in Sect.~\ref{sec:sel_sample}. The method, developed to derive parameters of spiral structure, is described in Sect.~\ref{sec:method}. In Sect.~\ref{sec:results}, we provide results of our analysis and discuss them in Sect.~\ref{sec:discussion}. We summarise our main conclusions in Sect.~\ref{sec:conclusion}.

\section{The data}
\label{sec:sel_sample}
In this section we describe a selection process of spiral galaxies to be analysed in our work. Also, we present details on image preparation for the selected galaxies. We classify their spiral pattern and briefly comment on the general properties of the selected sample.

\subsection{Selection of the sample}
\label{sec:sel_of_sample}

To create a sample of face-on spiral galaxies whose angular size would be large enough for the purposes of this study, we make use of the following projects, based on the Sloan Digital Sky Survey \citep[SDSS,][]{2000AJ....120.1579Y}. First, from the GalaxyZoo sample\footnote{Table 2, \url{https://data.galaxyzoo.org/}} \citep{2008MNRAS.389.1179L}, we selected objects which fulfil the following criteria based on the fraction of user votes: \texttt{P\_CW}~$\geq0.9$ (clockwise spirals) \texttt{ OR P\_ACW}~$\geq0.9$ (anticlockwise spiral) \texttt{AND P\_EL}~$\leq0.1$ (non-elliptical galaxies) \texttt{AND P\_EDGE}~$\leq0.1$ (non-edge-on galaxies). This query yielded 19102 galaxies, the vast majority of which are too tiny for the purposes of this paper.
Therefore, all selected galaxies were processed via the HyperLeda database \citep{2014A&A...570A..13M} to identify their names and photometric parameters. From this preliminary sample, we chose galaxies with an optical diameter (calculated from the $logd25$ value from HyperLeda, as the diameter at the isophote 25\,mag\,arcsec$^{-2}$ in the $B$ band) larger than 50\arcsec. The choice of a 50$\arcsec$ diameter is arbitrary -- it was chosen to be large enough to provide a statistically significant number of galaxies of all types and, at the same time, provide a sufficient resolution for measuring the width of the spiral arms in galaxies: as we show in Sect.~\ref{sec:armwidth}, a galaxy with an optical diameter of 50$\arcsec$ has, on average, an arm width of $3.5\arcsec$, which is 2.7 times larger than the average FWHM $\approx 1.3\arcsec$ of the point spread function (PSF) in the SDSS $r$ band (see Sect.~\ref{sec:data_prep}). However, taking into account that the arm width is usually growing outwards from the center (see Sect.~\ref{sec:armwidth}), it may have a somewhat smaller width at the beginning of the arm. Therefore, to make sure that the image resolution is sufficient for measuring the arm width at virtually all radii, we selected galaxies with HyperLeda diameters larger than $50\arcsec$. This yielded 1611 galaxies in total.

After that, we additionally revisited the EFIGI \citep{2011A&A...532A..74B} catalogue searching for large (nearby) spiral galaxies using the following criteria: $1\leq T \leq8$  \texttt{AND Arm\_Strength}~$\geq 0.25$ \texttt{AND Arm\_Curvature}~$\geq 0.5$ \texttt{AND diam}~$ > 50\arcsec$ \texttt{AND ax\_ratio}~$ < 1.3$. This added another 376 objects to our preliminary sample.
The joined sample was examined by eye 3 times to remove highly inclined galaxies, galaxies contaminated by foreground stars, overlapping galaxies, or strongly interacting galaxies. Also, in our final sample we selected only those galaxies for which at least one spiral arm can be traced (i.e. the spiral pattern is not too flocculent, not wound up in a ring and not too weak to be analysed by our method, see Sect.~\ref{sec:method_outline}).
Finally, we settled  on a sample of 155 (85 galaxies are found in the EFIGI sample) spiral galaxies which are appropriate for a further analysis (we should note, however, that we started with an initial sample of approximately 200 galaxies, but some of them were then removed because their spiral arms were too faint or hard to be analysed or classified). The final sample is listed in Table~\ref{general_pars}\footnote{The whole table is available online.}, along with some general parameters which are described in Sect.~\ref{sec:classification} and~\ref{sec:gen_properties}.

The sample suffers from some obvious selection effects which we discuss in detail in Sect.~\ref{sec:discussion}.

\subsection{Data preparation}
\label{sec:data_prep}

For our photometric analysis of galaxies, we use imaging from the SDSS DR13 \citep{2017ApJS..233...25A} in the three photometric bands $g$, $r$, and $i$. The corrected frames in each band were retrieved from the SDSS Science Archive Server (SAS) using a specially developed Python script\footnote{\url{https://github.com/latrop/sdss_downloader}}. The mosaics out of several SDSS fields (if a galaxy covers more than one field) were prepared using the {\small SWARP} code \citep{2002ASPC..281..228B}. Also, all frames were converted from NMgy back to original ADU, in order to use the signal-to-noise, the CCD gain, and the initially subtracted sky value $I_\mathrm{sky,ini}$ (which can be found in the supplementary SDSS tables) in our further fitting.

Although the downloaded frames have been bias subtracted, flat-fielded, and sky-subtracted, we re-estimated the sky background around each galaxy using an initial mask of all detected objects in each frame provided by \textsc{sextractor} \citep{1996A&AS..117..393B}. The unmasked area was fitted with a 2nd polynomial (with the average value $I_\mathrm{sky,re}$) and subtracted from the original frame. The final sky-subtracted value is thus $I_\mathrm{sky} = I_\mathrm{sky,ini}+I_\mathrm{sky,re}$, with the standard deviation $\sigma_\mathrm{sky}$. Also, we cut out the original image: the final galaxy image should encompass the isophote at the $2 \sigma_\mathrm{sky}$ level enlarged by a factor of 1.5. The initial mask of contaminating objects was revisited by eye to mask objects which had not been detected by \textsc{sextractor} or  were masked erroneously\footnote{All these steps were performed using the special Python package \url{https://github.com/latrop/pipeline}}.

To create a PSF image for each galaxy frame, we used the {\small PSFEX} package \citep{2011ASPC..442..435B}, which searches for isolated non-saturated field stars near the target
galaxy to compute a combined PSF image. As a rule, there were 10--20 such PSF stars, which allowed us to reliably determine an average PSF FWHM and to estimate its
dispersion $\sigma_\mathrm{FWHM}$. The average values of the FWHM for each band are: $1.41\pm0.16\arcsec$ (the $g$ band), $1.35\pm0.16\arcsec$ (the $r$ band), and $1.33\pm0.14\arcsec$ (the $i$ band).

For applying our fitting method of spiral structure, which we describe in detail in Sect.~\ref{sec:method}, we should correct the images for the projection effect. This is also important for the classification of the spiral arms (see Sect.~\ref{sec:classification}). Galaxy discs may
have an arbitrary orientation in space, which is described by the inclination $\mathfrak{i}$ and position $\mathrm{PA}$ angles.
To correct an image for the projection effect, it is necessary to rotate it by a value of $\mathrm{PA}$, so
that the major axis of the disc orients along the $x$-axis of the image, and then stretch the image along the $y$-axis
by the value of the major-to-minor axes ratio $1/q$ of the best-fit elliptical isophote. To derive the values $q$ and $\mathrm{PA}$ (as well as some other general galaxy parameters, see Sect.~\ref{sec:gen_properties}),
we used an iterative isophote fitting procedure \citep{1987MNRAS.226..747J}. Since the shape of the isophotes of spiral galaxies can be influenced by non-axisymmetric
inner components, such as bars or spirals, it is preferable to use the outermost isophotes for our de-projection analysis.
In this work, we use median values of $q$ and $\mathrm{PA}$ for ten isophotes equally spaced between the 24-th and
25-th isophotes in the $i$ band (in this band the impact of the spiral pattern on the parameters of the outermost isophotes
is minimal among the $g$, $r$ and $i$ bands under consideration). We denote this median value of $q$ as $q_{25}$.

\subsection{General properties of the sample galaxies}
\label{sec:gen_properties}

\begin{figure*}
\centering
\includegraphics[width=5.5cm, angle=0, clip=]{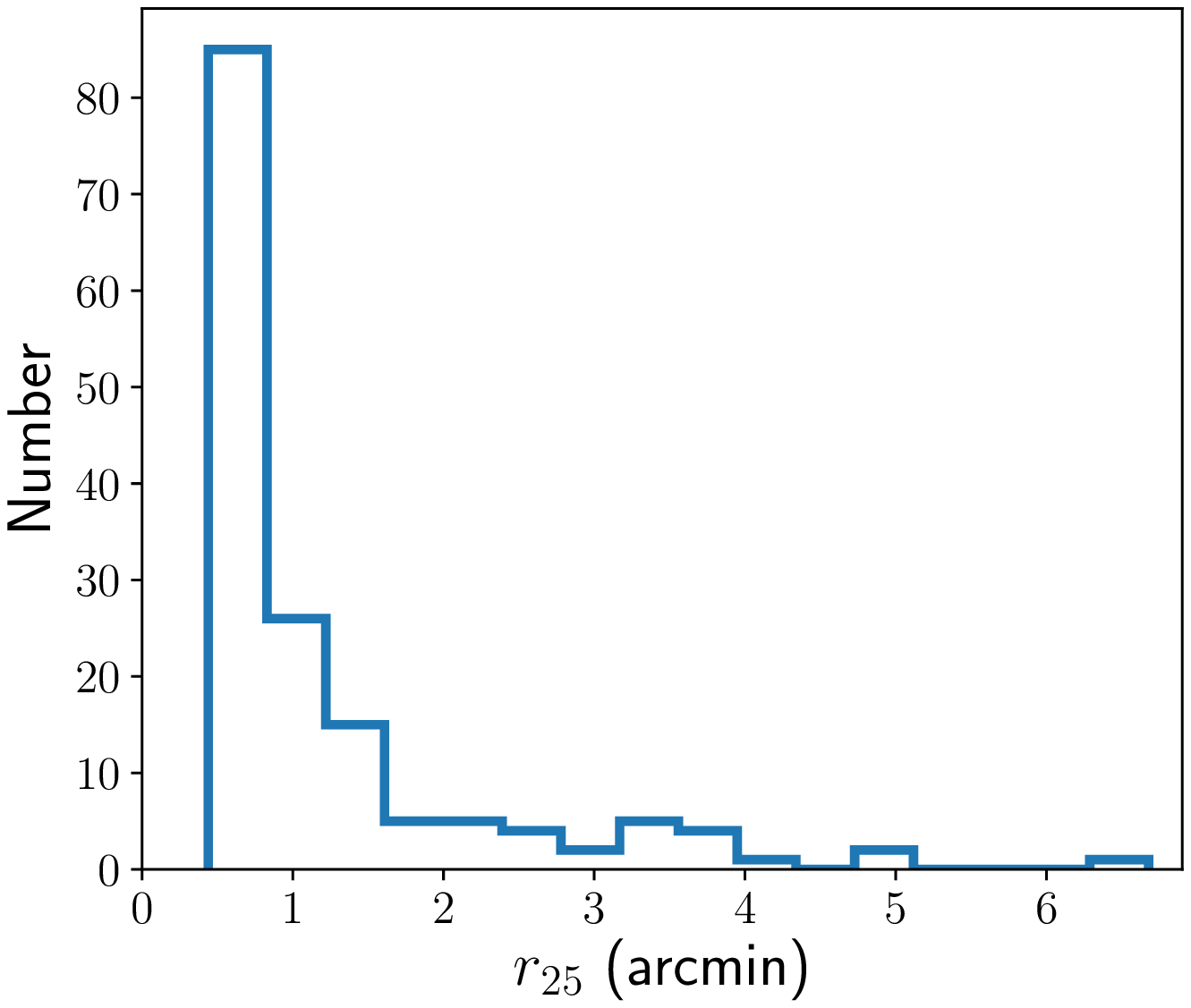}
\includegraphics[width=5.55cm, angle=0, clip=]{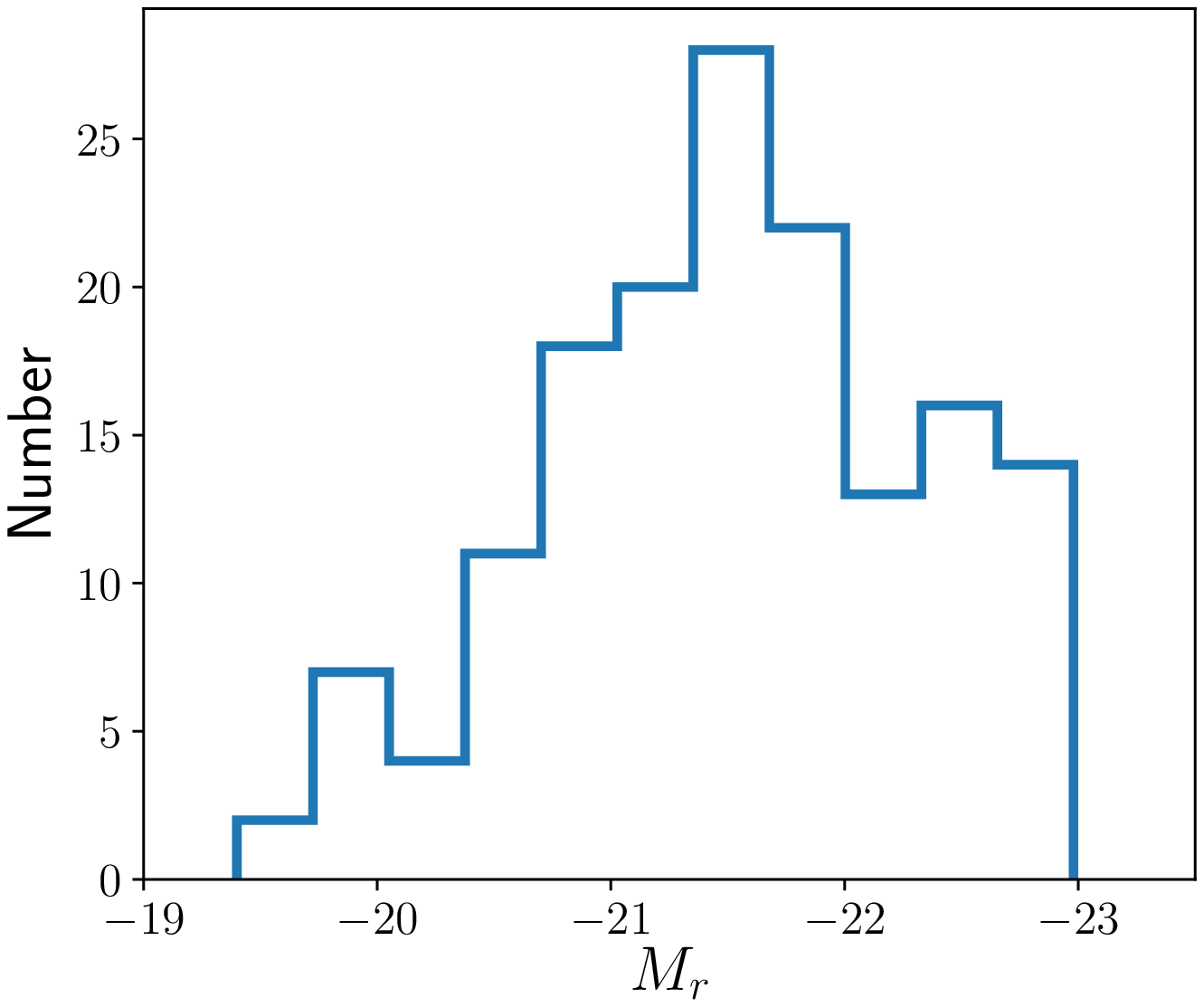}
\includegraphics[width=5.6cm, angle=0, clip=]{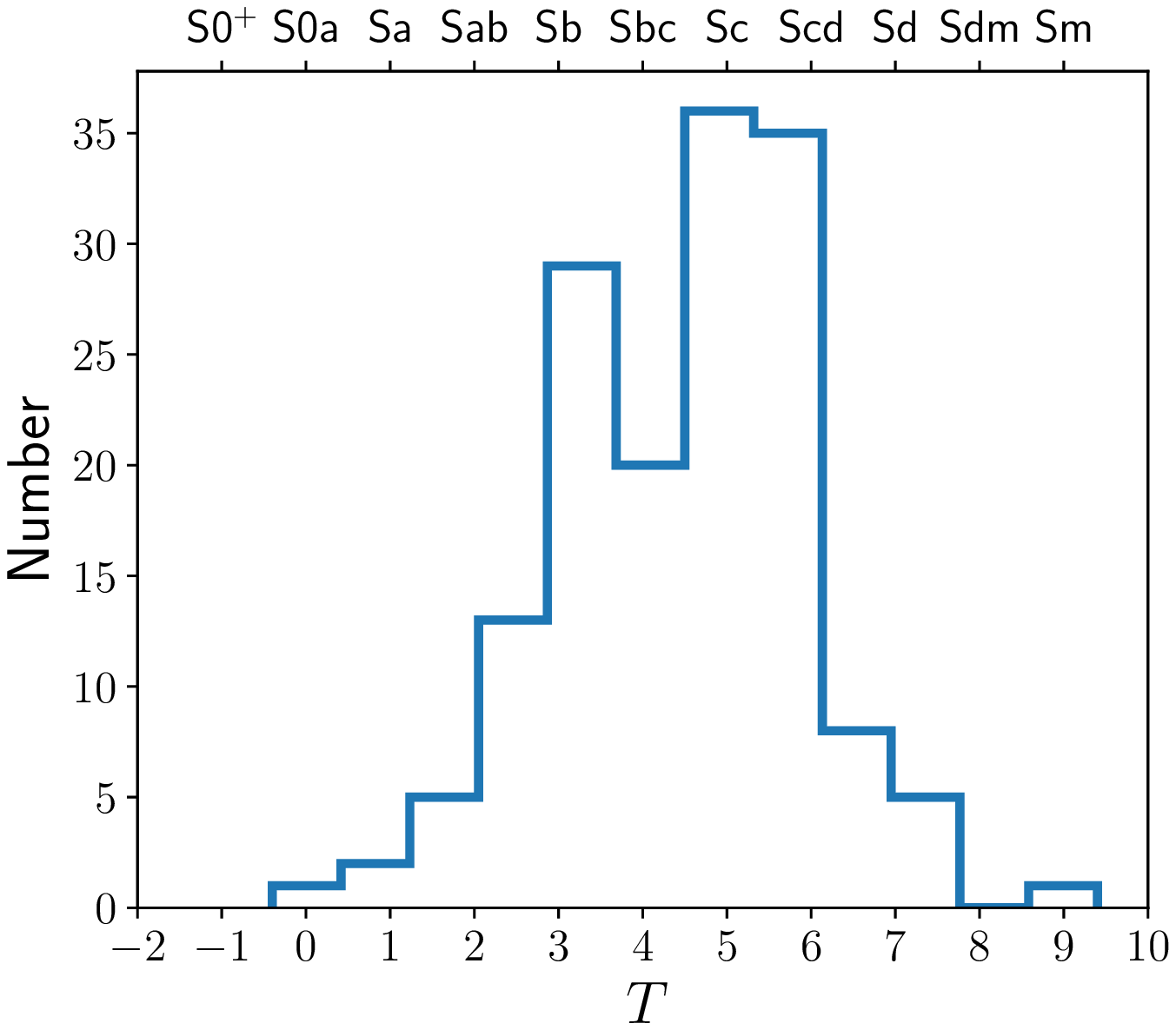}
\caption{Distribution of the sample galaxies as a function of optical radius (leftmost panel), luminosity (middle panel), and numerical (Hubble) type (rightmost panel).}
\label{general_distr}
\end{figure*}

Our sample includes nearby spiral galaxies, such as PGC\,5974 (M\,74), and rather distant galaxies, up to $D=290$~Mpc for PGC\,53625 ($z=0.067$). The mean distance for the sample is $90.0\pm57.7$~Mpc. In Fig.~\ref{general_distr} we show the distribution of the sample galaxies by the optical radius $r_{25}$ in the $r$ band (leftmost panel), which we estimated using the isophote fitting (see Sect.~\ref{sec:data_prep}). As one can see, most galaxies have a radius less than 1.5\arcmin  (the mean value $\langle r_{25} \rangle = 1.2\pm1.0\arcmin$).

The selected galaxies are mostly viewed face-on (the mean apparent flattening derived in Sect.~\ref{sec:data_prep} $\langle q_{25} \rangle = 0.82\pm0.13$), so only minor de-projection to the true face-on orientation was needed.

To estimate the total magnitude $m$, we created a growth curve for each galaxy image based on our isophote fitting results. Asymptotic magnitudes in each band were estimated by extrapolating the dependence of the gradient $\mathrm{d}m/\mathrm{d}r$ on magnitude $m$ to $\mathrm{d}m/\mathrm{d}r=0$ (by definition, this gives the asymptotic magnitude of the galaxy, see e.g. \citealt{2015ApJS..219....3M}). From the total magnitudes we computed the absolute magnitudes in each waveband and colours $g-r$ and $r-i$, each corrected for the Galactic extinction \citep{2011ApJ...737..103S} taken from the NASA/IPAC Extragalactic Database (NED)\footnote{\url{http://ned.ipac.caltech.edu/}}. The distribution of the sample galaxies as a function of luminosity is shown in the middle panel of Fig.~\ref{general_distr}. The absolute magnitude of our sample ranges from -19 to -23 in the $r$ band, with the average $-21.5\pm0.8$ --- in our sample we consider regular spiral galaxies with a typical luminosity \citep[see e.g.][]{2011A&A...532A..75D}.

The right panel of Fig.~\ref{general_distr} shows the distribution of our sample galaxies by the numerical Hubble type according to the HyperLeda database.
As one can see, most numerous galaxies in our sample are of types Sb--Sc ($\langle T \rangle=3.8\pm1.6$), with some fraction of lenticular galaxies with a faint yet detectable spiral structure and a few Sa galaxies with tightly wound spiral arms.
According to the visual inspection of the galaxy images, in our sample we have 60 unbarred galaxies versus 95 galaxies with a bar.

Numerical simulations show (for example, \citealt{2000MNRAS.319..377S}, \citealt{2010MNRAS.403..625D})
that tidal interactions can be related to the formation of spiral structure in disc galaxies.
In order to investigate a possible relation between the properties of spiral structure and the spatial
environment of their hosts, we collected information on the spatial environment of galaxies from the NED database (specified as \textit{hierarchy}) and from \citet{2012A&A...540A.106T}.  All galaxies, which belong to a pair, a triple, a galaxy group or cluster, we classify as non-isolated, the rest galaxies -- as isolated.  In total, our sample comprises 45 isolated galaxies and 110 non-isolated galaxies.

\begin{table*}
 \centering
 \begin{minipage}{180mm}
  \centering
  \parbox[t]{150mm} {\caption{General characteristics of the sample galaxies. This table is published in its entirety in the electronic version of the MNRAS.}
  \label{general_pars}}
	\begin{tabular}{lcccccccccccc}
  \hline
  \hline
		Name & $D$    & Scale                & $r_{25}$ & $q_{25}$     & $m_r$ & $M_r$      & $(g-r)_0$& $(r-i)_0$ & $T$ & Env & Bar \\
                 &  (Mpc)& (kpc/arcsec)     & (arcmin)   &                    & (mag)  & (mag)       &     (mag)  &    (mag)   &        &       &        \\
           (1) &     (2) &          (3)          & (4)           &         (5)      &    (6)   &     (7)        &        (8)   &      (9)     & (10)&(11) & (12) \\
		\hline
PGC303 & 73.5 & 0.344 & 1.01 & 0.59 & 13.13 & -20.91 & 0.51 & 0.27 & 3.1 & g & --- \\
PGC2182 & 84.3 & 0.394 & 0.93 & 0.67 & 13.02 & -21.42 & 0.60 & 0.36 & 4.1 & p & + \\
PGC2440 & 84.3 & 0.394 & 0.57 & 0.81 & 13.51 & -20.91 & 0.49 & 0.33 & 3.2 & p? & --- \\
PGC2600 & 64.6 & 0.304 & 1.01 & 0.88 & 12.02 & -21.76 & 0.50 & 0.14 & 5.2 & i & --- \\
PGC2901 & 68.9 & 0.324 & 1.87 & 0.89 & 11.12 & -22.78 & 0.73 & 0.46 & 1.6 & g & + \\
...            & ...     & ...       &  ...    & ...      &  ...      & ...         & ...     & ...     & ...  & ... & ...        \\
  \hline\\
  \end{tabular}
\end{minipage}

   \parbox[t]{160mm}{ Columns: \\
   (1) PGC name from HyperLeda, \\
   (2), (3) distance and scale calculated from the best distance modulus from HyperLeda,  \\
   (4) the semi-major axis of the isophote 25\,mag\,arcsec$^{-2}$ in the $r$ band,  \\
   (5) median flattening between the 24-25\,mag\,arcsec$^{-2}$ isophotes in the $r$ band, \\
   (6) and (7) apparent asymptotic and absolute magnitudes in the $r$ band calculated from the growth curve, corrected for the Milky Way extinction \citep{2011ApJ...737..103S},  \\
   (8) and (9) colours calculated from the corresponding asymptotic magnitudes and corrected for the Milky Way extinction \citep{2011ApJ...737..103S},  \\
   (10) numerical morphological type from HyperLeda,\\
   (11) environment: G or g -- belongs to a group, P or p -- belongs to a pair, T or t -- belongs to a triple, I or i -- isolated galaxy; C or c -- belongs to a cluster. Capital and small letter mean that the information is taken from NED or \citet{2012A&A...540A.106T}, respectively. `?' after the letter means that classification is found neither in NED nor in \citet{2012A&A...540A.106T} and we classified the galaxy environment on our own using the SDSS SkyServer Navigate tool on the basis of visual closeness to the target galaxy and known spectroscopic redshift.     \\
   (12) Bar presence according to the visual inspection.}
\end{table*}

\subsection{Classification of spiral structure}
\label{sec:classification}

Galaxies can be very different in their morphological properties and thus are usually classified according to one or another schema. Since we investigate spiral arms, it is natural to use their features as a basis for our classification. The most well-known classification system for this purpose was introduced in \citet{1987ApJ...314....3E} and represents 10 classes (AC1-AC12, AC10 and AC11 are no longer in use). However, these classes can also be combined in three main classes \citep{2011ApJ...737...32E,2018MNRAS.481..566K}: flocculent (`F', Arm Classes AC1-AC3), multiple arm galaxies (`M', Arm Classes A4C-AC9), and grand design galaxies (`G', Arm Classes AC12).

Only 11 galaxies in our sample are found in the sample of \citet{1987ApJ...314....3E} and few in \citet{2017A&A...603A.113S} and \citet{2018MNRAS.481..566K}, where galaxies were already classified. Also, our sample includes 29 common galaxies with the S${^4}$G survey \citep{2010PASP..122.1397S,2015ApJS..217...32B}. As the number of sample galaxies with available arm classes is low, we decided to classify all of them ourselves using a major voting schema between all authors of this study, similar to that used in \citet{2017A&A...603A.113S} and \citet{2018MNRAS.481..566K}. For the voting evaluation, we used the images prepared in Sect.~\ref{sec:data_prep}. We decided to use a combined approach with the 3 major classes, since our sample is not that big and the usage of the original \citet{1987ApJ...314....3E} classification will lead to a situation, when most of the classes will be underpopulated. Each author of this paper independently assigned one of the F, M, G classes to the galaxy spiral pattern, based on the de-projected images and using the following criteria. Flocculent (F) galaxies demonstrate spiral arms which are chaotic, fragmented, asymmetric or uniformly distributed around the centre. Grand design (G) galaxies have two distinct long symmetric arms dominating the disc. The remaining multiple arm (M) class includes the following situations: the galaxy exhibits only one prominent arm (the others are flocculent), more than two prominent continuous arms, or two symmetric arms in the inner part and multiple irregular outer arms (plums), or vice versa. We consider the classification of a galaxy to be reliable if at least three votes have the same class.

For the 155 galaxies from the initial sample at least three authors voted for the same class and for approximately half of them all four authors were unanimous in their classification. Galaxies with unreliable classification mostly demonstrate transition between F and M classes and were removed from the initial sample of around 200 galaxies, see Sect.~\ref{sec:sel_of_sample}. The final sample comprises of 20 F, 100 M and 35 G galaxies.

It is difficult to directly compare our classification results to other works, since our sample is obviously biased to galaxies with distinct spiral pattern and thus should not represent any unbiased distribution. In \citet{2017A&A...603A.113S}, the authors studied a sample of spiral galaxies in the SDSS $g+u$ bands. They used two classes and reported 45 galaxies classified as flocculent and 18 as grand design galaxies. Our classification holds near the same ratio if we consider M-class galaxies closer to flocculent. In \citet{2015ApJS..217...32B}, a classification is given for 1114 S${^4}$G spiral galaxies in the $3.6\mu$\,m band, based on the same classification scheme as we use. From these galaxies, 29 are common with our sample. For half of them, \citet{2015ApJS..217...32B} listed the same class as we assign, but for the other 13 galaxies, their class is different. However, in all cases the reported classes are close i.e. G vs M or M vs F (the second combination is more often). This comparison supports, in some way, our classification, besides the fact that the sample in \citet{2015ApJS..217...32B} is more biased to the F class (50\% F, 32\% M, and 18\% G cases, but the ratio G:M+F is again close to ours) and besides the fact that some galaxies can demonstrate a very different view of the spiral pattern at different wavelengths (the S${^4}$G survey is a near-infrared survey, whereas here we consider the $gri$ bands). For example, NGC\,5055 and NGC\,2841 can be classified as F in the optical and G in the near-infrared \citep{1996ApJ...469L..45T}.

We should stress here that the flocculent galaxies in our sample have better-defined spiral arms because of the described in Sect.~\ref{sec:sel_of_sample} selection criteria than usually analysed in the literature. Therefore, our category of flocculent galaxies does not map one-to-one onto those used by other authors.

\begin{figure*}
\centering
\includegraphics[width=5.5cm, height=5.5cm, angle=0]{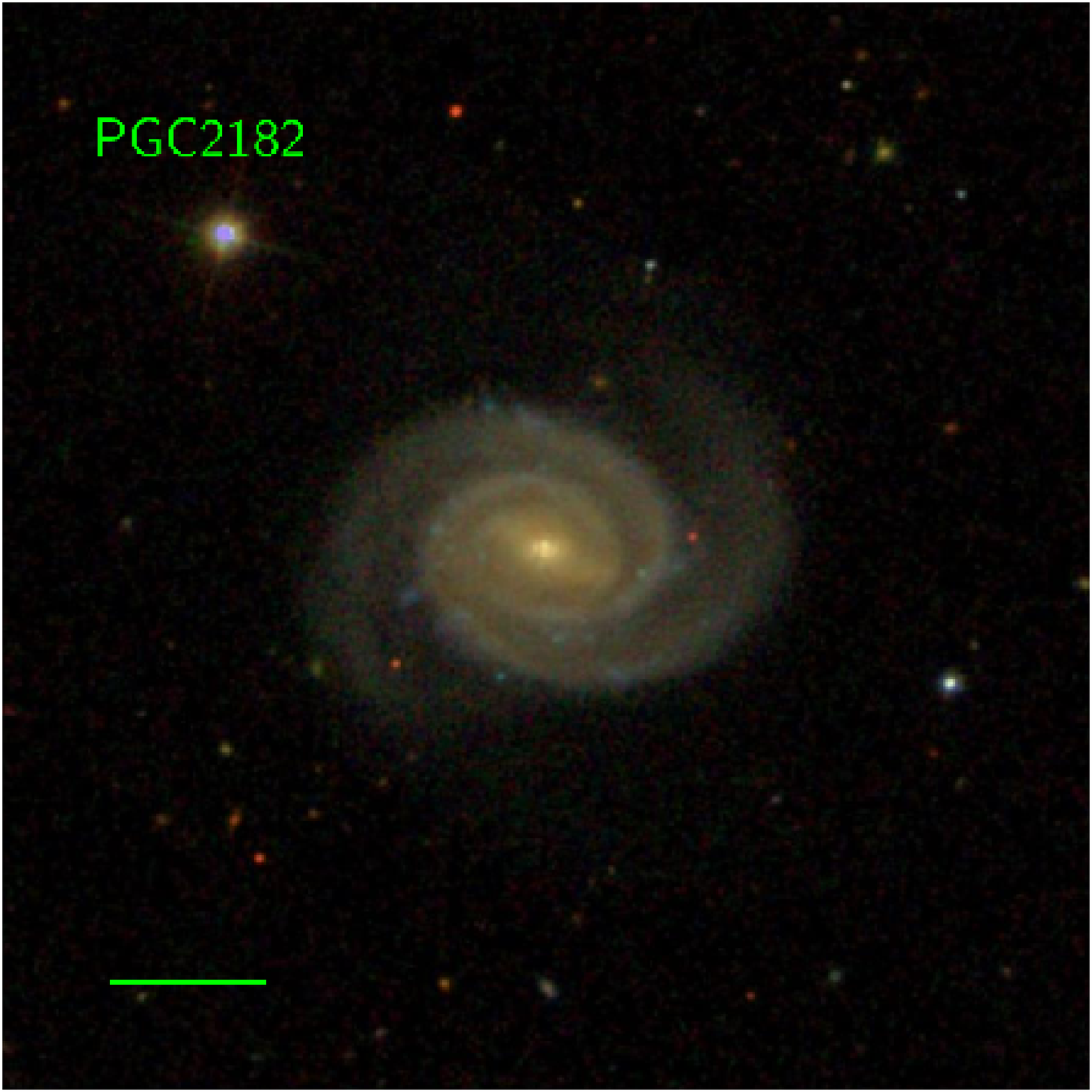}
\includegraphics[width=5.5cm, height=5.5cm, angle=0]{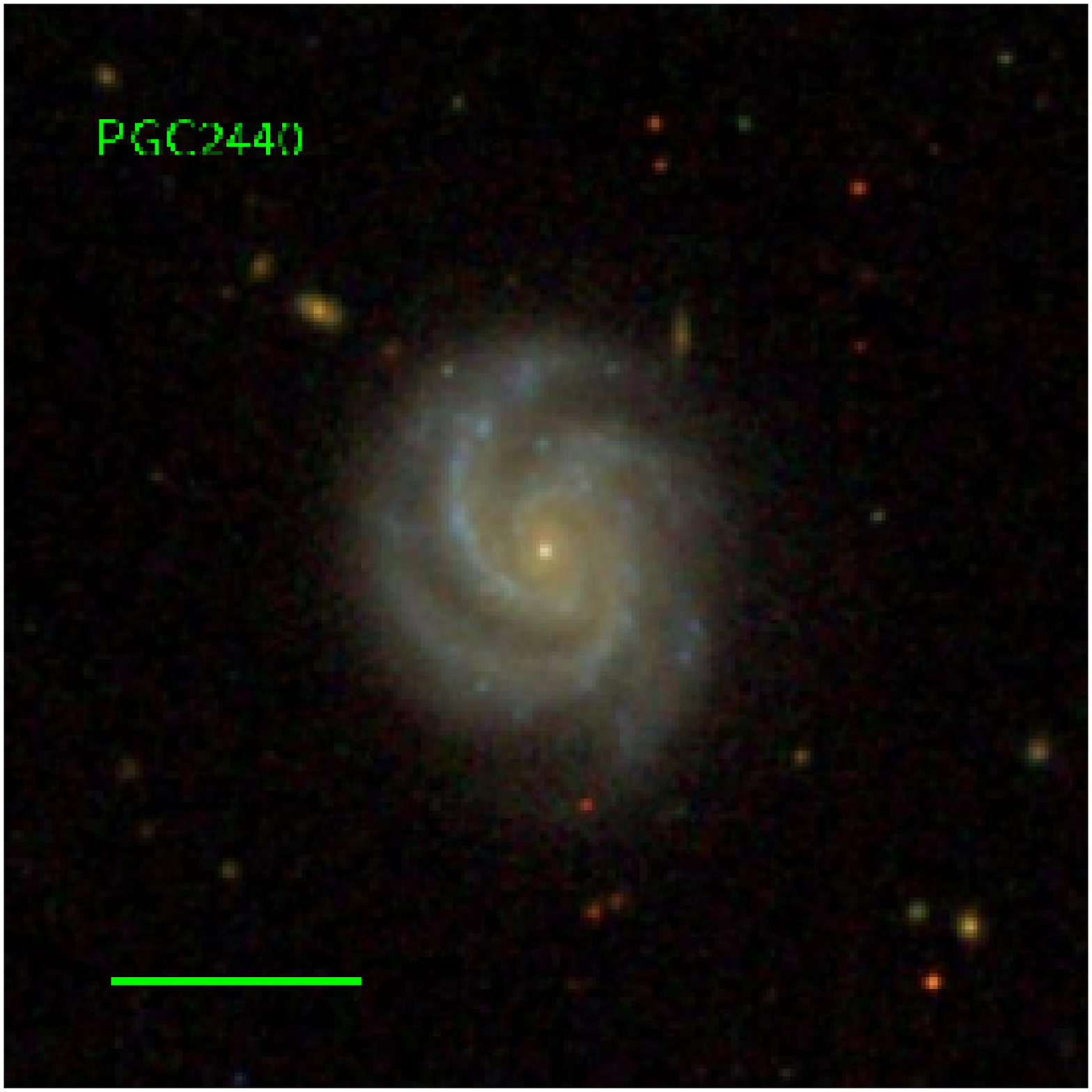}
\includegraphics[width=5.5cm, height=5.5cm, angle=0]{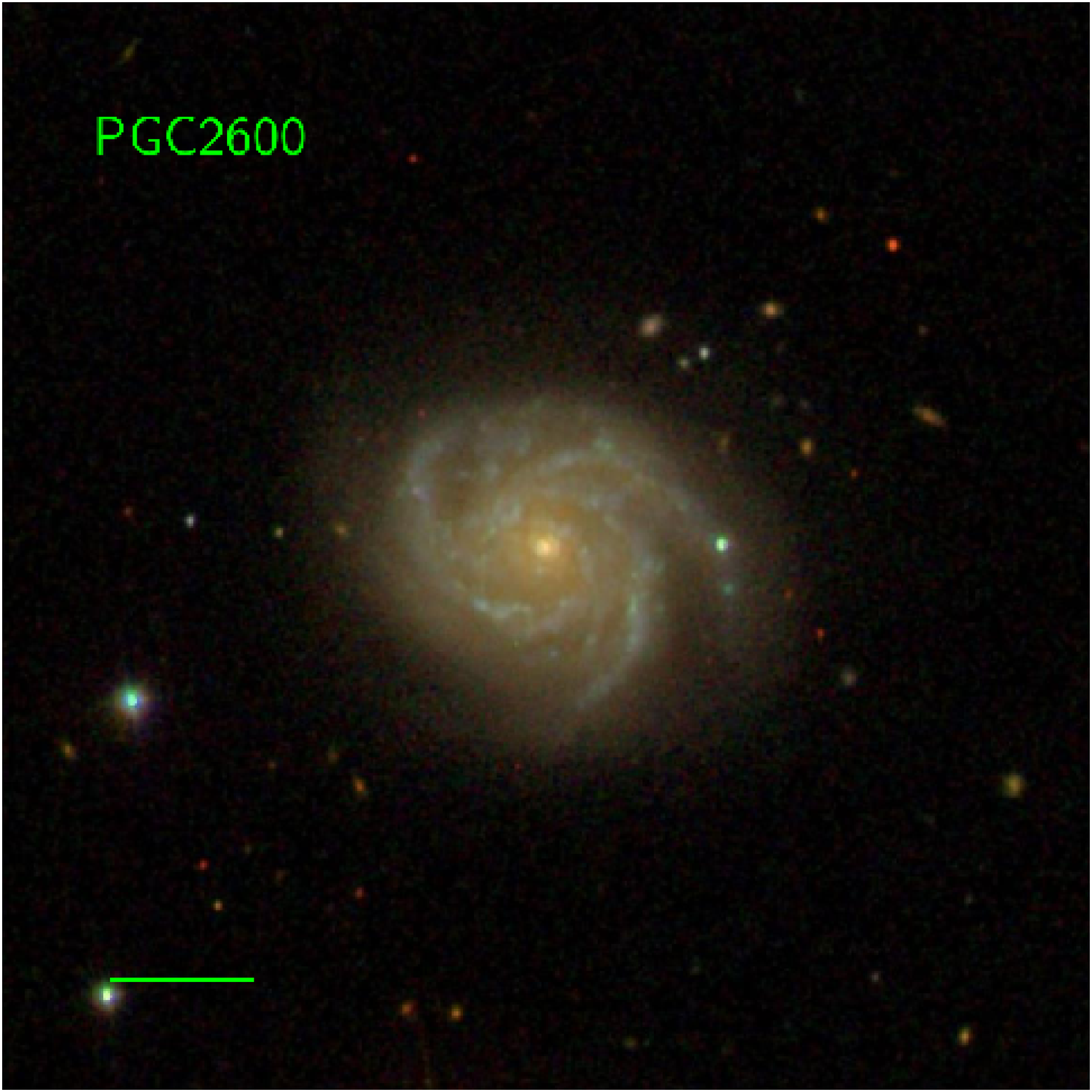}
\includegraphics[width=5.5cm, height=5.5cm, angle=0]{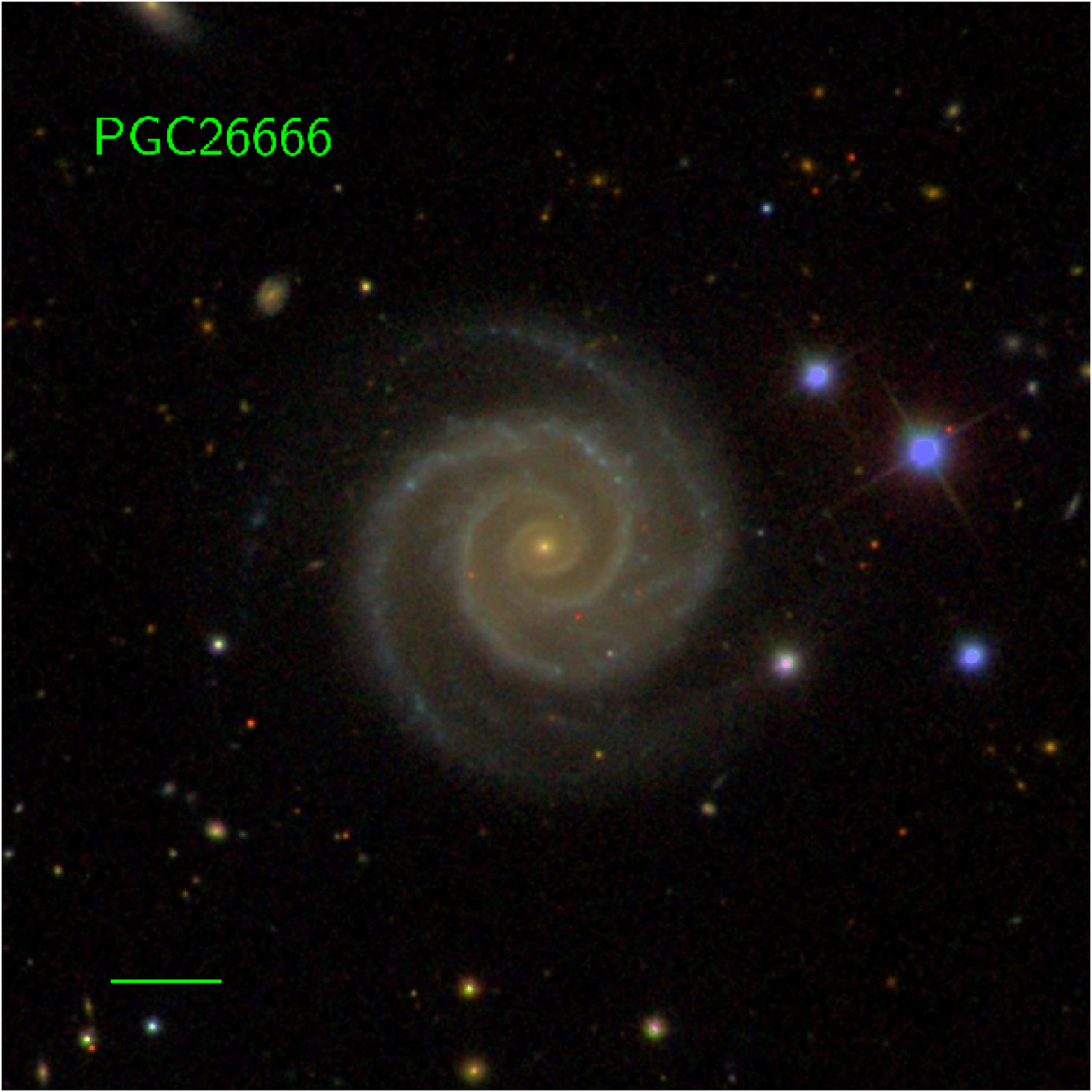}
\includegraphics[width=5.5cm, height=5.5cm, angle=0]{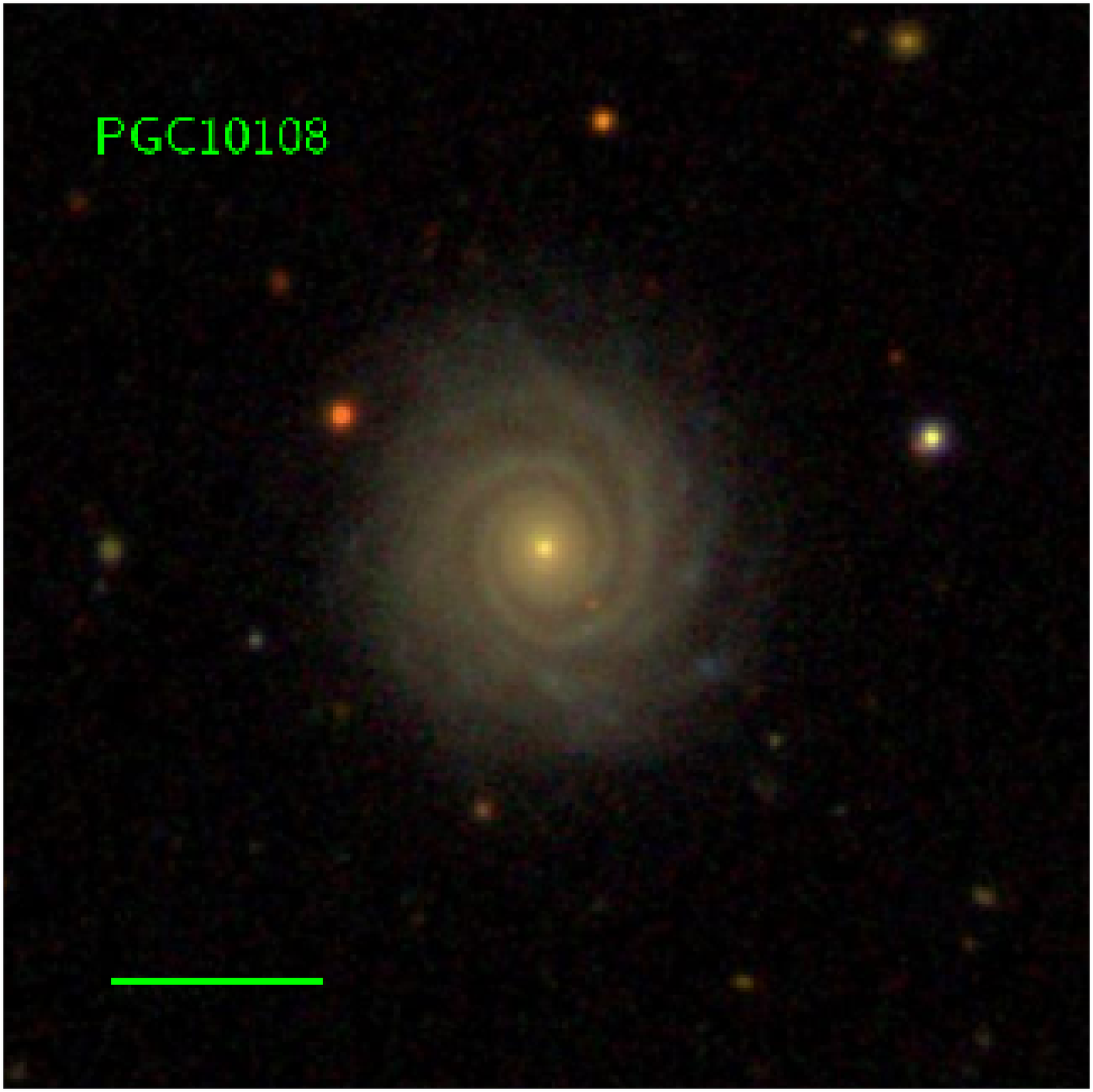}
\includegraphics[width=5.5cm, height=5.5cm, angle=0]{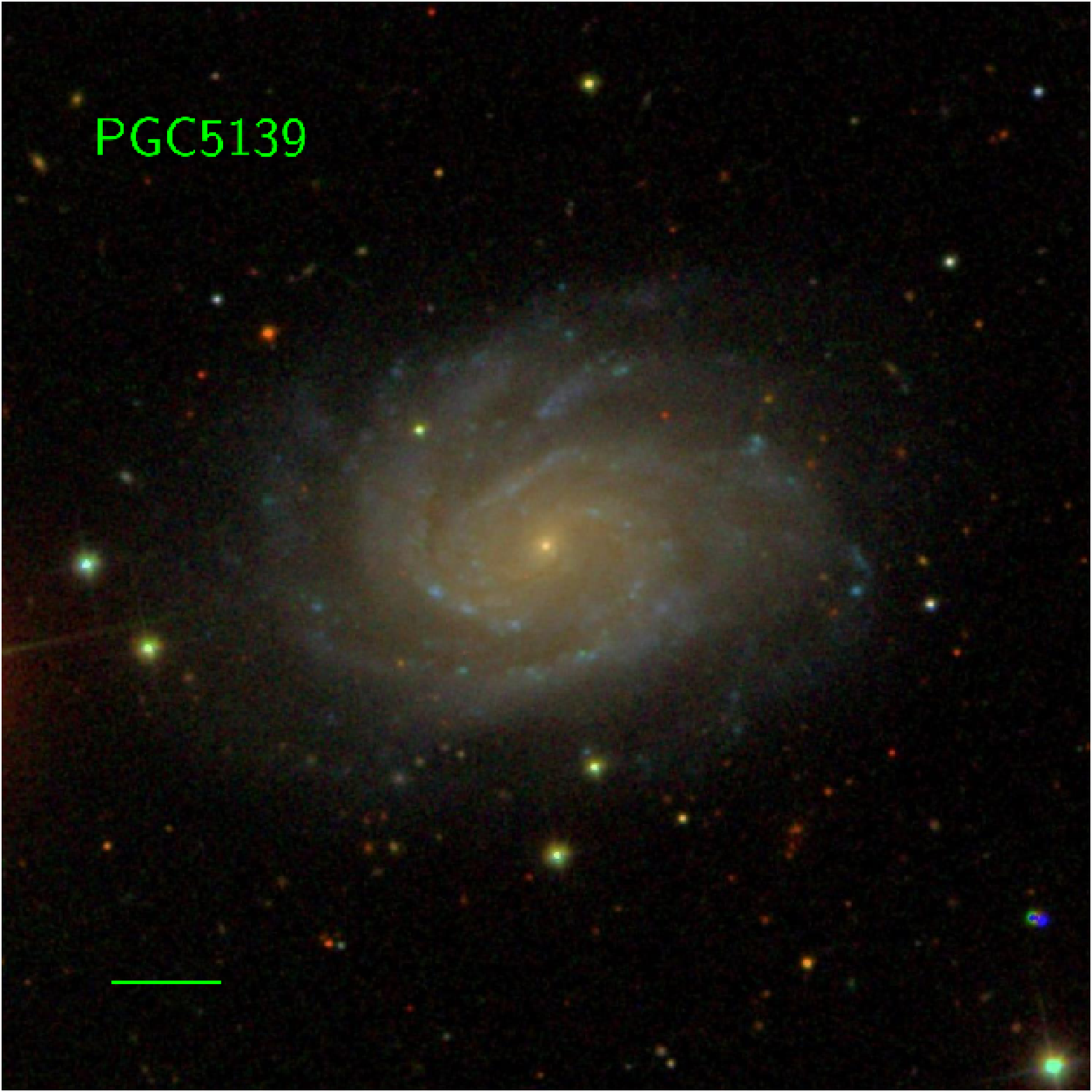}
\includegraphics[width=5.5cm, height=5.5cm, angle=0]{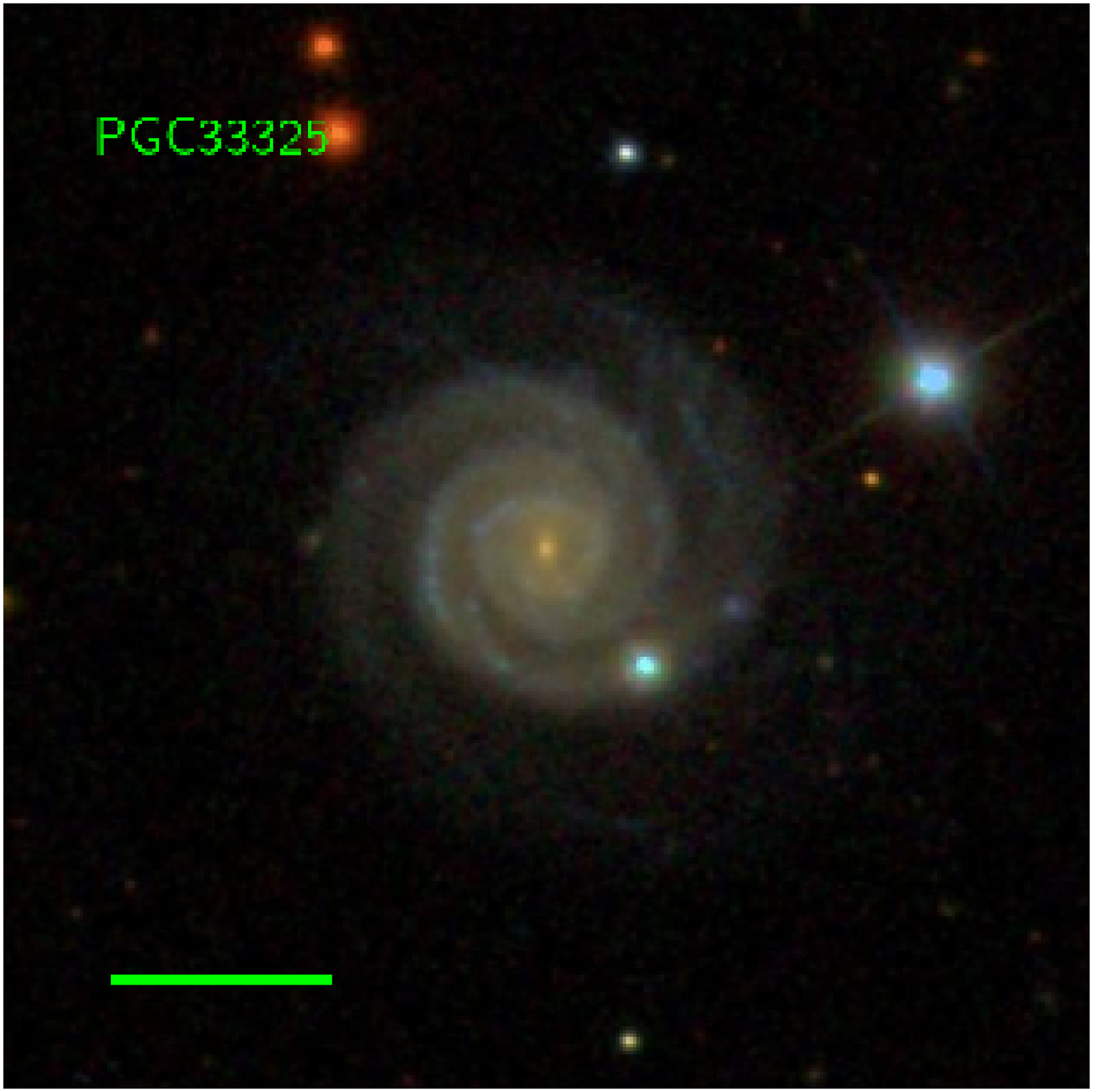}
\includegraphics[width=5.5cm, height=5.5cm, angle=0]{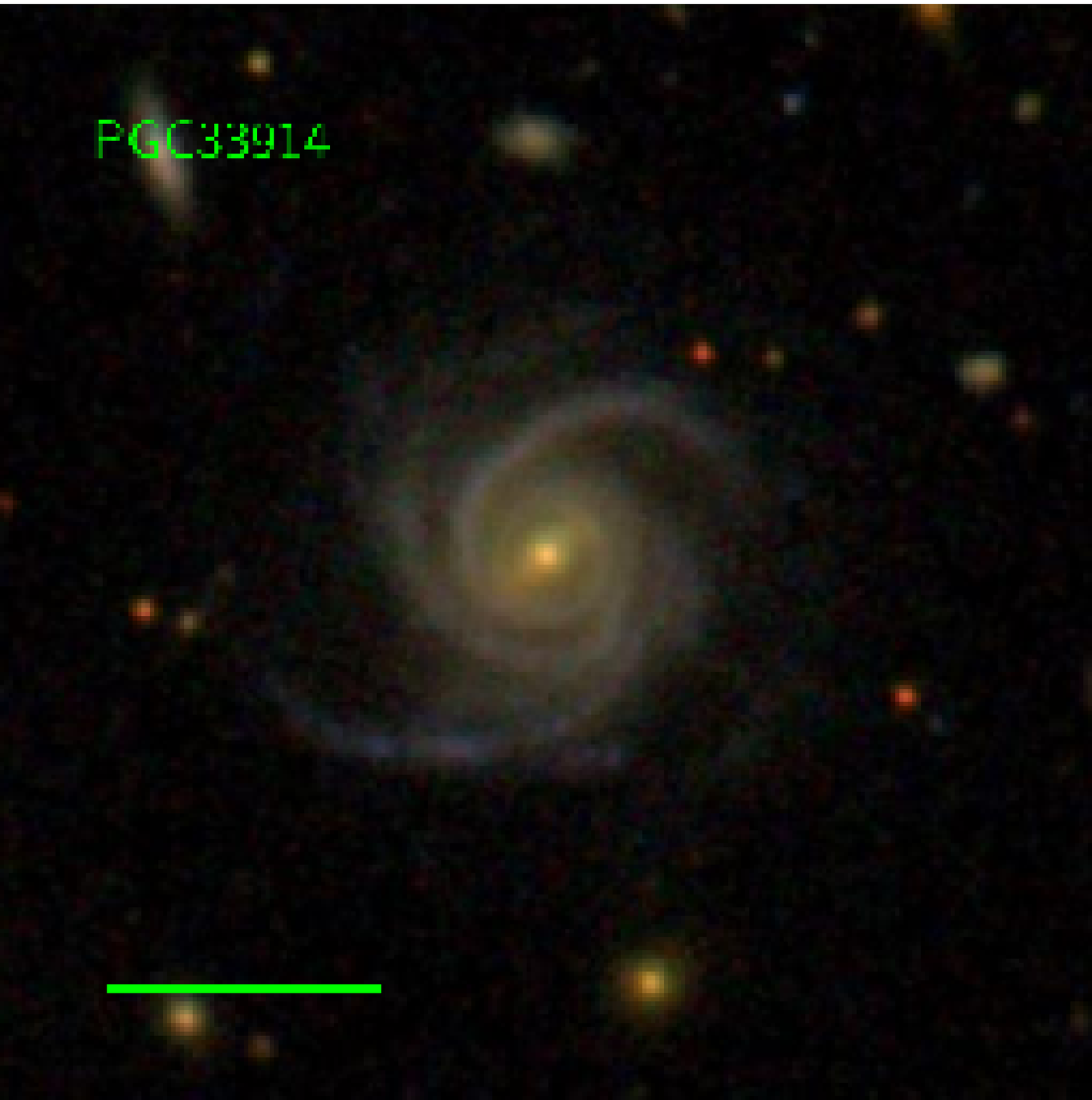}
\includegraphics[width=5.5cm, height=5.5cm, angle=0]{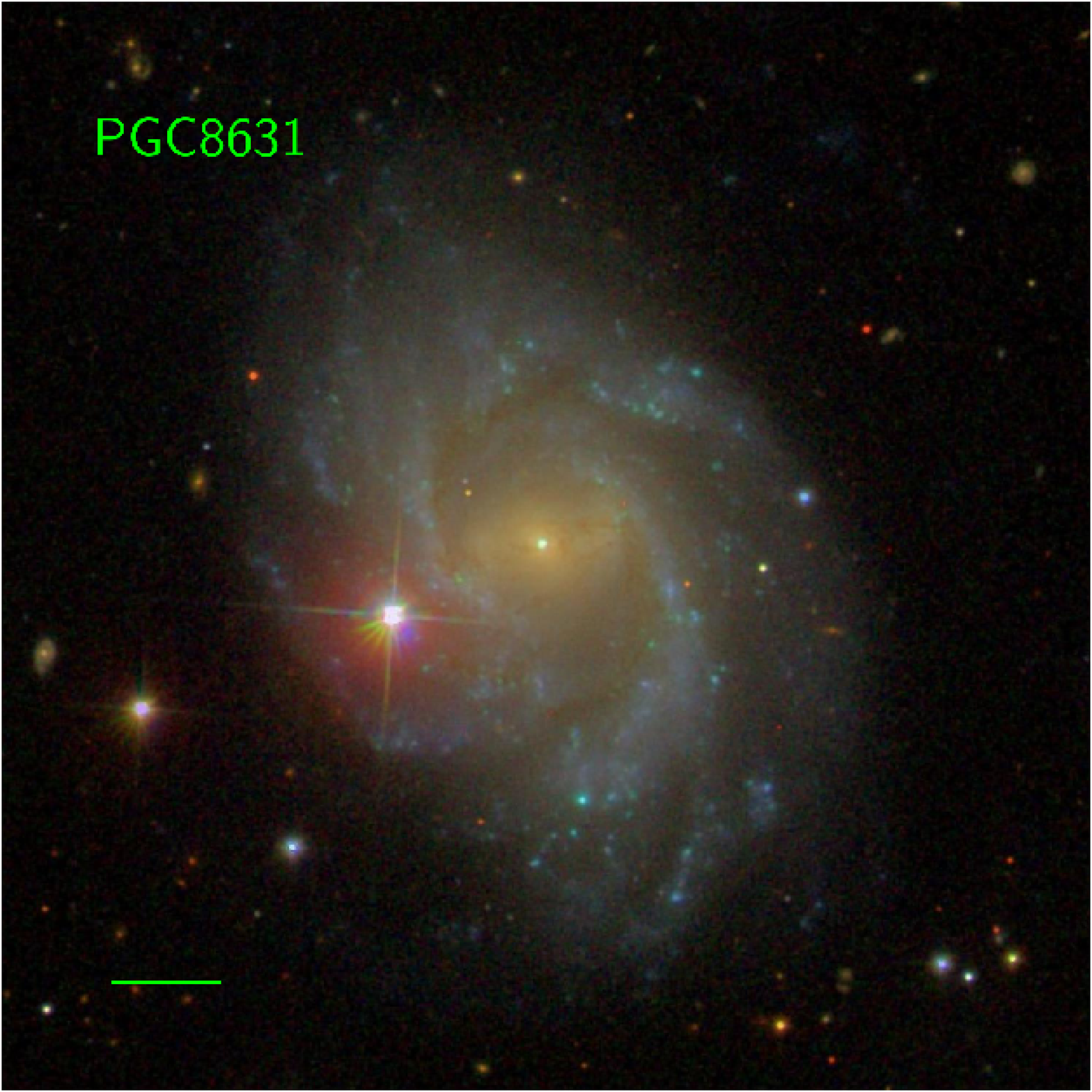}

\caption{SDSS thumbnail images for nine typical spiral galaxies in our sample: left column are of G-class
    (PGC2182, PGC26666, PGC33325), middle column -- M-class (PGC2440, PGC10108, PGC33914), and the right one shows
    F-class spirals (PGC2600, PGC5139, PGC8631). A bar in the bottom left corner of each image shows one arcminute scale.
    The images are oriented such that the north direction is up and the west is right.}
\label{ex_image}
\end{figure*}

In Fig.~\ref{ex_image} we show nine\footnote{All images of our sample are available at \url{https://vo.astro.spbu.ru/node/129}}
 typical spiral galaxies in our sample of the G, M and F classes.

\begin{figure}
  \centering
  \includegraphics[width=\columnwidth]{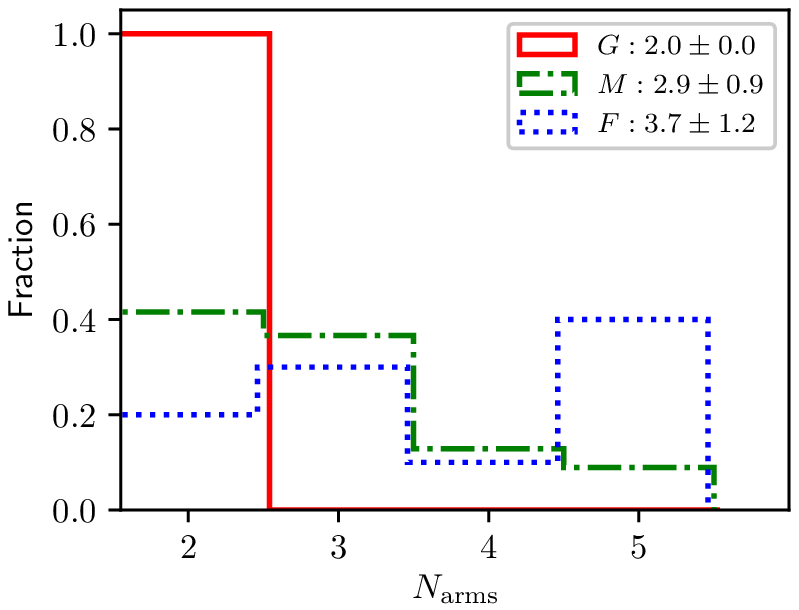}
  \caption{Distribution of grand design (red solid line), multi-armed (green dash-dotted line) and flocculent
    (blue dotted line) galaxies in our sample by the number of spiral arms.}
  \label{fig:num_of_arms_hist}
\end{figure}

Along with the arm classification, we counted the number of well-distinct spiral arms (though not all of them can be analysed with our method because of their faintness or visible bifurcations, see Sect.~\ref{sec:method}) as an indirect test of the classification correctness. If the number of spiral arms is larger than 5 or cannot not be counted (for flocculent galaxies), we assign to this number `>5'.
Fig.~\ref{fig:num_of_arms_hist} demonstrates the distribution of grand design, multi-armed and flocculent galaxies of our sample by the number of spiral arms. One can see, that all grand design
galaxies in our sample have a two-armed spiral structure, whereas multi-armed and flocculent galaxies may have different number of spiral arms, from
2 to >5. Flocculent galaxies tend to have, on average, a larger number of spirals than multi-armed galaxies.

\section{The method}
\label{sec:method}
In this section we outline our method for analysing spiral structure in galaxies. We start with some additional image preparation which
is required for our method to work. Then we describe the method itself. At the end of this section, we illustrate the results of the method applied to a galaxy from our sample.

\subsection{Additional image reduction}
\label{sec:image_prep}
One additional step of the image preparation has not yet been done -- removing an axisymmetric component from the galaxy image. The spiral arms are observed as an additional component over the stellar disc, the light from which can interfere with the spiral arms when their parameters are measured. To remove the light of the disc from the galaxy image, we subtracted an azimuthal profile
of the galaxy from the original image. To compute the profile, we found the mean flux value along a set of concentric ellipses of different sizes with
the ellipticity fixed to $\cos \mathfrak{i}$ and position angle to $\mathrm{PA}$ of the galaxy (see Sect.~\ref{sec:image_prep}). To suppress small scale flux variations, we
 applied an iterative sigma clipping procedure before the averaging. Our azimuthal profile is axially symmetric, so by subtracting it we remove an axially symmetric component (the disc) from the image and leave in the image only non-axially symmetric components, such as spirals, a bar, star-forming regions etc.

After this additional step, the galaxy images are ready to be processed with our analysis.

\subsection{Method outline}
\label{sec:method_outline}

The main idea of our method is to analyse a set of photometric cuts made across spiral arms at various points
of the spiral structure. The local parameters of a spiral arm (such as the width of the arm at a given point)
are derived from the properties of the corresponding slice, whereas the global parameters (such as the pitch
angle and the overall width variation along the arm) can be inferred from the analysis of the whole set of such slices.
The slices should cover the entire arm, from the beginning to the end with some given step.

It is proven to be extremely difficult \citep[see e.g.][]{2014ApJ...790...87D}
to determine automatically if a given point of an image belongs to a spiral structure, and to group such points into
separate spirals: local kinks, discontinuities, forks, and background objects lead to splitting the arms or
to joining separate arms together. In this work we decided to visually inspect each galaxy image and manually put some points
along the spiral arms to use them as a first-guess tracer of the spiral arm for our method. To do so, we used the {\small DS9}\footnote{\url{http://ds9.si.edu/site/Home.html}}
package to display a de-projected galaxy image and place several circular regions along each prominent spiral arm. The radii of these regions were adjusted
such that the regions spanned approximately to the middle of the interarm area covering the full width of the arm.
The radii of the regions were used to calculate the lengths of the slices. If an arm is splitting into two arms at some
point, we chose the brightest branch (which also usually follows the general direction of the parent arm). If an arm is
splitting into two arms of roughly the same amplitude, or has more than one splitting point, we discard it.

In the next step, the algorithm fills in gaps between the sparse user-specified points to trace the whole spiral arm, with a step of 2 pixels.
For every pair of adjacent points $i$ and $j$, a local value of the pitch angle is estimated as
$$
\psi_{ij} = \atan \left( \frac{\log r_j - \log r_i}{\phi_j - \phi_i}\right)\,,
$$
where $r$ and $\phi$ -- are polar coordinates of the points. The space between these two points is then filled using
an interpolation by a logarithmic spiral with a constant value of the pitch angle:
$$
r_k = r_i  \exp \left(\tan\psi\left(\phi_i-\phi_k\right)\right)\,,
$$
where values of $\phi_k$ are evenly spaced between $\phi_i$ and $\phi_j$ such that the new points are located
at a distance of about 2 pixels from each other.

After that the algorithm makes photometric cuts perpendicular to the arm at every point. The slope of
a line perpendicular to a logarithmic spiral with the pitch angle $\psi$ at the point with
azimuth $\phi$ can be found as
$$
\xi = - \frac{\tan\psi\cos\cos\phi-\sin\phi}{\tan\psi\sin\phi+\cos\phi}\,.
$$

\begin{figure}
  \centering
 \includegraphics[width=0.9\columnwidth]{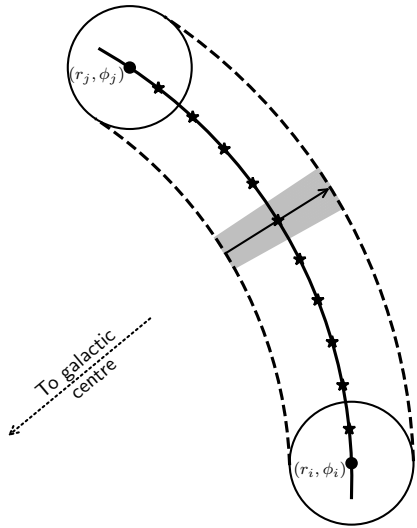}
 \caption{A scheme of an orthogonal cut construction. The curved solid line depicts a spiral arm, with two circles to be
   user specified points $i$ and $j$ (see text). The stars show locations of the interpolated points at
   which the cuts are made. One of the cuts is shown as a solid line with an arrow with a
   shaded region to be the cut width. The dashed lines show the limits of the cuts.}
 \label{fig:scheme}
\end{figure}

To enhance the signal-to-noise ratio, we make several cuts placed with a small shifts along the arm and then
compute an averaged cut using a median filter. This averaging also allows us to filter out small-scale flux variations, such as compact H{\sc ii} regions
and unmasked stars: the flux variation across the cut is higher in such regions, so the corresponding pixels
can be assigned with smaller weights during the subsequent analysis. At this stage we also use the mask of the
background objects (see Sect.~\ref{sec:data_prep}) to exclude the masked pixels. Fig.~\ref{fig:scheme} demonstrates schematically the construction of a cut.

\begin{figure}
  \centering
 \includegraphics[width=\columnwidth]{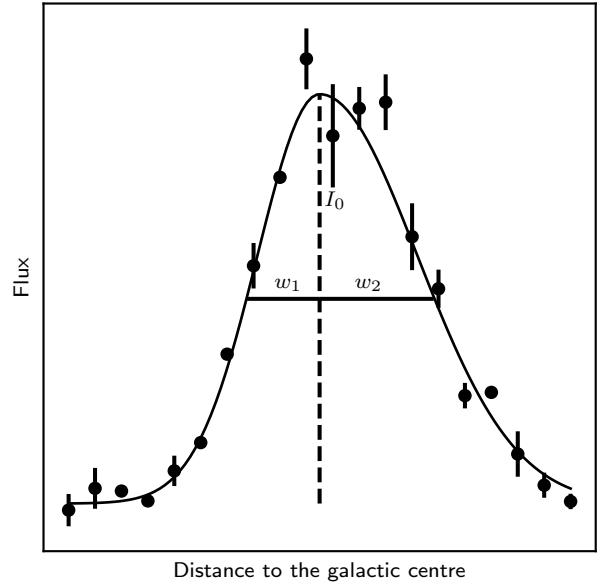}
 \caption{A schematic view for a single slice. The dots are data points, the solid line is a fit by a lopsided Gaussian.
   The peak location is shown by the vertical dashed line and the amplitude and both widths are marked
   as $I_0$, $w_1$ and $w_2$. The errorbars show the flux variation across the cut.}
 \label{fig:dummy_slice}
\end{figure}

When a photometric cut is made at a given point of an arm, we fit it with an analytical function.
Schematically, a cut across a spiral arm should appear as a bell-shaped curve with a peak located near the centerline of the spiral
and gradually decreasing brightness at both sides of the peak (Fig.~\ref{fig:dummy_slice}).

To fit such a bell-shaped curve and also to take into account possible asymmetry of a spiral arm, we adopted an asymmetric
Gaussian function as a fitting function:
\begin{equation}
  I_{\mathrm{model}}(r) = I_0 \times
  \begin{cases}
    \exp \left( - \frac{ \left[ r - r_{\mathrm{peak}} \right]^2 }{w_1^2} \right), & r < r_{\mathrm{peak}} \\
    \exp \left( - \frac{ \left[ r - r_{\mathrm{peak}} \right]^2 }{w_2^2} \right), & r > r_{\mathrm{peak}}\,,
  \end{cases}
\end{equation}
where $I_0$ is the amplitude, $r_{\mathrm{peak}}$ is the peak location, and $w_1$ and $w_2$ are the half-widths of the spiral
arm ``inward'' (in the direction to the galactic centre) and ``outward'' (away from the centre), accordingly.
When $w_1=w_2$, the brightness distribution is symmetric around the peak at this point of the arm. If $w_1<w_2$, the inner side of
the arm is steeper than the outer one (as in Fig.~\ref{fig:dummy_slice}), and vice versa. We also added a free
 baseline level to the fitting function as a linear trend. The fitting function has, therefore, six free
 parameters ($I_0$, $r_{peak}$, $w_1$,  $w_2$ and two parameters for a baseline level).

Since the observed image of a galaxy is a real image convolved with a point spread function,
a direct measurement of the parameters of the spiral structure via the fitting procedure will yield distorted values
(a lower amplitude and larger widths). To solve this problem, one should convolve the fitting function
with the PSF before comparing it with the observed flux inside the fitting procedure. Therefore, the real
fitting function has to be written as:
$$
I_{\mathrm{fitting}} = I_{\mathrm{model}} * \mathrm{PSF}\,,
$$
and the fitting parameters should be inferred from this function. If the derived value of the FWHM
of the (unconvolved) cut is lower than the PSF FWHM, the spiral at this point is too thin to be resolved
in the image, and the corresponding width value should be considered as unreliable and has to be excluded
from the further analysis. Another reason for discarding a slice from the analysis is a low signal-to-noise
ratio. If the measured amplitude $I_0$ is lower than the background variation, we also consider this slice as unreliable.
The outermost slice of each spiral arm is, therefore, the last slice with the amplitude above the background variation level.
We consider a slice with the amplitude equal to the background variation level as still reliable because the slice is constructed
as a mean of $\sim 10$ adjacent slices, so the effective noise variation for it is lower, and the outermost slices are
detected at the $\sim 3\sigma$ level.

To estimate the uncertainties of the fitting parameters within a slice, we run a Monte-Carlo simulation. We repeated our fitting
process of a cut many times randomising the flux values of the cut points according to their uncertainties. The errors
of the fitting parameters are then estimated as standard deviations of their values within these randomised runs.

\subsection{The output values}
\label{sec:method_output}
Here we describe the parameters of the spiral structure which are derived by means of our algorithm for an example galaxy, PGC~2182 (Fig.~\ref{ex_image}, left panel). The results of the application of the algorithm to the
entire sample are shown in Sect.~\ref{sec:results}.

As a main output of the algorithm at every point of the galactic spiral structure, we derive:
\begin{itemize}
\item coordinates of the peak position,
\item surface brightness of the spiral arm at this point,
\item characteristic values of spiral widths, ``inward'' and ``outward''.
\end{itemize}

\begin{figure}
  \centering
 \includegraphics[width=\columnwidth]{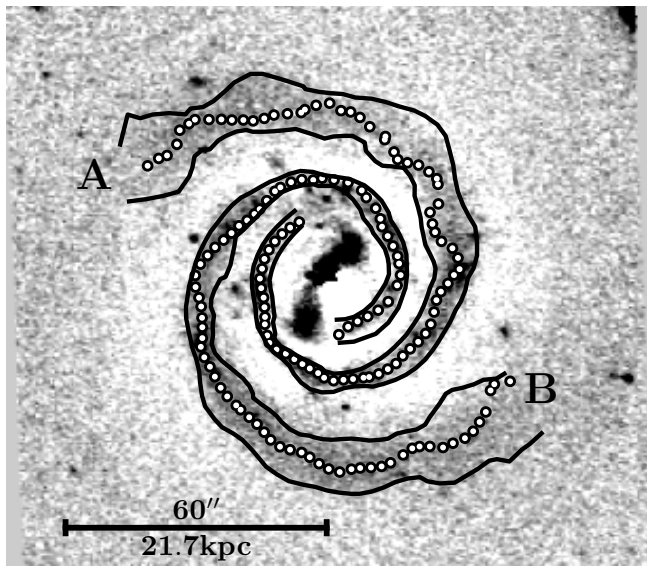}
 \caption{Results of the fitting of the spiral arms for PGC~2182 in the $r$ band. The white points show locations of the cut peaks, the solid
   lines show inward and outward widths of spirals. The line in the bottom left corner shows a 60\arcsec scale for the image
   (which is 21.7~kpc at the distance to the galaxy).}
 \label{fig:pgc_2182_spirals}
\end{figure}

Fig.~\ref{fig:pgc_2182_spirals} illustrates results of our fitting process applied to PGC~2182 and plotted over the image, which was
created as a residual between the de-projected galaxy image and azimuthally averaged model in the $r$ band. The white circles show locations
of the positions of the cut peaks, computed from the $r_{\mathrm{peak}}$ values; the solid lines show inner and outer edges of the arms computed
from $w_1$ and $w_2$ values, correspondingly. We use a Savitzky-Golay filter to smooth the locations of both peak positions and
arm edges before plotting them since they are influenced by the overall flocculence of the spiral structure on small scales whereas we are interested
in large scale variations of the arm parameters. The letters A and B mark spirals for further references.

\begin{figure}
  \centering
 \includegraphics[width=\columnwidth]{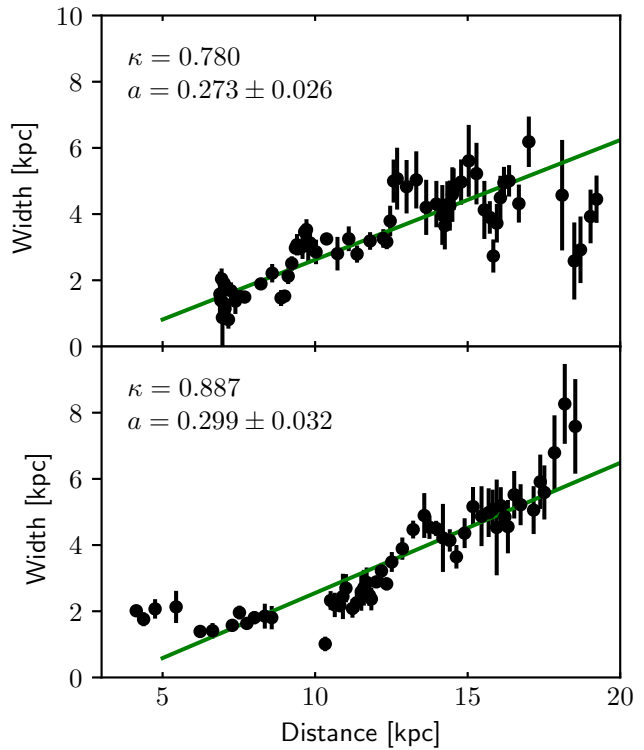}
 \caption{Values of the full width of the spirals A (top) and B (bottom) for PGC~2182 as a function of
    radius. In top left corners of each panel, the values of the Pearson coefficient $\kappa$
   and the slope of the best-fit linear fit are shown. The fit itself is shown as a green solid line.
   The gaps in the data are due to the exclusion of the unresolved slices (see text).}
 \label{fig:pgc_2182_widths}
\end{figure}

One can see from Fig.~\ref{fig:pgc_2182_spirals} that the width of the spirals changes with radius.
Fig.~\ref{fig:pgc_2182_widths} demonstrates this in a form of width--radius relations:
the top panel shows the full width ($w = w_1 + w_2$) of the arm A as a function of radius, the
same relation for the arm B is shown in the bottom panel. One can see that in the case of PGC~2182, the widths of both arms systematically
increase with the distance from the galaxy centre. The values of $\kappa$ in top left corners of each panel
show a Pearson correlation coefficient for these width -- radius relations, and the values of $a$ are slopes
of best linear fits (i.e. average rates of the width change with radius in kpc per kpc).

\begin{figure}
  \centering
 \includegraphics[width=\columnwidth]{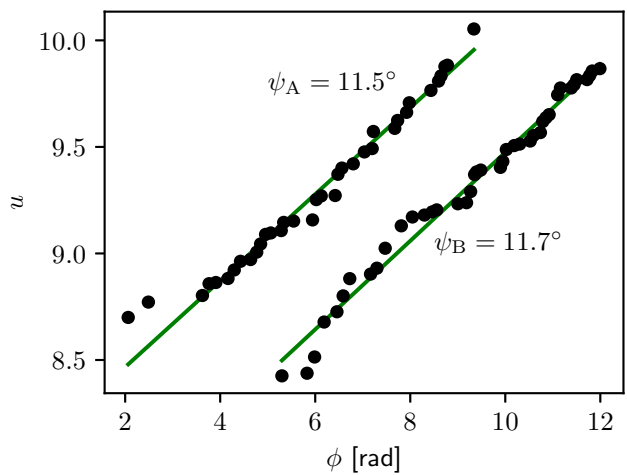}
 \caption{Pitch angle estimation for the two arms of PGC~2182: the points show the spiral arms in the log-polar coordinate system,
   and the green solid lines are the best linear fits to the data.}
 \label{fig:pgc_2182_pitch}
\end{figure}

Having values of $w_1$ and $w_2$ for each point of the spiral structure, we can compute some measure of the
asymmetry of the spiral arm. In this work, we decided to compute the asymmetry as
\begin{equation}
  A = \frac{w_2 - w_1}{w_2}\,.
  \label{eq:asymmetry}
\end{equation}
Positive values of $A$ mean that the inner side of an arm has a steeper slope then the outer one (as on Fig.~\ref{fig:dummy_slice}).

Another value that can be derived from our analysis is a pitch angle value. If a spiral
arm can be described by a logarithmic spiral, it will appear as a straight line in the log-polar coordinates. The
slope of this line is determined by the pitch angle value of the arm:
$$
u = u_0 + \tan \psi \cdot \phi\,,
$$
where $u=\log r$ and $u_0=\log u_0$. The pitch angle, therefore, can be found as a slope of a linear fit of the points of
the spiral structure in the log-polar coordinates \citep{1981AJ.....86.1847K, 2019MNRAS.482.5362F}. Fig.~\ref{fig:pgc_2182_pitch}
demonstrates the pitch angle determination for PGC~2182.
The solid lines show linear fits to the data, and the value of the pitch angle is shown for both spirals. The advantage
of this approach is that the value of the pitch angle can be derived individually for all spirals of the galaxy, in contrast
to the widely used Fourier analysis applied to a galaxy image \citep{1982A&A...111...28C}, which yields only some average value
for the entire spiral structure.

Our further analysis of the pitch angle includes an investigation of the pitch angle variation
with radius. To compute local values of the pitch angle at
different galactocentric distances, we used a spatial window. This window isolates a part of
the arm located at a given range of the radii, so that the pitch angle inference can be applied
only for this part of the arm. Moving the window along the arm, one can measure how
the pitch angle changes with distance to the galaxy centre. In this work, we use a window width
equal to one-third of the full arm extent. This window size allows one to capture a general trend
of the pitch angle variation while still having a high signal-to-noise ratio.

Since our method yields the values of the pitch angle for all spiral arms separately, it is possible to
estimate the value of the interarm pitch variation $\frac{\Delta\psi_{\mathrm{arms}}}{\left<\psi\right>}$, i.e. the value that
shows the variation of the pitch angle value between the galaxy spirals. We define this value as a difference between
the maximal and minimal pitch angles for all arms divided by the mean value of the pitch angle for the galaxy
(so $\frac{\Delta\psi_{\mathrm{arms}}}{\left<\psi\right>}=0$ indicates that all spirals in the galaxy have the same value of the pitch angle).

Using the peak positions and width values, it is possible to construct an image which contains the spiral mask. This
spiral mask is an image of the same size as the original galaxy image and has pixel values equal to 1 if
the corresponding pixel belongs to the spiral structure (i.e. it lies between the solid lines in Fig.~\ref{fig:pgc_2182_spirals}),
and 0 otherwise.

The constructed spiral masks can be used in several ways. At first, we use them to rectify azimuthal profile subtraction
(Sect.~\ref{sec:image_prep}). The problem is that when we compute an azimuthal profile, the spirals also contribute to it, even if
 a sigma clipping procedure is applied. This leads to an overestimation of the disc brightness which, in its turn, leads to
negative intensities in the interarm regions of the residual image (the result can be seen in Fig.~\ref{fig:pgc_2182_spirals}
for PGC~2182: the interarm regions are brighter than the background regions which surround the galaxy and have the zero mean flux). This effect does
not alter the results related to the positions and scales (the peak positions, widths, pitch angles), but can introduce
a systematic shift in the results related to the brightness (spiral arm colours and fluxes).

To fix this problem, we performed azimuthal averaging once again, but with the additional mask of the spiral arms this time. When the spirals
are masked out, they do not make a contribution to the azimuthal profile, so the true disc background can be computed and
subtracted from the galaxy image. This, however, can be achieved only when a galaxy has a regular spiral pattern and
all the spirals can be reasonably covered by a mask. Some galaxies in our sample (mostly of the classes M and F) have
splitting spirals (bifurcations) or contain fragments of spirals, so their masks do not cover the entire spiral structure.
We therefore selected a subsample of galaxies with a regular spiral structure and performed a further analysis
which requires a proper azimuthal subtraction only for this subsample. Hereafter, we call this subsample as ``stage 2''
galaxies, to stress that the analysis of the spiral structure was applied two times for them: the first time -- with a default azimuthal
profile and the second time -- with a profile, for creating of which a mask of all spiral arms was taken into account. In total, our sample
contains 67 of such stage 2 galaxies, which include 29 grand design spirals, 33 multi-armed and 5 flocculent galaxies.

Once the azimuthal profile is corrected for the presence of the spiral structure, the ``original minus model'' residual
image should mainly contain the light coming from non-axisymmetric galaxy components. We then use the spiral
mask to isolate the spirals and compute a total flux coming from the spiral pattern of the galaxy. This flux value
can be then used to compute the arms-to-total ratio $f_{\mathrm{arms}}$ by dividing it by the total aperture flux of the
galaxy.  This presents a new method for the ``arms strength'' estimation, completely independent of existing techniques
such as $m=2, 3, 4$ Fourier modes strength (for example, \citealt{2011MNRAS.414..538K}, \citealt{2018ApJ...862...13Y}).
Comparing the spiral flux in different passbands, we can also compute a colour of the spiral structure.

\begin{figure}
  \centering
 \includegraphics[width=\columnwidth]{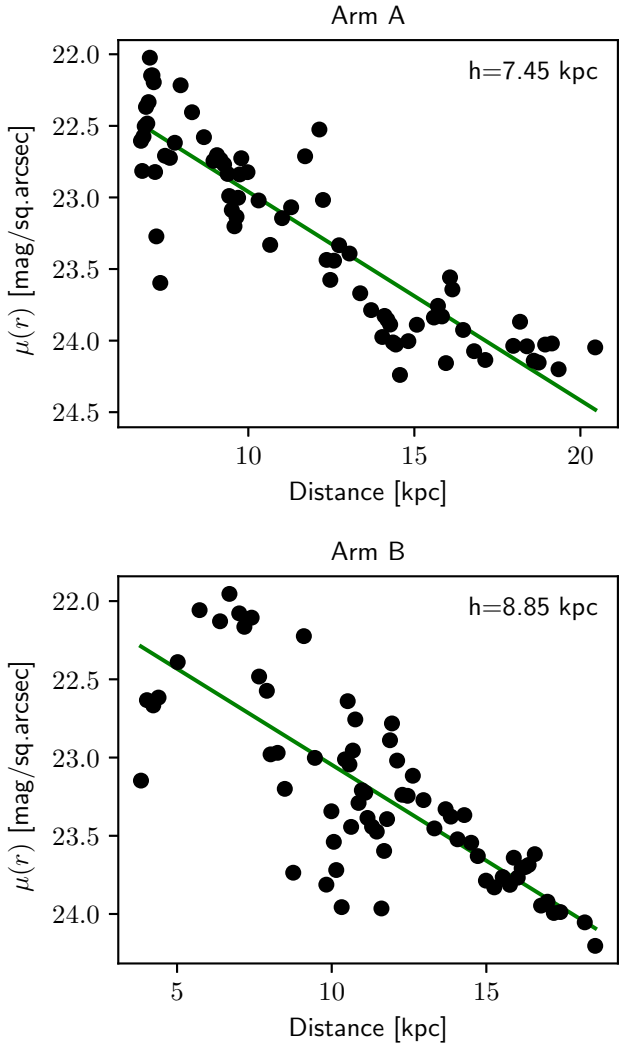}
 \caption{Surface brightness distribution (computed from the fitted $I_0$)} for the spirals of PGC~2182 as a function of radius.
 The solid line in each panel is a linear fit of the corresponding surface brightness distribution. The derived values of the exponential scale are shown in top right corners of each panel.
 \label{fig:pgc_2182_surf_bri}
\end{figure}

The radial brightness profile of a spiral is another property of the spiral structure which can be obtained using our algorithm. Having a
set of points along the arm, one can find how the arm surface brightness changes with radius.
Fig.~\ref{fig:pgc_2182_surf_bri} shows the radial surface brightness distribution for the two spiral arms of PGC~2182.
The spirals of this galaxy show approximately exponential radial brightness profiles. An exponential scale of this
profile can be derived via a linear fit of the data. The solid line in Fig.~\ref{fig:pgc_2182_surf_bri} shows results of a linear fit and
the values in the top right panels are the derived values of the exponential scale. This arm exponential scale length will be discussed in
a future work, where we are about to perform a photometric decomposition of our sample galaxies and study the dependence between
 different structural parameters and spiral structure.

\section{Results}
\label{sec:results}
In this section we present the results of our analysis applied to the whole sample of the selected galaxies.
In Table~\ref{fit_pars}, we list the main fitting and other important parameters of the spiral structure for the entire sample\footnote{The whole table is available online. The animations, which demonstrate the process of our spiral structure analysis for all galaxies of our sample, are available at \url{https://vo.astro.spbu.ru/node/126}}.
 In this section we only consider results for the $r$ waveband, as this band is the deepest among the other SDSS bands and we have checked that the results in the other bands are similar to what we derived for this band (see Sect.~\ref{sec:dis:bands}).

\begin{table*}
 \centering
 \begin{minipage}{180mm}
  \centering
  \parbox[t]{150mm} {\caption{Results of the fitting for the $r$ band. This table is published in its entirety in the electronic version of the MNRAS.}  \label{fit_pars}}
  \begin{tabular}{lcccccccccc}
    \hline
    \hline
    Name & $\langle \psi \rangle$    & $\frac{\delta \psi}{\langle \psi \rangle}$ & $w$          & $a$ & $A$ & $f_\mathrm{arms}$ & $(g-r)_\mathrm{arms}$ & $T_\mathrm{arms}$ &  $N_\mathrm{arms}$ \\
         &  (deg)     &                                                               & ($r_{25}$)    &        &        &                   &                         &      &  \\
    (1) &     (2)     &                             (3)                              &      (4)       &  (5) & (6)  &      (7)        &           (8)          & (9) & (10) \\
    \hline
PGC303 & 23.0 & 0.4 & 0.13 & -0.007 & 0.27 & 0.15 & 0.44 & M & 4 \\
PGC2182 & 11.5 & 0.2 & 0.14 & 0.286 & 0.12 & 0.25 & 0.46 & G & 2 \\
PGC2440 & 17.2 & --- & 0.14 & 0.108 & -0.17 & --- & 0.50 & M & 3 \\
PGC2600 & 23.9 & 0.3 & 0.10 & 0.179 & 0.02 & 0.13 & 0.27 & F & 3 \\
PGC2901 & 9.1 & 1.0 & 0.10 & 0.153 & 0.21 & 0.13 & 0.58 & G & 2 \\
    ...     &  ... & ... & ...  &  ...   &   ... &  ... &  ... & ...   \\
    \hline\\
  \end{tabular}
\end{minipage}

   \parbox[t]{160mm}{ Columns: \\
   (1) PGC name from HyperLEDA, \\
   (2) mean pitch angle averaged for all spiral arms we traced, \\
   (3) relative pitch angle variation along the radius, \\
   (4) mean width of spiral arms (in units of $r_{25}$) averaged for all spiral arms we traced,\\
   (5) slope of the best linear fit to the radial width variation (positive values mean that the width increases with radius),\\
   (6) mean asymmetry of the cuts (positive values mean that the inner slope of the cuts is steeper),\\
   (7) fraction of the spiral pattern in the total galaxy flux in the $r$ band,\\
   (8) $(g-r)$ colour of the spiral pattern,\\
   (9) armclass: G -- grand design, M -- multi-armed, F -- flocculent,\\
   (10) total number of arms we counted for each galaxy.}
\end{table*}

\subsection{Reliability of the results}
\label{sec:reliability}

In this section we test the reliability of the obtained results in several possible ways. To do so, we validate some characteristics we measured in this study, such as the pitch angles, disc inclinations, total magnitudes, compared to those from other available sources. In addition to that, we discuss individual slices of our fitting, uncertainties obtained from Monte-Carlo simulations and do an indirect comparison of the arm widths derived from other works and known techniques for extracting information on the spiral arms.
It is important to mention here that two tests have been already done: the quality of our classification was evaluated in Sect.~\ref{sec:classification} and all measured arm widths are greater than the corresponding PSF FWHM by design (see Sect.~\ref{sec:method_outline}): 23\% of all slices were removed from the analysis due to insufficient resolution
(most of them are in the inner regions of the galaxies).

To verify that our galaxy de-projection is correct, we compared the apparent disc flattening $q_{25}$, estimated in Sect.~\ref{sec:data_prep}, with that given in the HyperLeda database as $logr25$ (measured in the $B$ band) and converted to the the minor-to-major axis ratio. We found that these parameters correlate well with the Pearson correlation coefficient $\kappa=0.69$, the slope 1.02 and the zero-intercept. Therefore, we can be assured that our de-projection is reliable.

In order to validate our photometry data, we find an intersection with the Catalog Archive Server (CAS) database from the SDSS DR13. For 146 galaxies in this intersection, we compared our total magnitudes from Table~\ref{general_pars} to the Petrosian magnitude \texttt{petroMag} from the SDSS database. Besides the fact that these magnitudes are not equally comparable, the coincidence between the two magnitudes is good with a small number of exceptions (PGC\,38240, PGC\,51592, and PGC\,43161). For two of these exceptions, the reason for this discrepancy can be an erroneous galaxy center location in the CAS database. After removing the outliers, the average Pearson correlation coefficient for all three bands equals 0.97. In all cases, the values of the Petrosian magnitude are slightly larger as compared to our measurements, which means that galaxy is fainter and which is natural according to the definition of the Petrosian magnitude. The average difference in the magnitudes for all galaxies is around 0.5 in the $g$ band and 0.3 in the $r$ and $i$ bands.

\begin{figure*}
\centering
\includegraphics[width=2\columnwidth]{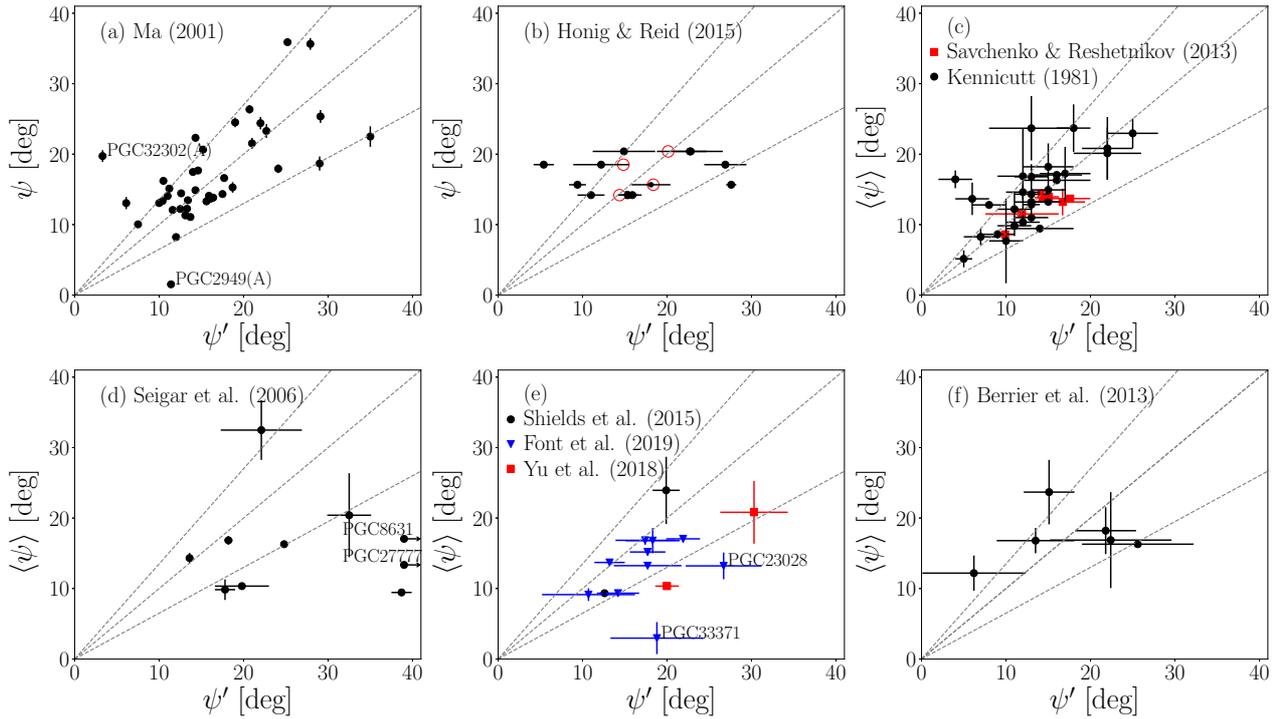}
\caption{Comparison of the pitch angles found in this work with other sources. The dashed lines show a diagonal and 30\%-margin error. The pitch angles are shown for the $r$ band. For galaxies from the other works with a single pitch angle found, we plot  $\langle \psi \rangle$ (see Table~\ref{fit_pars}) and show the scatter between different arms as the uncertainty bar. The sources used are (a) \citet{2001ChJAA...1..395M}; (b) individual arcs from \citet{2015ApJ...800...53H}, the circles represent an average value between the arcs for one arm; (c) \citet{2013MNRAS.436.1074S} in $g$ band (red squares) and \citet{1981AJ.....86.1847K} (black circles); (d) \citet{2006ApJ...645.1012S} for the $B$ band, arrows show two galaxies with both $\psi^{\prime}$ and uncertainties $>40\degr$; (e) \citet{2015arXiv151106365S} (black circles) in the $g$ (PGC24531) and $B$ band (PGC2600), \citet{2019MNRAS.482.5362F} (blue triangles), \citet{2018ApJ...862...13Y} (red squares) in the $R$ band for 2DFFT; (f) \citet{2013ApJ...769..132B}.}
\label{fig:pitch_comparison}
\end{figure*}

The pitch angle of spiral structure has been measured for a significant number of galaxies in different works. For common galaxies, we compare our measures with those derived in these works
 (denoted as $\psi^{\prime}$). The result of this comparison is presented in Fig.~\ref{fig:pitch_comparison}. The comparison in almost all cases is indirect because we find angles separately for each spiral arm detected, while in most  other works only one angle is measured for the whole pattern, usually according to some Fourier-based technique. Only in \citet{2001ChJAA...1..395M} the pitch angles were determined for each arm separately and can be compared directly.

We have 26 galaxies in intersection with \citet{2001ChJAA...1..395M}, from which 12 galaxies have only one common spiral arm measured and 14 galaxies -- two spiral arms. The upper left subplot (a) in Fig.~\ref{fig:pitch_comparison} shows a good agreement between all values except two cases, which can be a result of the different de-projection method used in \citet{2001ChJAA...1..395M}. In \citet{2015ApJ...800...53H}, the authors carried out a detailed analysis of H\,{\sc ii} regions for four galaxies, in which they measured pitch angles for individual arcs in an arm. Two of these galaxies, M\,74 (PGC\,5974) and NGC\,3184 (PGC\,30087), are included in our analysis, both with two spiral arms measured. The comparison in subplot (b) in Fig.~\ref{fig:pitch_comparison} demonstrates that individual parts of the arms can differ from our measured $\psi$, but their average value is close to it. As the remaining (c)\,--\,(f) subplots show, in all other cases the agreement is good and the errors are within a 30\% margin, even besides the variety of methods used for the pitch angle estimation, whether it is by fitting a line to the $\phi$ vs $\log r$ data, or using one-dimensional Fast Fourier Transforms (1DFFT) and two-dimensional Fast Fourier Transforms (2DFFT), or by the so-called {\sc spirality} method \citep{2015arXiv151106365S}. One exception is shown in the subplot (d) with a severe disagreement with the values taken from \citet{2006ApJ...645.1012S}. However, it has been recently reported that the pitch angles, determined in \citet{2006ApJ...645.1012S}, can have a significant discrepancy with other works (see, for example, fig. 8 and 9 and related discussion in \citealp{2019ApJ...871..194Y}). Finally, in subplot (e), the pitch angle found in \citet{2019MNRAS.482.5362F} for PGC\,33371 is much larger than we measure here, which can be an effect of different parts of the spiral arm used for estimating its pitch angle (the same is true for PGC\,23028, but with a lower significance). In total, we compared our pitch angles with the literature for 48 galaxies, and the comparison demonstrates a good agreement with the other works.


As to the width of spiral structure, there are few works with measured widths and in all cases only several individual objects were studied \citep{2010ApJ...725..534F,2014AJ....148..133P,2015ApJ...800...53H}, so it is impossible to compare our results with these sources due to a lack of common objects.

To estimate the goodness of the slices fitting procedure, we find reduced $\chi^2$ statistic for all individual slices. The mean value for $\chi^2$ does not differ much between the $gri$ bands and varies from 1.13 in the $i$ band to 1.07 in the $r$ band. These values, which are close to unity, indicate that the match between the observations and estimates is in accord with the error variance. Thus, we can conclude that, on average, the usage of an asymmetric Gaussian function leads to a good, non-overfitted representation for the slices.

As mentioned in Sect.~\ref{sec:method_outline}, we run a Monte-Carlo simulation with 25 realisations in order to validate the global stability of the procedure used and to estimate uncertainties on the derived output values. We find that all three $gri$ bands show similar results in these simulations. The coordinates of the peak position show both average and median uncertainty around 10\% of the slice length. The surface brightness of the spiral arm at this point varies even less than that. The standard deviations of the ``inward'' and ``outward'' spiral widths increase and are slightly less than 20\%, on average. These uncertainties become bigger for faint parts of the arm, which are usually located in the external regions of the galaxy. However, estimating the arms-to-total ratio for a spiral mask of an individual Monte-Carlo realisation gives an uncertainty of just several percent. For the pitch angle $\psi$ and slope $a$, the uncertainty between the realisations is less than the difference between the individual arms. In total, our simulations demonstrate good stability for the obtained results and that all findings we present in the next sections should be valid if we take into consideration the uncertainties from these Monte-Carlo simulations.

Finally, it is interesting to compare results, which were obtained with automated arm-detection techniques, to our approach.  \citet{2010ApJ...725..534F} and \citet{2014ApJ...790...87D} developed such methods which should be compared with in this paper. \citet{2014ApJ...790...87D} apply a sophisticated computer vision algorithm called {\sc SpArcFiRe}, which detects arm segments and extracts structural information about a spiral arm (the code is publicly available\footnote{\url{http://sparcfire.ics.uci.edu}}). We attempted to apply this code to our galaxy images, but the output results are always covered by many detected arcs (see also a note on this in \citealt{2018ApJ...862...13Y}), which obviously do not belong to the spiral pattern. A possible reason for that is a need in additional image processing or more proper tuning various parameters of their algorithm which, unfortunately, cannot be done by the unprepared user. Another method, which was developed in \citet{2010ApJ...725..534F}, is based on a Fourier transform where arms are detected as all pixels above some threshold in a reconstructed image. Unfortunately, it has been shown that besides discarding asymmetric features and including emission features as part of the arms themselves, the strongest mode in DFFT does not always fit the spiral pattern (see discussion in \citealt{2018ApJ...862...13Y}) and, thus, cannot be reliable. The detailed comparison with other 1DFFT and 2DFFT methods, which can only detect an arm itself but not its boundaries, is beyond the scope of this paper.

\subsection{Pitch angle}
\label{sec:pitch}

\begin{figure}
  \centering
 \includegraphics[width=\columnwidth]{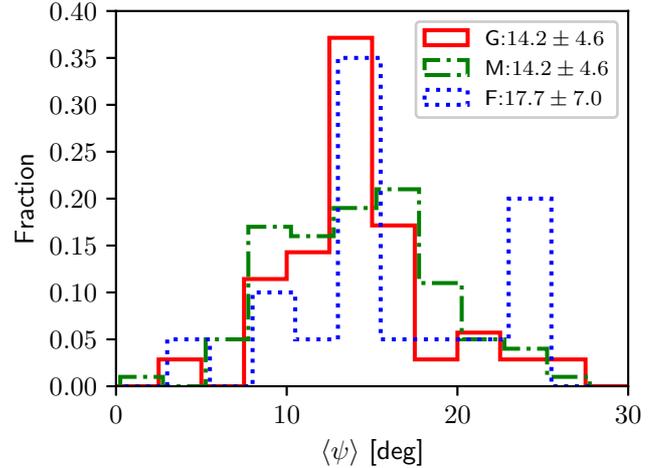}
 \caption{Distribution of grand design (red solid line), multi-armed (green dash-dotted line) and flocculent
   (blue dotted line) galaxies by the pitch angle value.}
 \label{fig:arm_all_pitch(r)_distrib}
\end{figure}

Fig.~\ref{fig:arm_all_pitch(r)_distrib} shows the distribution of galaxies of different arm classes by the mean pitch angle value $\langle \psi \rangle$, averaged for all arms we traced in our analysis. The
mean value of the pitch angle for all sample galaxies is $14.8 \pm 5.3$\textdegree\, which is close to those found in the literature \citep[see e.g.][]{1981AJ.....86.1847K,2001ChJAA...1..395M,2013MNRAS.436.1074S}. Galaxies of all three spiral arm classes demonstrate peaks around 15\textdegree\,, although we notice a second peak around 25\textdegree\, for flocculent galaxies. In general, we do not see a statistically significant difference between the distributions by the pitch angle for the different arm classes. This result is in line with \citet{1992A&AS...93..469P}, who found that there is no correlation between the pitch angle and arm class. However, \citet{2017MNRAS.472.2263H} studied a large stellar mass-complete sample of spiral galaxies and concluded that multi-armed spirals are looser (by 2\textdegree\,) than two-armed spirals. A similar conclusion can be made from the results by \citet{2019arXiv190804246D}, based on the imaging for spiral galaxies from the S$^4$G survey (see their table~2 and fig.~12): grand design spirals have lower pitch angles (by several degrees) than flocculent and multi-armed ones. Though we do not see this difference for our much smaller sample, the average difference in winding of several degrees between the different arm classes is rather small and definitely lies within the standard deviation for each class (see the legend in Fig.~\ref{fig:arm_all_pitch(r)_distrib}) and within the uncertainty of pitch angle estimation. In this and other studies we can clearly see that grand design and multi-armed spirals may almost equally have very small or vice versa very large pitch angles. Therefore, the pitch angle is definitely not a good characteristic to discriminate between different arm classes, and, potentially, arm generation mechanisms.

\begin{figure}
  \centering
 \includegraphics[width=\columnwidth]{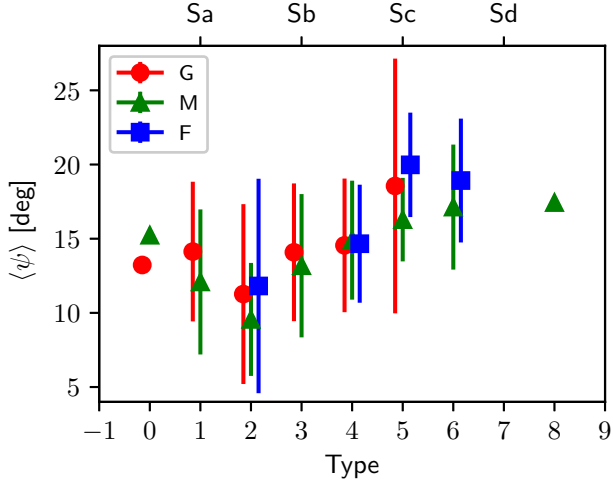}
 \caption{Mean pitch angle values as a function of morphological type.}
 \label{fig:arm_all_pitch(r)_vs_type}
\end{figure}
Fig.~\ref{fig:arm_all_pitch(r)_vs_type} shows the mean values of the pitch angle
as a function of Hubble type. Although there is a weak general trend of increasing pitch
angle with Hubble stage, the overall dispersion is quite large such that each type
has a significant overlapping in the measured pitch angles with other types. Here again we cannot see a difference between the arm classes within a bin of morphological stage.

\begin{figure}
  \centering
 \includegraphics[width=\columnwidth]{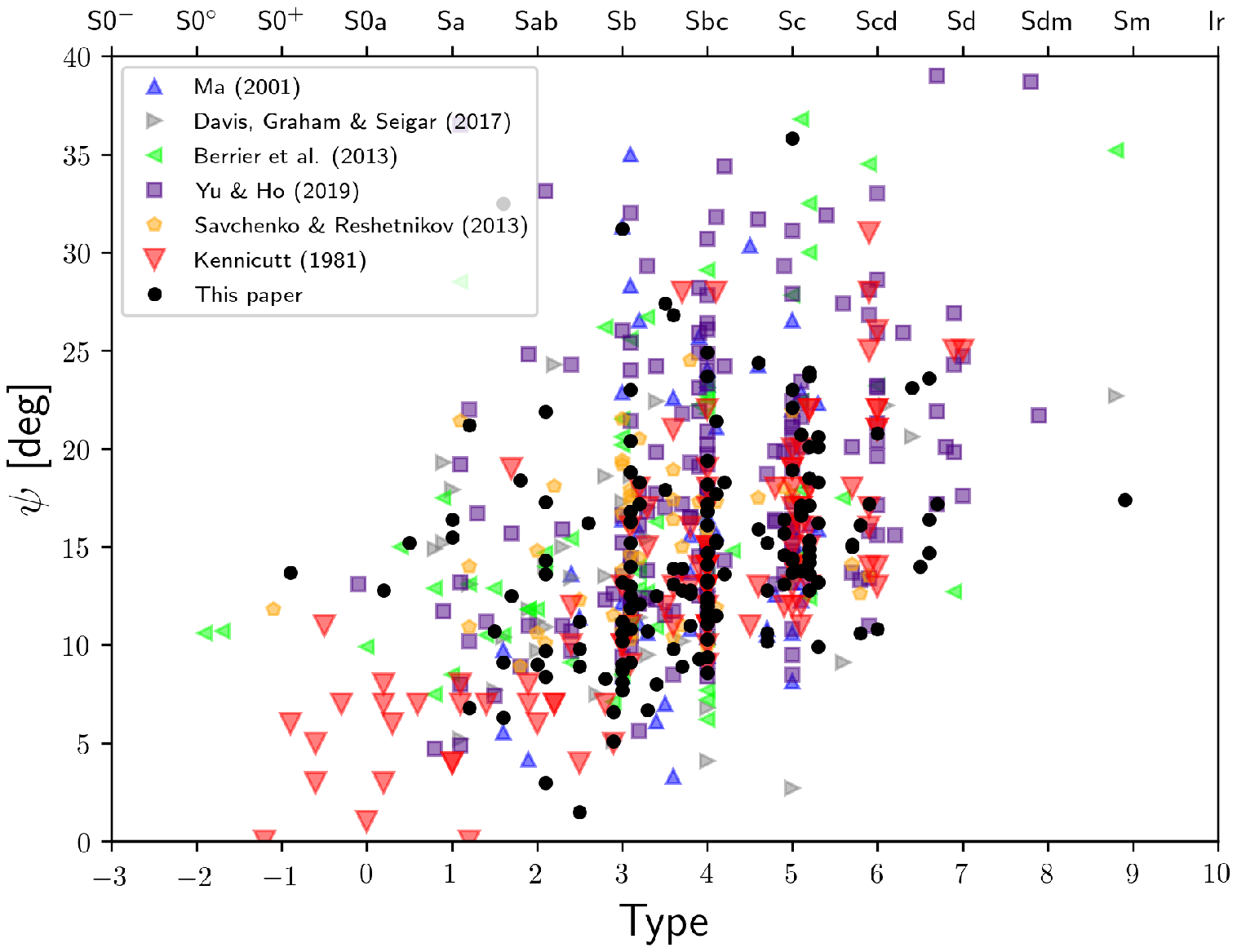}
 \caption{Correlation between the pitch angle and Hubble stage for different results taken from the literature: \citet{2001ChJAA...1..395M}, \citet{2017MNRAS.471.2187D}, \citet{2013ApJ...769..132B}, \citet{2019ApJ...871..194Y}, \citet{2013MNRAS.436.1074S}, \citet{1981AJ.....86.1847K}, and this work.}
 \label{fig:T_pitch}
\end{figure}

We collected estimates of the pitch angle from different sources (most representative samples were used). We show the correlation between the pitch angle and morphological type in Fig.~\ref{fig:T_pitch} (the sources are listed in the caption to the Figure). As one can see, the correlation is very weak for almost all samples. Yet, the trend is still visible: early-type spirals (Sa) are, on average, most tightly wound, whereas late-type spirals (Sd) have a tendency to be open; intermediate spirals may have various pitch angles. It is fair to note here that the results from \citet{1981AJ.....86.1847K} show a better correlation as compared to the other results. We selected common galaxies in each pair from the sources used and found that the results of \citet{1981AJ.....86.1847K} are significantly different from the other sources. If we exclude the Kennicutt's sample from the consideration, the average difference in the pitch angles between the remaining sources is $13\pm14.6$\% of the average pitch angle calculated for these sources; if we compare the Kennictutt's sample alone with the other sources (by pairs), this turns out to be $-19\pm51$\%, with many early-type spirals having a too small pitch angle as compared to the other sources, see Fig.~\ref{fig:T_pitch}, the left bottom corner. The reason for this discrepancy is not clear and, probably, is hidden in the method used for measuring the pitch angle in \citet{1981AJ.....86.1847K}. Also, the weak and scattered trend between the pitch angle and morphology may signify that the modern morphological classification of spiral galaxies is more related to the bulge prominence rather than to the spiral structure (see the discussion in \citealt{2019MNRAS.487.1808M}).

\begin{figure}
  \centering
 \includegraphics[width=\columnwidth]{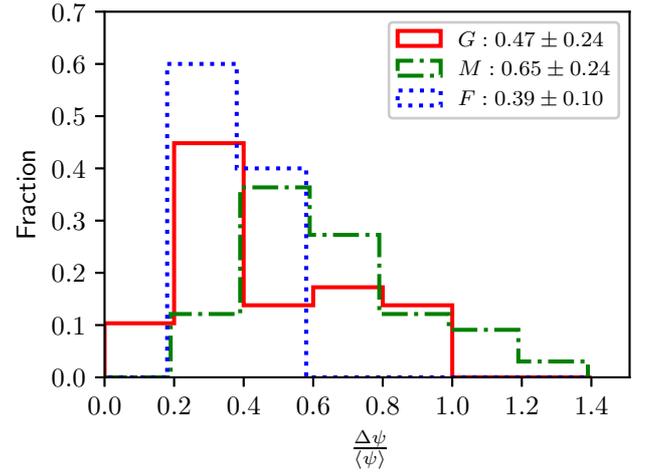}
 \caption{Distribution of the pitch angle variations for grand design (red solid line),
   multi-armed (green dash-dotted line) and flocculent (blue dotted line) classes.}
 \label{fig:arm_all_pitch_variation(r)}
\end{figure}

Fig.~\ref{fig:arm_all_pitch_variation(r)} shows the distribution of the sample galaxies based on the
relative pitch angle variation defined as a standard deviation of all pitch angle values along the spiral arm
divided by the mean value of the pitch angle. Again, this is an average value for all arms under study. Only stage 2 galaxies (galaxies with the full coverage of their spiral
structure, see Sect.~\ref{sec:method_output}) were included in this Figure (the measurement of the relative pitch variation made by a part of the spiral arm would give a lower value).
 This distribution confirms the findings by \citet{2013MNRAS.436.1074S} and \citet{2019arXiv190804246D} that most
galaxies demonstrate significant pitch angle variations: the average value
of the relative pitch angle variation is $0.56 \pm 0.25$, which means that the pitch angle value may vary by more than 50\% along the radius. Grand design galaxies show somewhat
lower pitch angle variations ($0.47 \pm 0.24$) than galaxies with multi-armed spirals ($0.65 \pm 0.24$). The number of flocculent galaxies of stage 2 (only 5 galaxies) is insufficient for making a firm conclusion on the pitch angle variations for these spirals, though the selected flocculent spirals do not demonstrate such large variations as multi-armed galaxies. However, according to the results from \citet{2019arXiv190804246D}, flocculent galaxies also show slightly smaller variations of the pitch angle than grand design and multi-armed galaxies ($\sigma(\psi)=7.9\pm5.4\degr$ versus $8.6\pm4.6\degr$ and $10.1\pm4.0\degr$, respectively).

In total for the sample, we find that the dispersion of the pitch angle along the radius is $\sigma(\psi)=7.2\pm3.3\degr$ which is slightly less than found in \citet{2019arXiv190804246D} ($\sigma(\psi)=9.2\pm5.0\degr$, where they measured the standard deviation of the pitch angle of different spiral segments) for galaxies at 3.6~$\mu$m.  However, similar to \citet{2019arXiv190804246D}, we also find that the differences of the pitch angle along the radius for individual galaxies can be $>15-17\degr$.

\begin{figure}
  \centering
 \includegraphics[width=\columnwidth]{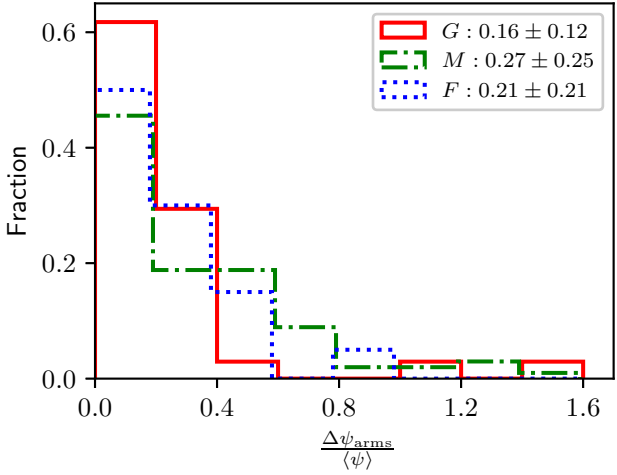}
 \caption{Distribution of the sample galaxies by the pitch angle variations between the arms.
   Grand design, multi-armed and flocculent galaxies are shown in red (solid line), green (dash-dotted line)
   and blue (dotted line) colours, respectively. }
 \label{fig:arm_all_interarm_pitch_variation(r)}
\end{figure}

Fig.~\ref{fig:arm_all_interarm_pitch_variation(r)} shows the distribution of the galaxies from our sample by the
relative difference between the average pitch angles for individual arms (divided by the average pitch angle for all spiral arms)
 $\frac{\Delta\psi_{\mathrm{arms}}}{\left<\psi\right>}$. One can see that for all arm-classes the
distribution looks similar: galaxies tend to have arms with close pitch angle values (the difference between the pitch angles for different arms is, on average, about 20-25\% of their mean value). However, there are also galaxies for which $\frac{\Delta\psi_{\mathrm{arms}}}{\left<\psi\right>}$ may differ up to 100\%.
The mean value of the pitch angle variation between the arms for grand design galaxies is somewhat smaller than for
multi-armed ones ($0.16 \pm 0.12$ and $0.27 \pm 0.25$, respectively), i.e. the spirals in grand design galaxies tend to
have closer values of the pitch angle than in galaxies with multi-armed spiral pattern.

\subsection{Width of spiral arms}
\label{sec:armwidth}

\begin{figure}
  \centering
 \includegraphics[width=\columnwidth]{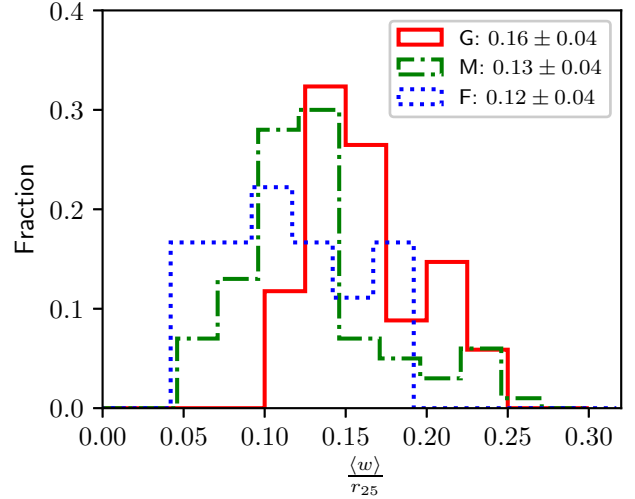}
 \caption{Distribution by average spiral width expressed in units of the optical radius $r_{25}$ for grand design (solid red line), multi-armed (green dash-dotted line) and flocculent galaxies (blue dotted line).}
 \label{fig:arm_all_width_mean(r)_in_r25}
\end{figure}

Fig.~\ref{fig:arm_all_width_mean(r)_in_r25} shows the distribution of galaxies by the mean value of the width of spiral structure. It is
calculated as a mean value of the full width $w=w_1+w_2$ at every point on the spiral structure and is represented here in units
 of the optical radius in the $r$ band (see Sect.~\ref{sec:data_prep}). The mean value of the width for the sample is $(0.14 \pm 0.05)\,r_{25}$. If we express the arm width in kpc, for grand design spirals $w=3.3\pm1.2$\,kpc, $2.5\pm0.9$\,kpc for multi-armed and $2.1\pm1.4$\,kpc for flocculent galaxies. Both quantities of $w$ show that grand design spirals have a slightly larger width, whereas flocculent galaxies have the narrowest spirals and multi-armed galaxies lie in between.

\begin{figure}
  \centering
 \includegraphics[width=\columnwidth]{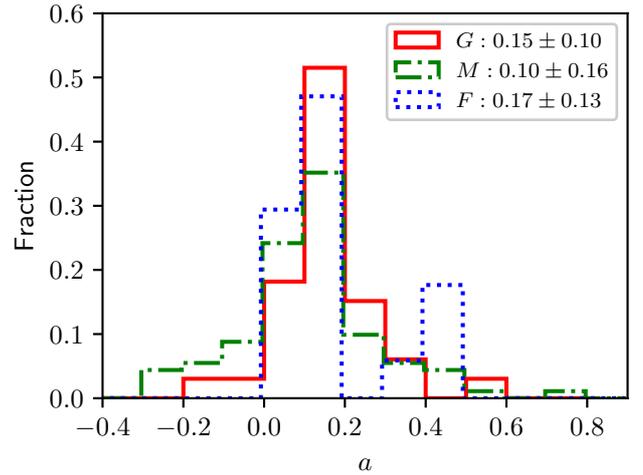}
 \caption{Distribution by the slope for the radius -- spiral width relation in the $r$ band.}
 \label{fig:arm_all_slope_mean(r)_stage2}
\end{figure}

Fig.~\ref{fig:arm_all_slope_mean(r)_stage2} shows the distribution of the galaxies by the slope $a$ for the dependence between the
galactocentric radius and spiral width (see Sect.~\ref{sec:method_output} and Fig.~\ref{fig:pgc_2182_widths}), where we can see how the arm width changes with radius.
One can see that most galaxies in our sample (85.8\%) demonstrate a positive slope which means that the width of their
spiral arms increases with galactocentric distance. This result is in agreement with \cite{2015ApJ...800...53H}
where for four galaxies it was shown that the width of all their major arms increases with radius.
However, in our sample we can also find galaxies with the spirals arms, which have a nearly constant width, and also
galaxies with the arms becoming thinner at the periphery. For example, PGC\,49881 ($a=-0.18$) has quite abrupt and unusual spirals which indeed do not look wider in the outer galaxy region. The multi-armed spiral galaxy PGC\,48478 ($a=-0.30$) also shows thinner spiral arms when going outwards. By contrast, PGC\,49401 exhibits much wider spiral arms ($a=0.78$) at large radii than in the inner region.
If we compare the distributions by this parameter for different arm classes, we can see that multi-armed galaxies have slightly lower average values of $a$ than grand design galaxies. However, statistically this difference is not significant.

\begin{figure}
  \centering
 \includegraphics[width=\columnwidth]{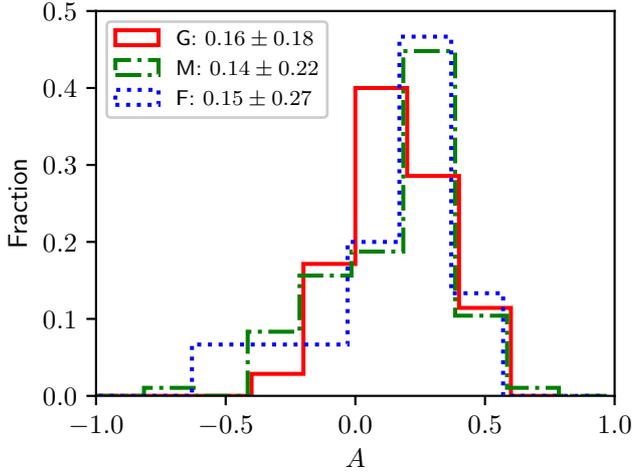}
 \caption{Distribution of the sample galaxies by the asymmetry values in the $r$ band.}
 \label{fig:arm_all_asymmetry(r)_stage2}
\end{figure}

Fig.~\ref{fig:arm_all_asymmetry(r)_stage2} shows the distribution of the sample galaxies by the value of the mean asymmetry
of the slices perpendicular to the spiral arms (see eq.~\ref{eq:asymmetry}). The mean asymmetry is computed as an average value for each slice
for all spirals in a galaxy. Most galaxies (75.3\%) in our sample have a positive asymmetry value, which means that the inner
side of their arms $w_1$ is narrower, on average, than the outer side $w_2$. The average value of the asymmetry $0.14\pm0.23$ means the $w_2$ half-width is approximately 16\% larger than the $w_1$ half-width. Numerical simulations predict \citep{1981ApJ...243..432Y} that
for the density-wave scenario this should be the case: inside of the co-rotation radius, the shock is formed behind the
spiral arm, and, therefore, we can expect that $w_2$ should be larger than $w_1$. Interestingly, all three spiral classes do not show any difference in the distributions by the width asymmetry.

\subsection{Luminosity and colour of spiral structure}
\label{sec:armLum}

\begin{figure}
  \centering
  \includegraphics[width=\columnwidth]{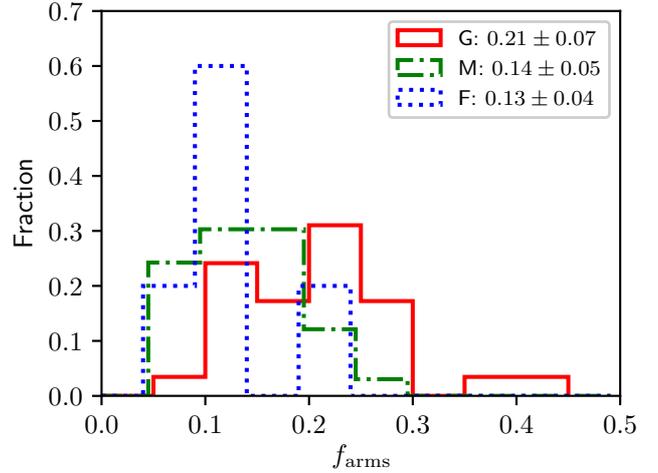}
  \caption{Distribution of the sample galaxies by their arms-to-total ratio. The red solid line depicts
  grand design galaxies, green dash-dotted line and blue dotted line show multi-armed and flocculent galaxies.}
  \label{fig:arm_all_arm-to-total_aper(r)_stage2}
\end{figure}

Fig.~\ref{fig:arm_all_arm-to-total_aper(r)_stage2} demonstrates the distribution of the galaxies by the arms-to-total
 ratio in the $r$ band. Only stage 2 galaxies were used to make this plot since we need
a full coverage of the spiral pattern by a spiral mask to compute a valid arms-to-total ratio (see Sect.~\ref{sec:method_output}).
The mean value for all galaxies is $\left<f_{\mathrm{arms}}\right> = 0.17\pm 0.07$. The
grand design galaxies show, on average, brighter spirals ($\left< f_{\mathrm{arms}} \right> = 0.21\pm 0.07$) than
multi-armed ($\left< f_{\mathrm{arms}} \right> = 0.14\pm 0.05$) and flocculent galaxies ($\left< f_{\mathrm{arms}} \right> = 0.13\pm 0.04$). Similar results were reported in
\citet{2011MNRAS.414..538K} for the NIR. They demonstrated that grand design galaxies
tend to have a higher non-axisymmetric power level value, which was used as a measure of the
spiral arms strength in the galaxy.

\begin{figure}
  \centering
  \includegraphics[width=\columnwidth]{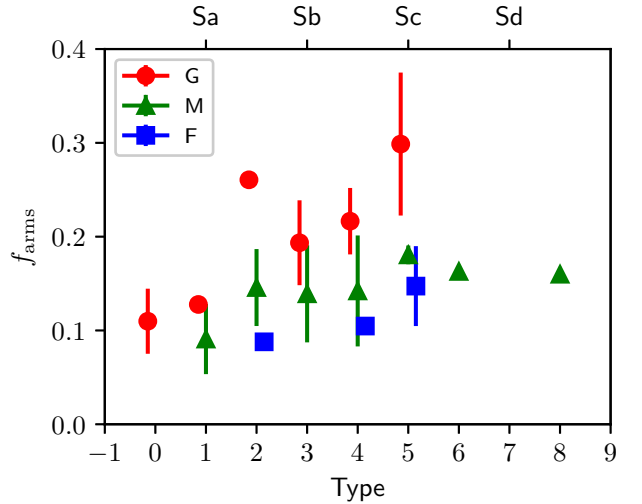}
  \caption{Dependence of the arms-to-total ratio on morphological type for grand design (red dots),
    multi-armed (green triangles) and flocculent (blue squares) galaxies.}
  \label{fig:arm_all_arm-to-total_aper(r)_stage2_vs_type}
\end{figure}

Fig.~\ref{fig:arm_all_arm-to-total_aper(r)_stage2_vs_type} shows the relation between the arms-to-total ratio
and Hubble type separately for grand design, multi-armed and flocculent galaxies. One can see that
in the case of grand design galaxies, there is a statistically significant correlation between these parameters:
the light fraction of the spiral pattern for later types is about two times larger than for earlier types. The corresponding
correlation for the multi-armed galaxies is considerably weaker (the Pearson correlation coefficient is only 0.24 versus
0.66 for the grand design galaxies). The multi-armed galaxies also have a lower arms-to-total ratio than the grand design galaxies
of the corresponding type.
We note that in \cite{2015MNRAS.446.4155K} it was reported that there is no
obvious correlation between the Hubble type and the spiral arm strength measured by the amplitude of the $m=2$ Fourier mode.

\begin{table}
  \centering
  \begin{tabular}[h]{cccc}
    \hline
    Band & $f_{\mathrm{arm}s}^a$ & $f_{\mathrm{arms}}^c$ & $\kappa$\\
    \hline
    $g$    &      $0.13 \pm 0.04$     &    $0.30 \pm 0.10$        &  0.64  \\
    $r$     &      $0.12 \pm 0.03$     &    $0.26 \pm 0.08$        &  0.66  \\
    $i$     &      $0.12 \pm 0.03$     &    $0.24 \pm 0.08$        &  0.61  \\
  \end{tabular}
  \caption{Arms-to-total ratios for early (Hubble type $a$, second column) and late (Hubble type
    $c$, third columns) spirals in the SDSS passbands $g$, $r$ and $i$. The last column is the Pearson correlation
    coefficient for $T-f_{\mathrm{arms}}$ relation.}
  \label{tab:arms_to_total_values}
\end{table}

Table~\ref{tab:arms_to_total_values} summarises the behaviour of arms-to-total -- type correlations in
different passbands. One can see that for the $g$ band, both the arms-to-total ratio and the difference between
early and late types is the highest. This result is in agreement with \citet{2018ApJ...862...13Y}, where they
show that the strength of Fourier modes, which they used as a spiral strength measure, is systematically larger
in bluer bands.

\begin{figure}
  \centering
  \includegraphics[width=\columnwidth]{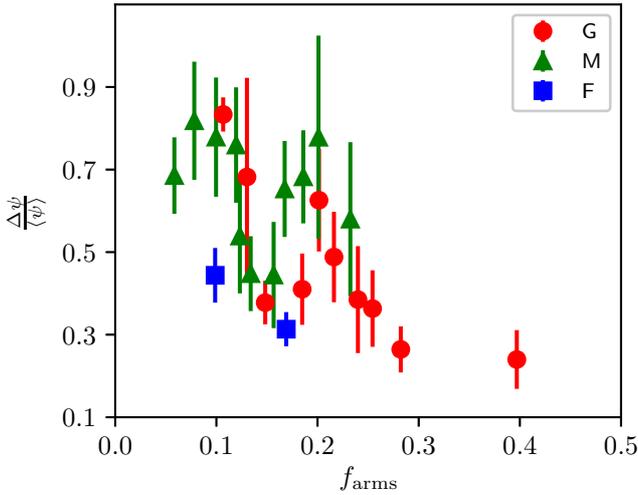}
  \caption{Pitch angle variation as a function of arms-to-total ratio in the $r$ band for grand design (red dots), multi-armed (green triangles) and
    flocculent (blue squares) classes of galaxies. Each point represents an averaging by three galaxies to enhance the signal-to-noise
    ratio.}
  \label{fig:arm-to-total_pitchvar}
\end{figure}

Fig.~\ref{fig:arm-to-total_pitchvar} presents the pitch angle variation as an arms-to-total ratio for grand design,
multi-armed and flocculent galaxies. One can see in the case of grand design spirals that there is a significant
anti-correlation (the Pearson coefficient $\kappa=-0.56$) between these two quantities: brighter spirals tend to have
a lower pitch angle variation (they are closer to a pure logarithmic shape) than weaker ones. On the other hand, multi-armed galaxies
do not show such a correlation.

\begin{figure}
  \centering
  \includegraphics[width=\columnwidth]{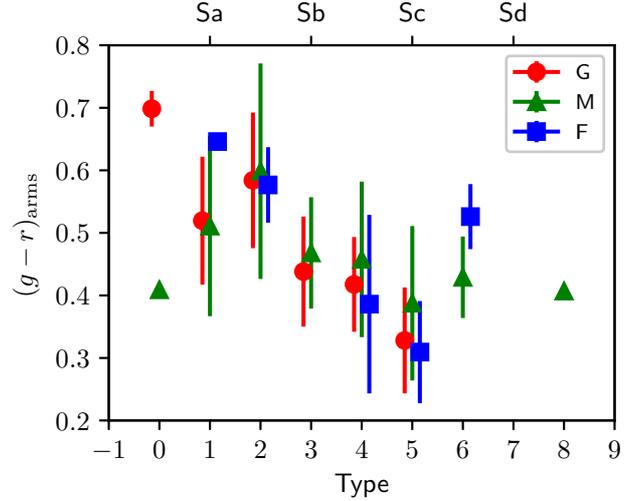}
  \caption{Colour of the spiral arms $g-r$ as a function of Hubble type for grand design (red dots),
    multi-armed (green triangles) and flocculent (blue squares) galaxies.}
  \label{fig:arm_all_aper_color(g-r)_stage2}
\end{figure}

Fig.~\ref{fig:arm_all_aper_color(g-r)_stage2} demonstrates the mean $g-r$ colour of the spiral arms as a function
of Hubble type. One can see that there is a statistically significant correlation between the
colour of spiral structure and Hubble type for all classes of spirals, although with some outlying points due to a poor
statistics in the case of flocculent galaxies. It is also worth noticing that within the same morphological type all three spiral classes
may have practically the same colour.
We conclude that the arms-to-total ratio and the colour of the arms have a more notable impact on the assigned Hubble
type of a galaxy than the pitch angle of its spiral structure. The fraction of light in the arms and their colour could explain much about the galaxy classification, which correlates with both the characteristics of the spiral arms and the ability to recognize them and sort their properties into classes.

\section{Discussion}
\label{sec:discussion}

\subsection{Important issues}
\label{sec:dis:caveats}
Here we consider how different factors may influence our results.

An important issue is how the uncertainty in the correction for the galaxy projection may potentially affect the measured quantities
of the spiral structure. First, the sample galaxies are viewed almost face-on (the average inclination for our sample is
$\langle \mathfrak{i} \rangle=36\pm13\degr$, as taken from HyperLeda). In addition to that, in Sect.~\ref{sec:reliability} we have checked that our measures of $q_{25}$ correspond well to those listed in HyperLeda. Therefore, we should not expect that a small uncertainty in the image de-projection would significantly
change the results of our fitting of the spiral structure. However, to estimate this effect, we made a series of de-projections with
changing the galaxy inclination within $\pm 8\degr$ and the position angle within $\pm 10\degr$ (which are conservative estimates of the errors on $\mathfrak{i}$ and $\mathrm{PA}$). We show, that the mean value of the pitch angle is almost not affected by the de-projection error: its rms value is only
$\sim 2\%$. The mean dispersion of the total magnitude of the arm structure is $\approx 0.05$, the mean width varies at about 7\% level.
The most affected parameter is the pitch angle variation with an rms of 17\%.
In general, our test gave us the confidence that the variation of the estimated parameters of the spiral structure due to the de-projection errors is not significant.

As in this study we work with the optical data, the role of dust must be considered. The dust opacity is greater within spiral arms \citep{2005A&A...444..109H} than in the inter-arm or outside-arm disc regions. This is related to a higher surface density of molecular clouds in spiral arms and associated star formation. Notwithstanding the higher dust extinction in the spiral arms, measuring the arm pitch angle should not be affected by it as we fit the maxima on the photometric cuts perpendicular to the spiral arms and assume that most dust obscuration is located in narrow lanes alongside the inner edge of the spiral arm. Thus, we expect that there should not be a shift of the maximum emission on this cut due to the dust absorption. As our estimates of the pitch angle are consistent with the works where different approaches for tracing spiral arms are used (see Sect.~\ref{sec:reliability}), we are confident that dust does not significantly influence our measured pitch angles.

As to the arm width, arm fraction and arm colours, this statement needs to be validated. To estimate the effect of internal extinction, we adopt the following correction for the apparent disc magnitude from \citet{2008ApJ...678L.101D} and use this value as a lower estimate for the arms: $m_\mathrm{disc,obs} - m_\mathrm{disc,intrin} = d_1+d_2\,[1-\cos(\mathfrak{i})]^{d_3}\,,$
where $\mathfrak{i}$ is estimated in Sect.~\ref{sec:data_prep} and the coefficients in this equation depend on wavelength and are provided in \citet{2008ApJ...678L.101D} for various passbands, including the $g$, $r$ and $i$ wavebands. Thus, if we correct the observed total galaxy luminosity for the inclination-dependent extinction effect using $\Delta M_{r} = 1.27\,(\lg q_{25})^2$ from \citet{2008ApJ...687..976U}, we can estimate that the fraction of the the spiral arms is underestimated, on average, by at least 5\%. The corrected colour $g-r$ appears to be shifted by at least -0.06 mag.

The influence of dust attenuation on the measurements of the arm width should be considered in great detail and will be addressed by us in a further work. Here we can only assume that its influence cannot be neglected, at least for the $w_1$ half-width, but, at the same time, it should not change the main result of this paper: for most sample galaxies we clearly observe an increase of the arm width with radius.

We should also note that our sample inevitably suffers from different selection effects. For example, by the construction of the sample, we undercount flocculent galaxies of later types: in our sample the average type for grand design galaxies is $\langle T \rangle =3.5\pm1.3$, for multi-armed galaxies  $\langle T \rangle =4.0\pm1.3$, and for flocculent galaxies  $\langle T \rangle =4.2\pm1.3$. For comparison, \citet{2019arXiv190804246D} gives $T=3$, $T=5$ and $T=6$ for the same arm classes, but on the basis of 391 galaxies from the S$^4$G survey. Thus, the results regarding the flocculent arm class in our sample should be considered with caution.

\subsection{Influence of the environment on spiral structure}
\label{sec:dis:environment}
\begin{figure*}
  \centering
  \includegraphics[width=\columnwidth]{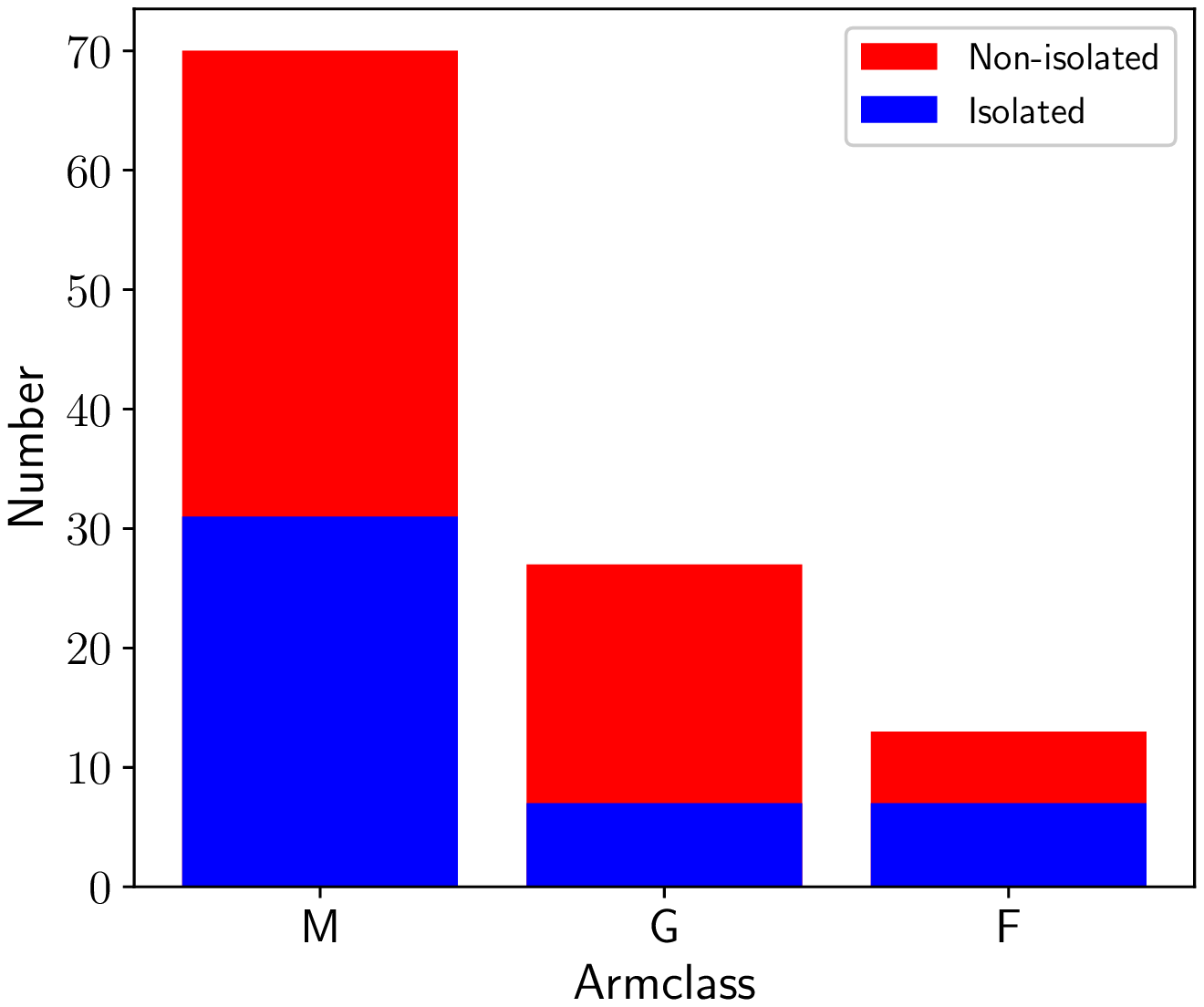}
  \includegraphics[width=\columnwidth]{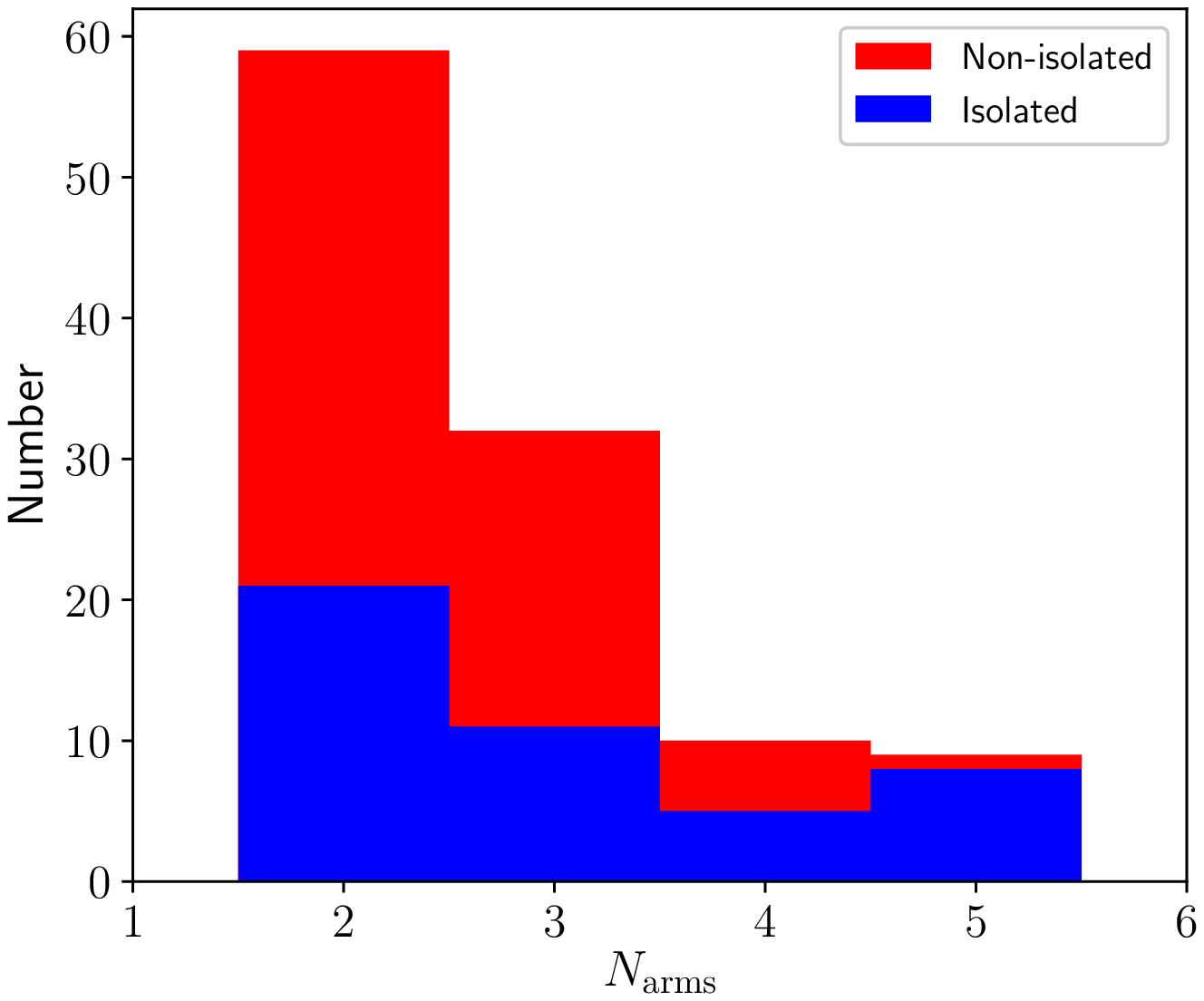}
  \caption{Distribution of the sample galaxies by arm class (leftmost panel) and arm number (rightmost panel) for isolated and non-isolated galaxies.}
  \label{fig:arm_environment}
\end{figure*}

Here we analyse how the parameters of spiral structure depend on the environment. In Table~\ref{general_pars} we listed the galaxy environment as belonging to a group, a triple, a group or a cluster, as well as being isolated. Here we simplify this classification to `isolated' (45 galaxies) or `non-isolated' (110 galaxies). In Fig.~\ref{fig:arm_environment} we show the distributions of the sample galaxies by arm class and number of spiral arms. As one can see, the percentage of grand design spirals among the isolated galaxies is lower as compared to the non-isolated galaxies (15\% versus 25\%, respectively). However, the percentage of $N_\mathrm{arms}=2$ spirals  in isolated galaxies is 47\% (21 galaxies) versus
53\% in non-isolated galaxies, with a slightly higher number of $N_\mathrm{arms}=3$ spirals (29\% vesrus 22\%). The lack of $N_\mathrm{arms}>3$ spirals can be explained by the selection effect, as we excluded many multi-armed and flocculent galaxies with indistinct spiral arms. We can conclude that in our sample the number of grand design spirals in non-isolated galaxies is slightly larger than in isolated galaxies. Interestingly, the same conclusion can be done for a larger sample of 391 galaxies from \citep{2019arXiv190804246D}: 17\% of their grand design galaxies are isolated versus 23\% of non-isolated grand design galaxies (20\% of isolated galaxies have $N_\mathrm{arms}=2$ versus 16\% of non-isolated galaxies). 
Also, for our sample we considered the other parameters of spiral structure (pitch angle, arm width etc.) depending on the environment, and did not find any statistically significant difference for any of them.

\subsection{Dependence on wavelength}
\label{sec:dis:bands}
We analysed how the retrieved parameters of spiral structure vary with waveband. As the $gri$ bands are located closely to each other, we did not find any statistical difference between these bands for the pitch angle $\psi$, width $w$, width asymmetry $A$, and width slope $a$. However, as was noted in Sect.~\ref{sec:armLum}, there is a decrease of the arms fraction going from the $g$ to $i$ passband:  $f_\mathrm{arms}^{r} = (0.91\pm0.42)\,f_\mathrm{arms}^{g}$ ($\kappa=0.98$) and $f_\mathrm{arms}^{i} = (0.85\pm0.51)\,f_\mathrm{arms}^{g}$ ($\kappa=0.96$).

\subsection{Dominant mechanisms for generating spiral arms}
\label{sec:dis:mechanisms}

In this paper we try to distinguish from observations between different spiral arm classes assuming that they correspond to different dominant models for generating the observed spiral structure. There have been proposed different observational tests for discriminating between, for example, the quasi-stationary density wave theory, dynamic spirals, bar-driven spirals and tidally-induced arms \citep[see a review by][]{2014PASA...31...35D}.

For explaining the large number of grand design galaxies, three possible scenarios are often discussed. In the quasi-stationary wave theory density waves are most likely to be stable in galaxies with two-armed spiral structure \citep{1967IAUS...31..313L}. \citet{1979ApJ...233..539K} suggested that spirals can be induced by tidal interactions or can be bar driven. \cite{1982MNRAS.201.1021E} found only 10\% of isolated galaxies having a grand design structure, whereas flocculent galaxies are most common among isolated and barred galaxies. In our sample, we do not see such a strong difference in spiral pattern between the isolated and non-isolated galaxies: grand design galaxies are almost equally found as in groups and binaries, as in isolation. Most likely, it is related to the selection effect, as we carefully selected galaxies with a rather traceable spiral pattern and without strong interactions. Also, it should be stressed that determining truly isolated galaxies and those that have not undergone a recent merger event is not a trivial task \citep{2007A&A...472..121V}, taking into account that grand design galaxies should retain their spiral structure for $\sim1$~Gyr after an interaction \citep{2008ApJ...683...94O}.

Although in our study we did not study the bar parameters and their link to spiral structure (we plan to do this in a subsequent paper), we found no influence of the bar presence on all parameters of the spiral structure under study. At the same time, we find that two out of 7 isolated grand design galaxies have bars, whereas 12 galaxies out of 27 non-isolated grand design galaxies demonstrate a bar structure.

One of the most interesting results of this study is that the spiral arms in grand design galaxies have a lower difference between their overall pitch angles than in multi-armed galaxies. In the case of recent merger, which could induce a grand design spiral pattern in one of the galaxies, we can expect to see a significant difference in the pitch angles of the individual arms shortly after the interaction \citep[see e.g. the results of modelling for M\,51,][]{2010MNRAS.403..625D}. However, all our grand design galaxies do not show a strong interaction, and, thus, even if a merger took place in the past, this difference could merely vanish with time. The steady-density wave theory states that all spiral arms should have the same pitch angle as the spiral pattern has the same angular speed everywhere in the galaxy. Even though the grand design galaxies in our sample show a lower difference in the pitch angles of their arms than the other two classes, this difference in significantly non-zero and is beyond the errors of the pitch angle estimation.  On the other hand, the swing amplification theory does predict that the spiral arms in a galaxy may have different pitch angles: due to differential rotation, one arm can become more tightly wound as time goes by, while new spiral arms with larger pitch angles start to grow \citep{2013ApJ...763...46B}.

The most important characteristic, we studied in this work, is the arm width. We found, that for the vast majority of the sample galaxies the arm width steadily grows with galactocentric distance, in agreement with \citep{2014ApJ...783..130R,2015ApJ...800...53H}. As can be seen from simulations by \citep{2010MNRAS.403..625D}, the arm width is indeed growing significantly with radius for tidally-induced spirals. On the other hand, the swing amplification scenario may also explain this trend, as `swinging inherently fans material outward' (\citealt{2015ApJ...800...53H}, see also \citealt{1981seng.proc..111T}). In the quasi-stationary density wave theory this question has not been addressed.

Therefore, based on the results we obtained in this work, we cannot determine for certain which mechanisms are responsible for the observed spiral structures. A multi-wavelength study for a large sample of galaxies is highly needed where different constituents in spirals (gas, dust, stars) can be distinguished and traced.

\section{Conclusions}
\label{sec:conclusion}
In this paper we used SDSS $gri$-band images to perform a detailed analysis of the spiral structure in the 155 face-on non-interacting spiral galaxies. Our sample is mostly composed of multi-armed and grand design galaxies, with a fraction of flocculent galaxies, in which at least one arm could be visually traced.

In our study we used an approach of fitting an arm profile by creating a series of photometric cuts perpendicular to its direction. For tracing the spiral structure, we start from the inner galaxy region (the ends of a bar or where spirals begin to be visible beyond the main body of a bulge) to the outermost regions to which we are able to trace the spirals. Along with the pitch angle and its variation with radius, we measured, for the first time based on the spiral view in the optical, the arm width and its variation with galactocentric distance, the fraction of the spiral structure to the total galaxy luminosity, and its colour. Coupling with the arm number and the general arm class (grand design, flocculent or multi-armed), these quantities define the photometric properties of the spiral structure in galaxies, which can in principal be used to distinguish between different mechanisms of generating spiral structure in galaxies.

The main results of this paper are as follows.
\begin{enumerate}
\item  The pitch angle does not exhibit a strong dependence on arm class: galaxies of all three classes may have a small or vice versa large pitch angle. Also, we did not find any correlation of the pitch angle with the bar presence or galaxy environment in the context of its belonging to a pair or a group, or being isolated. The difference of the pitch angle in the $gri$ bands is not significant.
\item The pitch angle demonstrates a weak correlation with morphological type in the sense that early spirals tend to be more tightly wound than the other types and late spirals may have a wide range of pitch angles. The pitch angle variation along the radius may be large (on average, approximately 50\% or 7.5 deg), which is in line with previous studies by \citet{2013MNRAS.436.1074S} and \citet{2019arXiv190804246D}. It seems smaller for flocculent and grand design galaxies and larger for multi-armed galaxies. We found that grand design galaxies show a smaller difference of pitch angle for individual arms than multi-armed galaxies.
\item The arm width, normalised by the optical radius, tends to be larger for grand design galaxies and smaller for flocculent galaxies, though the significance of this difference is weak. Also, we do not find a difference between the arm classes by the width asymmetry: on average, the inner half-width $w_1$ is less than the outer half-width $w_2$, which translates into a positive asymmetry value. Also, for most sample galaxies we observe an increase of the arm width with galactocentric distance, showing no dependence on the arm class. As for the pitch angle, we find no correlation of the arm width with the waveband considered, bar presence, and galaxy environment.
\item Grand design spirals, as one would expect, contribute slightly more light to the total galaxy luminosity than the other two spiral classes. This arm fraction decreases with increasing wavelength. The average fraction for grand design galaxies $f_\mathrm{arms}\approx0.2$, which in some cases may increase up to 0.4-0.5. Also, for grand design galaxies we see a strong correlation between the arms-to-total ratio and Hubble stage (late spirals have a larger fraction of the spiral pattern than early ones) and anti-correlation between the arms fraction and the pitch angle variation (more luminous spirals exhibit smaller variations of the pitch angle), whereas we do not see such a dependence for multi-armed spirals.
\item We find a tight correlation between the arm colour and morphological type: spirals become bluer in later-type galaxies for all spiral arm classes. Within one morphological type all three spiral classes do not differ significantly by their colour.

We stress that the conclusions on the flocculent class should be considered with caution as we significantly undercount flocculent spirals in our sample.

Observationally, based on only three optical bands and dividing galaxies into three classes by spiral pattern, we do not readily distinguish between these arm classes for the sample galaxies. The significant difference in them is only the average arm width and arm fraction, which merely reflect the scheme we used to classify the spirals. We assume that a study on a wider range of wavelengths would definitely help to find a strong observational difference between these three classes. Also a detailed look into the structure of galaxies (e.g. the view of spirals in the inner versus outer region) and the interplay between the spiral arms and main structural components (disc, bulge, bar) might shed light on this problem.

\end{enumerate}

\section*{Acknowledgements}
We thank the anonymous reviewer for the careful reading of our manuscript.

The work was supported by the RFBR grant 18-32-00194.
Funding for the SDSS has been provided by the Alfred P. Sloan
Foundation, the Participating Institutions, the National Science
Foundation, the US Department of Energy, the National Aeronautics
and Space Administration, the Japanese Monbukagakusho, the
Max Planck Society, and the Higher Education Funding Council for
England. The SDSS Web Site is \url{http://www.sdss.org/}.
This research makes use of the NASA/IPAC Extragalactic Database (NED) which is operated by the Jet Propulsion Laboratory, California Institute of Technology, under contract with the National Aeronautics and Space Administration, and the LEDA database (\url{http://leda.univ-lyon1.fr}).

This research made use of Astropy,\footnote{http://www.astropy.org} a community-developed core Python package for Astronomy \citep{astropy:2013, astropy:2018}.

\bibliographystyle{mnras}
\bibliography{article} 

\begin{thebibliography}{}
\makeatletter
\relax
\def\mn@urlcharsother{\let\do\@makeother \do\$\do\&\do\#\do\^\do\_\do\%\do\~}
\def\mn@doi{\begingroup\mn@urlcharsother \@ifnextchar [ {\mn@doi@}
  {\mn@doi@[]}}
\def\mn@doi@[#1]#2{\def\@tempa{#1}\ifx\@tempa\@empty \href
  {http://dx.doi.org/#2} {doi:#2}\else \href {http://dx.doi.org/#2} {#1}\fi
  \endgroup}
\def\mn@eprint#1#2{\mn@eprint@#1:#2::\@nil}
\def\mn@eprint@arXiv#1{\href {http://arxiv.org/abs/#1} {{\tt arXiv:#1}}}
\def\mn@eprint@dblp#1{\href {http://dblp.uni-trier.de/rec/bibtex/#1.xml}
  {dblp:#1}}
\def\mn@eprint@#1:#2:#3:#4\@nil{\def\@tempa {#1}\def\@tempb {#2}\def\@tempc
  {#3}\ifx \@tempc \@empty \let \@tempc \@tempb \let \@tempb \@tempa \fi \ifx
  \@tempb \@empty \def\@tempb {arXiv}\fi \@ifundefined
  {mn@eprint@\@tempb}{\@tempb:\@tempc}{\expandafter \expandafter \csname
  mn@eprint@\@tempb\endcsname \expandafter{\@tempc}}}

\bibitem[\protect\citeauthoryear{{Albareti} et~al.,}{{Albareti}
  et~al.}{2017}]{2017ApJS..233...25A}
{Albareti} F.~D.,  et~al., 2017, \mn@doi [\apjs] {10.3847/1538-4365/aa8992},
  \href {https://ui.adsabs.harvard.edu/abs/2017ApJS..233...25A} {233, 25}

\bibitem[\protect\citeauthoryear{{Astropy Collaboration} et~al.,}{{Astropy
  Collaboration} et~al.}{2013}]{astropy:2013}
{Astropy Collaboration} et~al., 2013, \mn@doi [\aap]
  {10.1051/0004-6361/201322068}, \href
  {http://adsabs.harvard.edu/abs/2013A%26A...558A..33A} {558, A33}

\bibitem[\protect\citeauthoryear{{Athanassoula}}{{Athanassoula}}{2012}]{2012MNRAS.426L..46A}
{Athanassoula} E.,  2012, \mn@doi [\mnras] {10.1111/j.1745-3933.2012.01320.x},
  \href {https://ui.adsabs.harvard.edu/abs/2012MNRAS.426L..46A} {426, L46}

\bibitem[\protect\citeauthoryear{{Athanassoula}, {Romero-G{\'o}mez}  \&
  {Masdemont}}{{Athanassoula} et~al.}{2009a}]{2009MNRAS.394...67A}
{Athanassoula} E.,  {Romero-G{\'o}mez} M.,   {Masdemont} J.~J.,  2009a, \mn@doi
  [\mnras] {10.1111/j.1365-2966.2008.14273.x}, \href
  {https://ui.adsabs.harvard.edu/abs/2009MNRAS.394...67A} {394, 67}

\bibitem[\protect\citeauthoryear{{Athanassoula}, {Romero-G{\'o}mez}, {Bosma}
  \& {Masdemont}}{{Athanassoula} et~al.}{2009b}]{2009MNRAS.400.1706A}
{Athanassoula} E.,  {Romero-G{\'o}mez} M.,  {Bosma} A.,   {Masdemont} J.~J.,
  2009b, \mn@doi [\mnras] {10.1111/j.1365-2966.2009.15583.x}, \href
  {https://ui.adsabs.harvard.edu/abs/2009MNRAS.400.1706A} {400, 1706}

\bibitem[\protect\citeauthoryear{{Athanassoula}, {Romero-G{\'o}mez}, {Bosma}
  \& {Masdemont}}{{Athanassoula} et~al.}{2010}]{2010MNRAS.407.1433A}
{Athanassoula} E.,  {Romero-G{\'o}mez} M.,  {Bosma} A.,   {Masdemont} J.~J.,
  2010, \mn@doi [\mnras] {10.1111/j.1365-2966.2010.17010.x}, \href
  {https://ui.adsabs.harvard.edu/abs/2010MNRAS.407.1433A} {407, 1433}

\bibitem[\protect\citeauthoryear{{Baba}, {Saitoh}  \& {Wada}}{{Baba}
  et~al.}{2013}]{2013ApJ...763...46B}
{Baba} J.,  {Saitoh} T.~R.,   {Wada} K.,  2013, \mn@doi [\apj]
  {10.1088/0004-637X/763/1/46}, \href
  {https://ui.adsabs.harvard.edu/abs/2013ApJ...763...46B} {763, 46}

\bibitem[\protect\citeauthoryear{{Baillard} et~al.,}{{Baillard}
  et~al.}{2011}]{2011A&A...532A..74B}
{Baillard} A.,  et~al., 2011, \mn@doi [\aap] {10.1051/0004-6361/201016423},
  \href {https://ui.adsabs.harvard.edu/abs/2011A&A...532A..74B} {532, A74}

\bibitem[\protect\citeauthoryear{{Berrier} et~al.,}{{Berrier}
  et~al.}{2013}]{2013ApJ...769..132B}
{Berrier} J.~C.,  et~al., 2013, \mn@doi [\apj] {10.1088/0004-637X/769/2/132},
  \href {https://ui.adsabs.harvard.edu/abs/2013ApJ...769..132B} {769, 132}

\bibitem[\protect\citeauthoryear{{Bertin}}{{Bertin}}{2011}]{2011ASPC..442..435B}
{Bertin} E.,  2011, in {Evans} I.~N.,  {Accomazzi} A.,  {Mink} D.~J.,   {Rots}
  A.~H.,  eds,  Astronomical Society of the Pacific Conference Series Vol. 442,
  Astronomical Data Analysis Software and Systems XX. p.~435

\bibitem[\protect\citeauthoryear{{Bertin} \& {Arnouts}}{{Bertin} \&
  {Arnouts}}{1996}]{1996A&AS..117..393B}
{Bertin} E.,  {Arnouts} S.,  1996, \mn@doi [\aaps] {10.1051/aas:1996164}, \href
  {https://ui.adsabs.harvard.edu/abs/1996A%26AS..117..393B} {117, 393}

\bibitem[\protect\citeauthoryear{{Bertin}, {Lin}, {Lowe}  \&
  {Thurstans}}{{Bertin} et~al.}{1989a}]{1989ApJ...338...78B}
{Bertin} G.,  {Lin} C.~C.,  {Lowe} S.~A.,   {Thurstans} R.~P.,  1989a, \mn@doi
  [\apj] {10.1086/167182}, \href
  {https://ui.adsabs.harvard.edu/abs/1989ApJ...338...78B} {338, 78}

\bibitem[\protect\citeauthoryear{{Bertin}, {Lin}, {Lowe}  \&
  {Thurstans}}{{Bertin} et~al.}{1989b}]{1989ApJ...338..104B}
{Bertin} G.,  {Lin} C.~C.,  {Lowe} S.~A.,   {Thurstans} R.~P.,  1989b, \mn@doi
  [\apj] {10.1086/167183}, \href
  {https://ui.adsabs.harvard.edu/abs/1989ApJ...338..104B} {338, 104}

\bibitem[\protect\citeauthoryear{{Bertin}, {Mellier}, {Radovich}, {Missonnier},
  {Didelon}  \& {Morin}}{{Bertin} et~al.}{2002}]{2002ASPC..281..228B}
{Bertin} E.,  {Mellier} Y.,  {Radovich} M.,  {Missonnier} G.,  {Didelon} P.,
  {Morin} B.,  2002, in {Bohlender} D.~A.,  {Durand} D.,   {Handley} T.~H.,
  eds,  Astronomical Society of the Pacific Conference Series Vol. 281,
  Astronomical Data Analysis Software and Systems XI. p.~228

\bibitem[\protect\citeauthoryear{{Binney} \& {Tremaine}}{{Binney} \&
  {Tremaine}}{2008}]{2008gady.book.....B}
{Binney} J.,  {Tremaine} S.,  2008, {Galactic Dynamics: Second Edition}.
Princeton University Press

\bibitem[\protect\citeauthoryear{{Buta} et~al.,}{{Buta}
  et~al.}{2015}]{2015ApJS..217...32B}
{Buta} R.~J.,  et~al., 2015, \mn@doi [\apjs] {10.1088/0067-0049/217/2/32},
  \href {https://ui.adsabs.harvard.edu/abs/2015ApJS..217...32B} {217, 32}

\bibitem[\protect\citeauthoryear{{Calzetti} et~al.,}{{Calzetti}
  et~al.}{2005}]{2005ApJ...633..871C}
{Calzetti} D.,  et~al., 2005, \mn@doi [\apj] {10.1086/466518}, \href
  {https://ui.adsabs.harvard.edu/abs/2005ApJ...633..871C} {633, 871}

\bibitem[\protect\citeauthoryear{{Cohen}, {Dame}  \& {Thaddeus}}{{Cohen}
  et~al.}{1986}]{1986ApJS...60..695C}
{Cohen} R.~S.,  {Dame} T.~M.,   {Thaddeus} P.,  1986, \mn@doi [\apjs]
  {10.1086/191101}, \href
  {https://ui.adsabs.harvard.edu/abs/1986ApJS...60..695C} {60, 695}

\bibitem[\protect\citeauthoryear{{Conselice}}{{Conselice}}{2006}]{2006MNRAS.373.1389C}
{Conselice} C.~J.,  2006, \mn@doi [\mnras] {10.1111/j.1365-2966.2006.11114.x},
  \href {https://ui.adsabs.harvard.edu/abs/2006MNRAS.373.1389C} {373, 1389}

\bibitem[\protect\citeauthoryear{{Considere} \& {Athanassoula}}{{Considere} \&
  {Athanassoula}}{1982}]{1982A&A...111...28C}
{Considere} S.,  {Athanassoula} E.,  1982, \aap, \href
  {https://ui.adsabs.harvard.edu/abs/1982A&A...111...28C} {111, 28}

\bibitem[\protect\citeauthoryear{{Dambis} et~al.,}{{Dambis}
  et~al.}{2015}]{2015AstL...41..489D}
{Dambis} A.~K.,  et~al., 2015, \mn@doi [Astronomy Letters]
  {10.1134/S1063773715090017}, \href
  {https://ui.adsabs.harvard.edu/abs/2015AstL...41..489D} {41, 489}

\bibitem[\protect\citeauthoryear{{Davis} \& {Hayes}}{{Davis} \&
  {Hayes}}{2014}]{2014ApJ...790...87D}
{Davis} D.~R.,  {Hayes} W.~B.,  2014, \mn@doi [\apj]
  {10.1088/0004-637X/790/2/87}, \href
  {https://ui.adsabs.harvard.edu/abs/2014ApJ...790...87D} {790, 87}

\bibitem[\protect\citeauthoryear{{Davis}, {Graham}  \& {Seigar}}{{Davis}
  et~al.}{2017}]{2017MNRAS.471.2187D}
{Davis} B.~L.,  {Graham} A.~W.,   {Seigar} M.~S.,  2017, \mn@doi [\mnras]
  {10.1093/mnras/stx1794}, \href
  {https://ui.adsabs.harvard.edu/abs/2017MNRAS.471.2187D} {471, 2187}

\bibitem[\protect\citeauthoryear{{D{\'\i}az-Garc{\'\i}a}, {Salo}, {Knapen}  \&
  {Herrera-Endoqui}}{{D{\'\i}az-Garc{\'\i}a}
  et~al.}{2019}]{2019arXiv190804246D}
{D{\'\i}az-Garc{\'\i}a} S.,  {Salo} H.,  {Knapen} J.~H.,   {Herrera-Endoqui}
  M.,  2019, arXiv e-prints, \href
  {https://ui.adsabs.harvard.edu/abs/2019arXiv190804246D} {p. arXiv:1908.04246}

\bibitem[\protect\citeauthoryear{{Dobbs} \& {Baba}}{{Dobbs} \&
  {Baba}}{2014}]{2014PASA...31...35D}
{Dobbs} C.,  {Baba} J.,  2014, \mn@doi [\pasa] {10.1017/pasa.2014.31}, \href
  {https://ui.adsabs.harvard.edu/abs/2014PASA...31...35D} {31, e035}

\bibitem[\protect\citeauthoryear{{Dobbs}, {Theis}, {Pringle}  \&
  {Bate}}{{Dobbs} et~al.}{2010}]{2010MNRAS.403..625D}
{Dobbs} C.~L.,  {Theis} C.,  {Pringle} J.~E.,   {Bate} M.~R.,  2010, \mn@doi
  [\mnras] {10.1111/j.1365-2966.2009.16161.x}, \href
  {https://ui.adsabs.harvard.edu/abs/2010MNRAS.403..625D} {403, 625}

\bibitem[\protect\citeauthoryear{{Driver}, {Popescu}, {Tuffs}, {Graham},
  {Liske}  \& {Baldry}}{{Driver} et~al.}{2008}]{2008ApJ...678L.101D}
{Driver} S.~P.,  {Popescu} C.~C.,  {Tuffs} R.~J.,  {Graham} A.~W.,  {Liske} J.,
    {Baldry} I.,  2008, \mn@doi [\apjl] {10.1086/588582}, \href
  {https://ui.adsabs.harvard.edu/abs/2008ApJ...678L.101D} {678, L101}

\bibitem[\protect\citeauthoryear{{Elmegreen}}{{Elmegreen}}{1990}]{1990NYASA.596...40E}
{Elmegreen} B.~G.,  1990, \mn@doi [Annals of the New York Academy of Sciences]
  {10.1111/j.1749-6632.1990.tb27410.x}, \href
  {https://ui.adsabs.harvard.edu/abs/1990NYASA.596...40E} {596, 40}

\bibitem[\protect\citeauthoryear{{Elmegreen}}{{Elmegreen}}{2011}]{2011EAS....51...19E}
{Elmegreen} B.~G.,  2011, in {Charbonnel} C.,  {Montmerle} T.,  eds,  EAS
  Publications Series Vol. 51, EAS Publications Series. pp 19--30 (\mn@eprint
  {arXiv} {1101.3109}), \mn@doi{10.1051/eas/1151002}

\bibitem[\protect\citeauthoryear{{Elmegreen}}{{Elmegreen}}{2015}]{2015llg..book..455E}
{Elmegreen} D.~M.,  2015, {Galaxy Morphology at High Redshift}.
p.~455, \mn@doi{10.1007/978-3-319-10614-4_37}

\bibitem[\protect\citeauthoryear{{Elmegreen} \& {Elmegreen}}{{Elmegreen} \&
  {Elmegreen}}{1982}]{1982MNRAS.201.1021E}
{Elmegreen} D.~M.,  {Elmegreen} B.~G.,  1982, \mn@doi [\mnras]
  {10.1093/mnras/201.4.1021}, \href
  {https://ui.adsabs.harvard.edu/abs/1982MNRAS.201.1021E} {201, 1021}

\bibitem[\protect\citeauthoryear{{Elmegreen} \& {Elmegreen}}{{Elmegreen} \&
  {Elmegreen}}{1987}]{1987ApJ...314....3E}
{Elmegreen} D.~M.,  {Elmegreen} B.~G.,  1987, \mn@doi [\apj] {10.1086/165034},
  \href {https://ui.adsabs.harvard.edu/abs/1987ApJ...314....3E} {314, 3}

\bibitem[\protect\citeauthoryear{{Elmegreen}, {Sundin}, {Elmegreen}  \&
  {Sundelius}}{{Elmegreen} et~al.}{1991}]{1991A&A...244...52E}
{Elmegreen} D.~M.,  {Sundin} M.,  {Elmegreen} B.,   {Sundelius} B.,  1991,
  \aap, \href {https://ui.adsabs.harvard.edu/abs/1991A%26A...244...52E} {244,
  52}

\bibitem[\protect\citeauthoryear{{Elmegreen}, {Elmegreen}, {Rubin}  \&
  {Schaffer}}{{Elmegreen} et~al.}{2005}]{2005ApJ...631...85E}
{Elmegreen} D.~M.,  {Elmegreen} B.~G.,  {Rubin} D.~S.,   {Schaffer} M.~A.,
  2005, \mn@doi [\apj] {10.1086/432502}, \href
  {https://ui.adsabs.harvard.edu/abs/2005ApJ...631...85E} {631, 85}

\bibitem[\protect\citeauthoryear{{Elmegreen} et~al.,}{{Elmegreen}
  et~al.}{2011}]{2011ApJ...737...32E}
{Elmegreen} D.~M.,  et~al., 2011, \mn@doi [\apj] {10.1088/0004-637X/737/1/32},
  \href {https://ui.adsabs.harvard.edu/abs/2011ApJ...737...32E} {737, 32}

\bibitem[\protect\citeauthoryear{{Engargiola}, {Plambeck}, {Rosolowsky}  \&
  {Blitz}}{{Engargiola} et~al.}{2003}]{2003ApJS..149..343E}
{Engargiola} G.,  {Plambeck} R.~L.,  {Rosolowsky} E.,   {Blitz} L.,  2003,
  \mn@doi [\apjs] {10.1086/379165}, \href
  {https://ui.adsabs.harvard.edu/abs/2003ApJS..149..343E} {149, 343}

\bibitem[\protect\citeauthoryear{{Font}, {Beckman}, {James}  \&
  {Patsis}}{{Font} et~al.}{2019}]{2019MNRAS.482.5362F}
{Font} J.,  {Beckman} J.~E.,  {James} P.~A.,   {Patsis} P.~A.,  2019, \mn@doi
  [\mnras] {10.1093/mnras/sty2983}, \href
  {https://ui.adsabs.harvard.edu/abs/2019MNRAS.482.5362F} {482, 5362}

\bibitem[\protect\citeauthoryear{{Forgan}, {Ram{\'o}n-Fox}  \&
  {Bonnell}}{{Forgan} et~al.}{2018}]{2018MNRAS.476.2384F}
{Forgan} D.~H.,  {Ram{\'o}n-Fox} F.~G.,   {Bonnell} I.~A.,  2018, \mn@doi
  [\mnras] {10.1093/mnras/sty331}, \href
  {https://ui.adsabs.harvard.edu/abs/2018MNRAS.476.2384F} {476, 2384}

\bibitem[\protect\citeauthoryear{{Foyle}, {Rix}, {Walter}  \& {Leroy}}{{Foyle}
  et~al.}{2010}]{2010ApJ...725..534F}
{Foyle} K.,  {Rix} H.~W.,  {Walter} F.,   {Leroy} A.~K.,  2010, \mn@doi [\apj]
  {10.1088/0004-637X/725/1/534}, \href
  {https://ui.adsabs.harvard.edu/abs/2010ApJ...725..534F} {725, 534}

\bibitem[\protect\citeauthoryear{{Freeman}}{{Freeman}}{1970}]{1970ApJ...160..811F}
{Freeman} K.~C.,  1970, \mn@doi [\apj] {10.1086/150474}, \href
  {https://ui.adsabs.harvard.edu/abs/1970ApJ...160..811F} {160, 811}

\bibitem[\protect\citeauthoryear{{Georgelin} \& {Georgelin}}{{Georgelin} \&
  {Georgelin}}{1976}]{1976A&A....49...57G}
{Georgelin} Y.~M.,  {Georgelin} Y.~P.,  1976, \aap, \href
  {https://ui.adsabs.harvard.edu/abs/1976A%26A....49...57G} {49, 57}

\bibitem[\protect\citeauthoryear{{Gerola} \& {Seiden}}{{Gerola} \&
  {Seiden}}{1978}]{1978ApJ...223..129G}
{Gerola} H.,  {Seiden} P.~E.,  1978, \mn@doi [\apj] {10.1086/156243}, \href
  {https://ui.adsabs.harvard.edu/abs/1978ApJ...223..129G} {223, 129}

\bibitem[\protect\citeauthoryear{{Goldreich} \& {Tremaine}}{{Goldreich} \&
  {Tremaine}}{1978}]{1978ApJ...222..850G}
{Goldreich} P.,  {Tremaine} S.,  1978, \mn@doi [\apj] {10.1086/156203}, \href
  {https://ui.adsabs.harvard.edu/abs/1978ApJ...222..850G} {222, 850}

\bibitem[\protect\citeauthoryear{{Grabelsky}, {Cohen}, {Bronfman}, {Thaddeus}
  \& {May}}{{Grabelsky} et~al.}{1987}]{1987ApJ...315..122G}
{Grabelsky} D.~A.,  {Cohen} R.~S.,  {Bronfman} L.,  {Thaddeus} P.,   {May} J.,
  1987, \mn@doi [\apj] {10.1086/165118}, \href
  {https://ui.adsabs.harvard.edu/abs/1987ApJ...315..122G} {315, 122}

\bibitem[\protect\citeauthoryear{{Grosb{\o}l} \& {Dottori}}{{Grosb{\o}l} \&
  {Dottori}}{2012}]{2012A&A...542A..39G}
{Grosb{\o}l} P.,  {Dottori} H.,  2012, \mn@doi [\aap]
  {10.1051/0004-6361/201118099}, \href
  {https://ui.adsabs.harvard.edu/abs/2012A%26A...542A..39G} {542, A39}

\bibitem[\protect\citeauthoryear{{Harsoula} \& {Kalapotharakos}}{{Harsoula} \&
  {Kalapotharakos}}{2009}]{2009MNRAS.394.1605H}
{Harsoula} M.,  {Kalapotharakos} C.,  2009, \mn@doi [\mnras]
  {10.1111/j.1365-2966.2009.14427.x}, \href
  {https://ui.adsabs.harvard.edu/abs/2009MNRAS.394.1605H} {394, 1605}

\bibitem[\protect\citeauthoryear{{Hart} et~al.,}{{Hart}
  et~al.}{2016}]{2016MNRAS.461.3663H}
{Hart} R.~E.,  et~al., 2016, \mn@doi [\mnras] {10.1093/mnras/stw1588}, \href
  {https://ui.adsabs.harvard.edu/abs/2016MNRAS.461.3663H} {461, 3663}

\bibitem[\protect\citeauthoryear{{Hart}, {Bamford}, {Casteels}, {Kruk},
  {Lintott}  \& {Masters}}{{Hart} et~al.}{2017a}]{2017MNRAS.468.1850H}
{Hart} R.~E.,  {Bamford} S.~P.,  {Casteels} K.~R.~V.,  {Kruk} S.~J.,  {Lintott}
  C.~J.,   {Masters} K.~L.,  2017a, \mn@doi [\mnras] {10.1093/mnras/stx581},
  \href {https://ui.adsabs.harvard.edu/abs/2017MNRAS.468.1850H} {468, 1850}

\bibitem[\protect\citeauthoryear{{Hart} et~al.,}{{Hart}
  et~al.}{2017b}]{2017MNRAS.472.2263H}
{Hart} R.~E.,  et~al., 2017b, \mn@doi [\mnras] {10.1093/mnras/stx2137}, \href
  {https://ui.adsabs.harvard.edu/abs/2017MNRAS.472.2263H} {472, 2263}

\bibitem[\protect\citeauthoryear{{Holmberg}}{{Holmberg}}{1941}]{1941ApJ....94..385H}
{Holmberg} E.,  1941, \mn@doi [\apj] {10.1086/144344}, \href
  {https://ui.adsabs.harvard.edu/abs/1941ApJ....94..385H} {94, 385}

\bibitem[\protect\citeauthoryear{{Holwerda}, {Gonz{\'a}lez}, {van der Kruit}
  \& {Allen}}{{Holwerda} et~al.}{2005}]{2005A&A...444..109H}
{Holwerda} B.~W.,  {Gonz{\'a}lez} R.~A.,  {van der Kruit} P.~C.,   {Allen}
  R.~J.,  2005, \mn@doi [\aap] {10.1051/0004-6361:20053013}, \href
  {https://ui.adsabs.harvard.edu/abs/2005A%26A...444..109H} {444, 109}

\bibitem[\protect\citeauthoryear{{Honig} \& {Reid}}{{Honig} \&
  {Reid}}{2015}]{2015ApJ...800...53H}
{Honig} Z.~N.,  {Reid} M.~J.,  2015, \mn@doi [\apj]
  {10.1088/0004-637X/800/1/53}, \href
  {https://ui.adsabs.harvard.edu/abs/2015ApJ...800...53H} {800, 53}

\bibitem[\protect\citeauthoryear{{Hou}, {Han}  \& {Shi}}{{Hou}
  et~al.}{2009}]{2009A&A...499..473H}
{Hou} L.~G.,  {Han} J.~L.,   {Shi} W.~B.,  2009, \mn@doi [\aap]
  {10.1051/0004-6361/200809692}, \href
  {https://ui.adsabs.harvard.edu/abs/2009A%26A...499..473H} {499, 473}

\bibitem[\protect\citeauthoryear{{Hubble}}{{Hubble}}{1936}]{1936rene.book.....H}
{Hubble} E.~P.,  1936, {Realm of the Nebulae}

\bibitem[\protect\citeauthoryear{{Jedrzejewski}}{{Jedrzejewski}}{1987}]{1987MNRAS.226..747J}
{Jedrzejewski} R.~I.,  1987, \mn@doi [\mnras] {10.1093/mnras/226.4.747}, \href
  {https://ui.adsabs.harvard.edu/abs/1987MNRAS.226..747J} {226, 747}

\bibitem[\protect\citeauthoryear{{Julian} \& {Toomre}}{{Julian} \&
  {Toomre}}{1966}]{1966ApJ...146..810J}
{Julian} W.~H.,  {Toomre} A.,  1966, \mn@doi [\apj] {10.1086/148957}, \href
  {https://ui.adsabs.harvard.edu/abs/1966ApJ...146..810J} {146, 810}

\bibitem[\protect\citeauthoryear{{Kalnajs}}{{Kalnajs}}{1973}]{1973PASAu...2..174K}
{Kalnajs} A.~J.,  1973, \mn@doi [Proceedings of the Astronomical Society of
  Australia] {10.1017/S1323358000013461}, \href
  {https://ui.adsabs.harvard.edu/abs/1973PASAu...2..174K} {2, 174}

\bibitem[\protect\citeauthoryear{{Karapetyan}, {Hakobyan}, {Barkhudaryan},
  {Mamon}, {Kunth}, {Adibekyan}  \& {Turatto}}{{Karapetyan}
  et~al.}{2018}]{2018MNRAS.481..566K}
{Karapetyan} A.~G.,  {Hakobyan} A.~A.,  {Barkhudaryan} L.~V.,  {Mamon} G.~A.,
  {Kunth} D.,  {Adibekyan} V.,   {Turatto} M.,  2018, \mn@doi [\mnras]
  {10.1093/mnras/sty2291}, \href
  {https://ui.adsabs.harvard.edu/abs/2018MNRAS.481..566K} {481, 566}

\bibitem[\protect\citeauthoryear{{Kaufmann} \& {Contopoulos}}{{Kaufmann} \&
  {Contopoulos}}{1996}]{1996A&A...309..381K}
{Kaufmann} D.~E.,  {Contopoulos} G.,  1996, \aap, \href
  {https://ui.adsabs.harvard.edu/abs/1996A%26A...309..381K} {309, 381}

\bibitem[\protect\citeauthoryear{{Kendall}, {Kennicutt}, {Clarke}  \&
  {Thornley}}{{Kendall} et~al.}{2008}]{2008MNRAS.387.1007K}
{Kendall} S.,  {Kennicutt} R.~C.,  {Clarke} C.,   {Thornley} M.~D.,  2008,
  \mn@doi [\mnras] {10.1111/j.1365-2966.2008.13327.x}, \href
  {https://ui.adsabs.harvard.edu/abs/2008MNRAS.387.1007K} {387, 1007}

\bibitem[\protect\citeauthoryear{{Kendall}, {Kennicutt}  \& {Clarke}}{{Kendall}
  et~al.}{2011}]{2011MNRAS.414..538K}
{Kendall} S.,  {Kennicutt} R.~C.,   {Clarke} C.,  2011, \mn@doi [\mnras]
  {10.1111/j.1365-2966.2011.18422.x}, \href
  {https://ui.adsabs.harvard.edu/abs/2011MNRAS.414..538K} {414, 538}

\bibitem[\protect\citeauthoryear{{Kendall}, {Clarke}  \& {Kennicutt}}{{Kendall}
  et~al.}{2015}]{2015MNRAS.446.4155K}
{Kendall} S.,  {Clarke} C.,   {Kennicutt} R.~C.,  2015, \mn@doi [\mnras]
  {10.1093/mnras/stu2431}, \href
  {https://ui.adsabs.harvard.edu/abs/2015MNRAS.446.4155K} {446, 4155}

\bibitem[\protect\citeauthoryear{{Kennicutt}}{{Kennicutt}}{1981}]{1981AJ.....86.1847K}
{Kennicutt} R.~C. J.,  1981, \mn@doi [\aj] {10.1086/113064}, \href
  {https://ui.adsabs.harvard.edu/abs/1981AJ.....86.1847K} {86, 1847}

\bibitem[\protect\citeauthoryear{{Kormendy} \& {Norman}}{{Kormendy} \&
  {Norman}}{1979}]{1979ApJ...233..539K}
{Kormendy} J.,  {Norman} C.~A.,  1979, \mn@doi [\apj] {10.1086/157414}, \href
  {https://ui.adsabs.harvard.edu/abs/1979ApJ...233..539K} {233, 539}

\bibitem[\protect\citeauthoryear{{Law}, {Shapley}, {Steidel}, {Reddy},
  {Christensen}  \& {Erb}}{{Law} et~al.}{2012}]{2012Natur.487..338L}
{Law} D.~R.,  {Shapley} A.~E.,  {Steidel} C.~C.,  {Reddy} N.~A.,  {Christensen}
  C.~R.,   {Erb} D.~K.,  2012, \mn@doi [\nat] {10.1038/nature11256}, \href
  {https://ui.adsabs.harvard.edu/abs/2012Natur.487..338L} {487, 338}

\bibitem[\protect\citeauthoryear{{Lin} \& {Shu}}{{Lin} \&
  {Shu}}{1964}]{1964ApJ...140..646L}
{Lin} C.~C.,  {Shu} F.~H.,  1964, \mn@doi [\apj] {10.1086/147955}, \href
  {https://ui.adsabs.harvard.edu/abs/1964ApJ...140..646L} {140, 646}

\bibitem[\protect\citeauthoryear{{Lin} \& {Shu}}{{Lin} \&
  {Shu}}{1967}]{1967IAUS...31..313L}
{Lin} C.~C.,  {Shu} F.~H.,  1967, in {van Woerden} H.,  ed.,  IAU Symposium
  Vol. 31, Radio Astronomy and the Galactic System. p.~313

\bibitem[\protect\citeauthoryear{{Lintott} et~al.,}{{Lintott}
  et~al.}{2008}]{2008MNRAS.389.1179L}
{Lintott} C.~J.,  et~al., 2008, \mn@doi [\mnras]
  {10.1111/j.1365-2966.2008.13689.x}, \href
  {https://ui.adsabs.harvard.edu/abs/2008MNRAS.389.1179L} {389, 1179}

\bibitem[\protect\citeauthoryear{{Lynds}}{{Lynds}}{1970}]{1970IAUS...38...26L}
{Lynds} B.~T.,  1970, in {Becker} W.,  {Kontopoulos} G.~I.,  eds,  IAU
  Symposium Vol. 38, The Spiral Structure of our Galaxy. p.~26

\bibitem[\protect\citeauthoryear{{Ma}}{{Ma}}{2001}]{2001ChJAA...1..395M}
{Ma} J.,  2001, \mn@doi [\cjaa] {10.1088/1009-9271/1/5/395}, \href
  {https://ui.adsabs.harvard.edu/abs/2001ChJAA...1..395M} {1}

\bibitem[\protect\citeauthoryear{{Ma}}{{Ma}}{2002}]{2002A&A...388..389M}
{Ma} J.,  2002, \mn@doi [\aap] {10.1051/0004-6361:20020414}, \href
  {https://ui.adsabs.harvard.edu/abs/2002A%26A...388..389M} {388, 389}

\bibitem[\protect\citeauthoryear{{Makarov}, {Prugniel}, {Terekhova}, {Courtois}
   \& {Vauglin}}{{Makarov} et~al.}{2014}]{2014A&A...570A..13M}
{Makarov} D.,  {Prugniel} P.,  {Terekhova} N.,  {Courtois} H.,   {Vauglin} I.,
  2014, \mn@doi [\aap] {10.1051/0004-6361/201423496}, \href
  {https://ui.adsabs.harvard.edu/abs/2014A%26A...570A..13M} {570, A13}

\bibitem[\protect\citeauthoryear{{Masters} et~al.,}{{Masters}
  et~al.}{2010}]{2010MNRAS.405..783M}
{Masters} K.~L.,  et~al., 2010, \mn@doi [\mnras]
  {10.1111/j.1365-2966.2010.16503.x}, \href
  {https://ui.adsabs.harvard.edu/abs/2010MNRAS.405..783M} {405, 783}

\bibitem[\protect\citeauthoryear{{Masters} et~al.,}{{Masters}
  et~al.}{2019}]{2019MNRAS.487.1808M}
{Masters} K.~L.,  et~al., 2019, \mn@doi [\mnras] {10.1093/mnras/stz1153}, \href
  {https://ui.adsabs.harvard.edu/abs/2019MNRAS.487.1808M} {487, 1808}

\bibitem[\protect\citeauthoryear{{Mu{\~n}oz-Mateos} et~al.,}{{Mu{\~n}oz-Mateos}
  et~al.}{2015}]{2015ApJS..219....3M}
{Mu{\~n}oz-Mateos} J.~C.,  et~al., 2015, \mn@doi [\apjs]
  {10.1088/0067-0049/219/1/3}, \href
  {https://ui.adsabs.harvard.edu/abs/2015ApJS..219....3M} {219, 3}

\bibitem[\protect\citeauthoryear{{Oh}, {Kim}, {Lee}  \& {Kim}}{{Oh}
  et~al.}{2008}]{2008ApJ...683...94O}
{Oh} S.~H.,  {Kim} W.-T.,  {Lee} H.~M.,   {Kim} J.,  2008, \mn@doi [\apj]
  {10.1086/588184}, \href
  {https://ui.adsabs.harvard.edu/abs/2008ApJ...683...94O} {683, 94}

\bibitem[\protect\citeauthoryear{{Price-Whelan} et~al.,}{{Price-Whelan}
  et~al.}{2018}]{astropy:2018}
{Price-Whelan} A.~M.,  et~al., 2018, \mn@doi [\aj] {10.3847/1538-3881/aabc4f},
  \href {https://ui.adsabs.harvard.edu/#abs/2018AJ....156..123T} {156, 123}

\bibitem[\protect\citeauthoryear{{Puerari} \& {Dottori}}{{Puerari} \&
  {Dottori}}{1992}]{1992A&AS...93..469P}
{Puerari} I.,  {Dottori} H.~A.,  1992, \aaps, \href
  {https://ui.adsabs.harvard.edu/abs/1992A&AS...93..469P} {93, 469}

\bibitem[\protect\citeauthoryear{{Puerari}, {Elmegreen}  \& {Block}}{{Puerari}
  et~al.}{2014}]{2014AJ....148..133P}
{Puerari} I.,  {Elmegreen} B.~G.,   {Block} D.~L.,  2014, \mn@doi [\aj]
  {10.1088/0004-6256/148/6/133}, \href
  {https://ui.adsabs.harvard.edu/abs/2014AJ....148..133P} {148, 133}

\bibitem[\protect\citeauthoryear{{Regan} \& {Wilson}}{{Regan} \&
  {Wilson}}{1993}]{1993AJ....105..499R}
{Regan} M.~W.,  {Wilson} C.~D.,  1993, \mn@doi [\aj] {10.1086/116448}, \href
  {https://ui.adsabs.harvard.edu/abs/1993AJ....105..499R} {105, 499}

\bibitem[\protect\citeauthoryear{{Reid} et~al.,}{{Reid}
  et~al.}{2014}]{2014ApJ...783..130R}
{Reid} M.~J.,  et~al., 2014, \mn@doi [\apj] {10.1088/0004-637X/783/2/130},
  \href {https://ui.adsabs.harvard.edu/abs/2014ApJ...783..130R} {783, 130}

\bibitem[\protect\citeauthoryear{{Salo} \& {Laurikainen}}{{Salo} \&
  {Laurikainen}}{2000}]{2000MNRAS.319..377S}
{Salo} H.,  {Laurikainen} E.,  2000, \mn@doi [\mnras]
  {10.1046/j.1365-8711.2000.03650.x}, \href
  {https://ui.adsabs.harvard.edu/abs/2000MNRAS.319..377S} {319, 377}

\bibitem[\protect\citeauthoryear{{S{\'a}nchez-Menguiano}
  et~al.,}{{S{\'a}nchez-Menguiano} et~al.}{2017}]{2017A&A...603A.113S}
{S{\'a}nchez-Menguiano} L.,  et~al., 2017, \mn@doi [\aap]
  {10.1051/0004-6361/201630062}, \href
  {https://ui.adsabs.harvard.edu/abs/2017A%26A...603A.113S} {603, A113}

\bibitem[\protect\citeauthoryear{{Savchenko} \& {Reshetnikov}}{{Savchenko} \&
  {Reshetnikov}}{2013}]{2013MNRAS.436.1074S}
{Savchenko} S.~S.,  {Reshetnikov} V.~P.,  2013, \mn@doi [\mnras]
  {10.1093/mnras/stt1627}, \href
  {https://ui.adsabs.harvard.edu/abs/2013MNRAS.436.1074S} {436, 1074}

\bibitem[\protect\citeauthoryear{{Schlafly} \& {Finkbeiner}}{{Schlafly} \&
  {Finkbeiner}}{2011}]{2011ApJ...737..103S}
{Schlafly} E.~F.,  {Finkbeiner} D.~P.,  2011, \mn@doi [\apj]
  {10.1088/0004-637X/737/2/103}, \href
  {https://ui.adsabs.harvard.edu/abs/2011ApJ...737..103S} {737, 103}

\bibitem[\protect\citeauthoryear{{Seiden} \& {Gerola}}{{Seiden} \&
  {Gerola}}{1979}]{1979ApJ...233...56S}
{Seiden} P.~E.,  {Gerola} H.,  1979, \mn@doi [\apj] {10.1086/157366}, \href
  {https://ui.adsabs.harvard.edu/abs/1979ApJ...233...56S} {233, 56}

\bibitem[\protect\citeauthoryear{{Seigar} \& {James}}{{Seigar} \&
  {James}}{1998}]{1998MNRAS.299..685S}
{Seigar} M.~S.,  {James} P.~A.,  1998, \mn@doi [\mnras]
  {10.1046/j.1365-8711.1998.01779.x}, \href
  {https://ui.adsabs.harvard.edu/abs/1998MNRAS.299..685S} {299, 685}

\bibitem[\protect\citeauthoryear{{Seigar}, {Block}, {Puerari}, {Chorney}  \&
  {James}}{{Seigar} et~al.}{2005}]{2005MNRAS.359.1065S}
{Seigar} M.~S.,  {Block} D.~L.,  {Puerari} I.,  {Chorney} N.~E.,   {James}
  P.~A.,  2005, \mn@doi [\mnras] {10.1111/j.1365-2966.2005.08970.x}, \href
  {https://ui.adsabs.harvard.edu/abs/2005MNRAS.359.1065S} {359, 1065}

\bibitem[\protect\citeauthoryear{{Seigar}, {Bullock}, {Barth}  \&
  {Ho}}{{Seigar} et~al.}{2006}]{2006ApJ...645.1012S}
{Seigar} M.~S.,  {Bullock} J.~S.,  {Barth} A.~J.,   {Ho} L.~C.,  2006, \mn@doi
  [\apj] {10.1086/504463}, \href
  {https://ui.adsabs.harvard.edu/abs/2006ApJ...645.1012S} {645, 1012}

\bibitem[\protect\citeauthoryear{{Seigar}, {Kennefick}, {Kennefick}  \&
  {Lacy}}{{Seigar} et~al.}{2008}]{2008ApJ...678L..93S}
{Seigar} M.~S.,  {Kennefick} D.,  {Kennefick} J.,   {Lacy} C.~H.~S.,  2008,
  \mn@doi [\apjl] {10.1086/588727}, \href
  {https://ui.adsabs.harvard.edu/abs/2008ApJ...678L..93S} {678, L93}

\bibitem[\protect\citeauthoryear{{Sheth} et~al.,}{{Sheth}
  et~al.}{2010}]{2010PASP..122.1397S}
{Sheth} K.,  et~al., 2010, \mn@doi [\pasp] {10.1086/657638}, \href
  {https://ui.adsabs.harvard.edu/abs/2010PASP..122.1397S} {122, 1397}

\bibitem[\protect\citeauthoryear{{Shields} et~al.,}{{Shields}
  et~al.}{2015}]{2015arXiv151106365S}
{Shields} D.~W.,  et~al., 2015, arXiv e-prints, \href
  {https://ui.adsabs.harvard.edu/abs/2015arXiv151106365S} {p. arXiv:1511.06365}

\bibitem[\protect\citeauthoryear{{Skowron} et~al.,}{{Skowron}
  et~al.}{2019}]{2019Sci...365..478S}
{Skowron} D.~M.,  et~al., 2019, \mn@doi [Science] {10.1126/science.aau3181},
  \href {https://ui.adsabs.harvard.edu/abs/2019Sci...365..478S} {365, 478}

\bibitem[\protect\citeauthoryear{{Tempel}, {Tago}  \& {Liivam{\"a}gi}}{{Tempel}
  et~al.}{2012}]{2012A&A...540A.106T}
{Tempel} E.,  {Tago} E.,   {Liivam{\"a}gi} L.~J.,  2012, \mn@doi [\aap]
  {10.1051/0004-6361/201118687}, \href
  {https://ui.adsabs.harvard.edu/abs/2012A&A...540A.106T} {540, A106}

\bibitem[\protect\citeauthoryear{{Thornley}}{{Thornley}}{1996}]{1996ApJ...469L..45T}
{Thornley} M.~D.,  1996, \mn@doi [\apjl] {10.1086/310250}, \href
  {https://ui.adsabs.harvard.edu/abs/1996ApJ...469L..45T} {469, L45}

\bibitem[\protect\citeauthoryear{{Toomre}}{{Toomre}}{1969}]{1969ApJ...158..899T}
{Toomre} A.,  1969, \mn@doi [\apj] {10.1086/150250}, \href
  {https://ui.adsabs.harvard.edu/abs/1969ApJ...158..899T} {158, 899}

\bibitem[\protect\citeauthoryear{{Toomre}}{{Toomre}}{1981}]{1981seng.proc..111T}
{Toomre} A.,  1981, in {Fall} S.~M.,  {Lynden-Bell} D.,  eds, Structure and
  Evolution of Normal Galaxies. pp 111--136

\bibitem[\protect\citeauthoryear{{Toomre} \& {Toomre}}{{Toomre} \&
  {Toomre}}{1972}]{1972ApJ...178..623T}
{Toomre} A.,  {Toomre} J.,  1972, \mn@doi [\apj] {10.1086/151823}, \href
  {https://ui.adsabs.harvard.edu/abs/1972ApJ...178..623T} {178, 623}

\bibitem[\protect\citeauthoryear{{Unterborn} \& {Ryden}}{{Unterborn} \&
  {Ryden}}{2008}]{2008ApJ...687..976U}
{Unterborn} C.~T.,  {Ryden} B.~S.,  2008, \mn@doi [\apj] {10.1086/591898},
  \href {https://ui.adsabs.harvard.edu/abs/2008ApJ...687..976U} {687, 976}

\bibitem[\protect\citeauthoryear{{Vall{\'e}e}}{{Vall{\'e}e}}{2005}]{2005AJ....130..569V}
{Vall{\'e}e} J.~P.,  2005, \mn@doi [\aj] {10.1086/431744}, \href
  {https://ui.adsabs.harvard.edu/abs/2005AJ....130..569V} {130, 569}

\bibitem[\protect\citeauthoryear{{Verley} et~al.,}{{Verley}
  et~al.}{2007}]{2007A&A...472..121V}
{Verley} S.,  et~al., 2007, \mn@doi [\aap] {10.1051/0004-6361:20077481}, \href
  {https://ui.adsabs.harvard.edu/abs/2007A&A...472..121V} {472, 121}

\bibitem[\protect\citeauthoryear{{Vogel}, {Kulkarni}  \& {Scoville}}{{Vogel}
  et~al.}{1988}]{1988Natur.334..402V}
{Vogel} S.~N.,  {Kulkarni} S.~R.,   {Scoville} N.~Z.,  1988, \mn@doi [\nat]
  {10.1038/334402a0}, \href
  {https://ui.adsabs.harvard.edu/abs/1988Natur.334..402V} {334, 402}

\bibitem[\protect\citeauthoryear{{Wiklind}, {Rydbeck}, {Hjalmarson}  \&
  {Bergman}}{{Wiklind} et~al.}{1990}]{1990A&A...232L..11W}
{Wiklind} T.,  {Rydbeck} G.,  {Hjalmarson} A.,   {Bergman} P.,  1990, \aap,
  \href {https://ui.adsabs.harvard.edu/abs/1990A%26A...232L..11W} {232, L11}

\bibitem[\protect\citeauthoryear{{York} et~al.,}{{York}
  et~al.}{2000}]{2000AJ....120.1579Y}
{York} D.~G.,  et~al., 2000, \mn@doi [\aj] {10.1086/301513}, \href
  {https://ui.adsabs.harvard.edu/abs/2000AJ....120.1579Y} {120, 1579}

\bibitem[\protect\citeauthoryear{{Yu} \& {Ho}}{{Yu} \&
  {Ho}}{2019}]{2019ApJ...871..194Y}
{Yu} S.-Y.,  {Ho} L.~C.,  2019, \mn@doi [\apj] {10.3847/1538-4357/aaf895},
  \href {https://ui.adsabs.harvard.edu/abs/2019ApJ...871..194Y} {871, 194}

\bibitem[\protect\citeauthoryear{{Yu}, {Ho}, {Barth}  \& {Li}}{{Yu}
  et~al.}{2018}]{2018ApJ...862...13Y}
{Yu} S.-Y.,  {Ho} L.~C.,  {Barth} A.~J.,   {Li} Z.-Y.,  2018, \mn@doi [\apj]
  {10.3847/1538-4357/aacb25}, \href
  {https://ui.adsabs.harvard.edu/abs/2018ApJ...862...13Y} {862, 13}

\bibitem[\protect\citeauthoryear{{Yuan} \& {Grosbol}}{{Yuan} \&
  {Grosbol}}{1981}]{1981ApJ...243..432Y}
{Yuan} C.,  {Grosbol} P.,  1981, \mn@doi [\apj] {10.1086/158610}, \href
  {https://ui.adsabs.harvard.edu/abs/1981ApJ...243..432Y} {243, 432}

\bibitem[\protect\citeauthoryear{{de Lapparent}, {Baillard}  \& {Bertin}}{{de
  Lapparent} et~al.}{2011}]{2011A&A...532A..75D}
{de Lapparent} V.,  {Baillard} A.,   {Bertin} E.,  2011, \mn@doi [\aap]
  {10.1051/0004-6361/201016424}, \href
  {https://ui.adsabs.harvard.edu/abs/2011A&A...532A..75D} {532, A75}

\makeatother
\end{thebibliography}

\bsp	
\label{lastpage}
\end{document}